\documentclass[preprint,11pt]{elsarticle}

\usepackage{amssymb}
\usepackage{epsfig}
\usepackage{amsmath,amssymb}
\usepackage{graphicx}
\usepackage{slashed}
\usepackage[T1]{fontenc}
\usepackage{color}
\usepackage{hyperref}
\usepackage{times}
\usepackage{overpic}
\usepackage{rotating}

\journal{Prog. Part. Nucl. Phys.}

\begin{document}

\begin{frontmatter}



\title{Functional renormalization group studies \\of nuclear and neutron matter}


\author{Matthias Drews$^a$, Wolfram Weise$^{a,b}$}

\address[label1]{Physik-Department, Technische Universit\"at M\"unchen, D-85748 Garching, Germany}
\address[label2]{Kavli Institute for Theoretical Physics, University of California, Santa Barbara, \\CA 93106-4030,USA}

\begin{abstract}
Functional renormalization group (FRG) methods applied to calculations of isospin-symmetric and asymmetric nuclear matter as well as neutron matter are reviewed. The approach is based on a chiral Lagrangian expressed in terms of nucleon and meson degrees of freedom as appropriate for the hadronic phase of QCD with spontaneously broken chiral symmetry. Fluctuations beyond mean-field approximation are treated solving Wetterich's FRG flow equations. Nuclear thermodynamics and the nuclear liquid-gas phase transition are investigated in detail, both in symmetric matter and as a function of the proton fraction in asymmetric matter. The equations of state at zero temperature of symmetric nuclear matter and pure neutron matter are found to be in good agreement with advanced ab-initio many-body computations. Contacts with perturbative many-body approaches (in-medium chiral perturbation theory) are discussed.  As an interesting test case, the density dependence of the pion mass in the medium is investigated. The question of chiral symmetry restoration in nuclear and neutron matter is addressed.  A stabilization of the phase with spontaneously broken chiral symmetry is found to persist up to high baryon densities once fluctuations beyond mean-field are included. Neutron star matter including beta equilibrium is discussed under the aspect of the constraints imposed by the existence of two-solar-mass neutron stars.
\end{abstract}


\end{frontmatter}


\newpage
\section{Introductory survey: \\Scales, symmetry breaking patterns and strategies in low-energy QCD}
\label{intro}
Understanding nuclei and dense baryonic matter at the interface with the underlying theory of the strong interaction, quantum chromodynamics (QCD), is one of the pending challenges in nuclear many-body theory. The present review focuses on this issue, by combining basic symmetry principles and the symmetry breaking pattern of low-energy QCD with a functional renormalization group (FRG) approach. In this context, the FRG offers an efficient non-perturbative method for investigating important fluctuations beyond mean-field approximation. But before entering this main theme, it is useful for general guidance to start with a brief survey of the role that spontaneous chiral symmetry breaking plays in defining scales, identifying relevant degrees of freedom, designing instructive models and introducing chiral effective field theory for applications to nuclear systems.
 
QCD starts out with quarks and gluons as fundamental fermion and gauge boson degrees of freedom. At momentum scales exceeding several GeV (corresponding to short distance scales, $r < 0.1$ fm), QCD is indeed a theory of weakly-coupled quarks and gluons for which perturbative methods are applicable. On the other hand, in the low-energy limit and at momentum scales well below 1 GeV (corresponding to distances relevant for nuclear physics, $r > 1$ fm), QCD is governed by color confinement and dynamical (spontaneous) breaking of chiral symmetry, a global symmetry of QCD that is exact in the limit of massless quarks and explicitly broken by the non-zero quark masses generated at the Higgs scale.   

Investigations of lattice QCD thermodynamics at zero quark chemical potential have arrived at the following conclusions.  A transition from color confinement in hadrons to a deconfined quark-gluon phase is signalled by the behaviour of the Polyakov loop, accompanied by rapid changes of the energy density and of related quantities \cite{Petreczky2012,Bazavov2014,Borsanyi2014}. Furthermore the temperature dependence of the chiral (quark) condensate, $\langle\bar{q}q\rangle$, displays a transition from the unbroken (Wigner-Weyl) realisation of chiral symmetry to the Nambu-Goldstone realisation with spontaneously broken symmetry \cite{Borsanyi2010,Bazavov2012}. These two transitions (not phase transitions but continuous crossovers) take place in the same narrow temperature range, $T_c \sim (0.15 - 0.2)$ GeV. \footnote{The transition temperature for the chiral crossover with 2+1 quark flavours is reported to be $T_c\simeq 155$ MeV  \cite{Borsanyi2010}.}

Spontaneous chiral symmetry breaking implies the existence of pseudoscalar Nambu-Goldstone (NG) bosons. In the limit of exact chiral symmetry with zero-mass quarks these NG bosons are massless. For two quark flavors ($u$ and $d$) with nearly zero masses, the Goldstone bosons of chiral symmetry are identified with the isospin triplet of pions. Explicit chiral symmetry breaking by the non-zero but small $u$ and $d$ quark masses ($m_q = (m_u + m_d)/2 = 3.5^{+0.7}_{-0.2}$ MeV at a renormalization scale of 2 GeV \cite{PDG2014}) moves the pion mass to its physical value\footnote{... modulo electromagnetic effects.}, $m_{\pi^\pm} = 139.6$ MeV, with $m_\pi^2$ proportional to $m_q$ in leading order. Low-energy QCD in the matter-free vacuum is thus realized as a (chiral) effective field theory of the active, light degrees of freedom: the pions as pseudo-Goldstone bosons\footnote{The term {\it pseudo-Goldstone boson} is often used \cite{weinbergbook} to characterize a spontaneously broken {\it approximate} symmetry such as chiral symmetry in QCD. For simplicity we shall refer to pions simply as Nambu-Goldstone bosons in the following.}. In the low-energy, long-wavelength limit,
Nambu-Goldstone bosons have the property that their interactions with one another and with any
massive hadron are weak, governed by couplings proportional to space-time derivatives of the NG boson fields plus small symmetry-breaking mass terms. In this limit perturbative methods can be applied (chiral perturbation theory - ChPT), based on a systematic expansion in powers of momentum, energy and quark masses as "small" parameters \cite{Wei66/67,GL84}.

In the low-energy and low-density domain accessed by currently available experimental probes, chiral effective field theory as outlined above in terms of Goldstone boson fields {\it alone}
provides a sufficiently powerful framework also for a highly successful description of nuclear interactions \cite{evgenireview, hammerreview,machleidtreview}. This is the starting point for a systematic approach to nuclear
many-body dynamics and thermodynamics at densities and temperatures well within the
hadronic, "confined" phase of QCD \cite{hkwreview,hrw2016}\footnote{An alternative effective field theory incorporating hidden local symmetry and vector meson fields is reviewed and discussed in \cite{hrw2016}.}.

\subsection{\it Chiral symmetry and the pion}

Within this introductory survey, let us now briefly recall how the special role of the pion emerges through the Nambu-Goldstone mechanism of spontaneous chiral symmetry breaking in QCD. Historically, our understanding of the pion as an NG boson
emerged \cite{G61,NJL61} in the pre-QCD era of the 1960's and culminated in the current algebra approach \cite{AD68} (combined with the PCAC relation for the pion). Inspired by the BCS theory of superconductivity, Nambu and Jona-Lasinio (NJL) \cite{NJL61} developed a model that helped clarify the dynamics behind spontaneous chiral symmetry breaking and the formation of pions as Nambu-Goldstone bosons. Following the logic of the NJL model gives a hint at the link to the underlying QCD degrees of freedom. 

We start from the color current of quarks, ${\bf J}_\mu^a(x) = \bar{q}(x)
\gamma_\mu {\bf t}^a q(x)$, where $q(x)$ denotes the quark fields with $4N_c N_f$ components
representing their spin, color and flavor degrees freedom $(N_c=3, N_f=2)$ and
${\bf t}^a$ ($a = 1, ... ,8$) are the generators of the
$SU(3)_c$ color gauge group. One can make the additional ansatz that the distance over which color
propagates is restricted to a short correlation length $l_c$. Then the interaction between
low-momentum quarks, mediated by the coupling of the quark color current to the gluon fields,
can be schematically viewed as a local coupling between their color currents:
\begin{equation}
{\cal L}_{\rm int} = -G_c\,{\bf J}_\mu^a(x)\,{\bf J}^\mu_a(x)\,\, ,
\label{eq:Lint1}
\end{equation}
where $G_c \sim g_c^2\, l_c^2$ represents an effective coupling strength proportional to the square of the dimensionless QCD coupling, $g_c$, averaged over the relevant distance
scale\footnote{Such a scale is actually supposed to be related to the trace anomaly in QCD.}, in combination with the squared correlation length.

Given the local interaction in Eq.\,(\ref{eq:Lint1}), a model Lagrangian for the quark fields,
\begin{equation}
{\cal L} = \bar{q}(x)(i\gamma^\mu\partial_\mu - m_q)q(x) + {\cal L}_{\rm int}(\bar{q},q)~,
\label{eq:NJL}
\end{equation}
arises  by integrating out the gluon degrees of freedom and absorbing them in the local four-fermion interaction ${\cal L}_{\rm int}$. In this way the local $SU(3)_c$ gauge
symmetry of QCD is replaced by a global one. Confinement is lost but all other
symmetries of QCD are maintained. In Eq.\,(\ref{eq:NJL}) the mass matrix $m_q$
incorporates the small "bare" quark masses. In the limit $m_q \rightarrow 0$ the Lagrangian (\ref{eq:NJL}) has a chiral symmetry of left- and right-handed quarks,
$SU(N_f)_L\times SU(N_f)_R$, just like the original QCD Lagrangian for $N_f$ massless quark flavors.

Fierz transforming the color current-current interaction in Eq.\ (\ref{eq:Lint1}) produces a
set of exchange terms acting in quark-antiquark channels. For the case of $N_f = 2$:
\begin{equation}
{\cal L}_{\rm int} \rightarrow {G\over 2}\left[(\bar{q}q)^2 + (\bar{q}\,i\gamma_5\,
\boldsymbol{\tau} \,q)^2\right] + ... \,\, , 
\label{eq:Lint2}
\end{equation}
where $\boldsymbol{\tau} = (\tau_1,\tau_2,\tau_3)$ represents the vector of isospin Pauli matrices.
Included in Eq.\ (\ref{eq:Lint2}) but not shown explicitly is a series of terms involving
vector and axial vector currents, both in color-singlet and color-octet channels. The new constant
$G$ is proportional to the color coupling strength $G_c$. The ratio $G/G_c$ is uniquely
determined by the number of colors and flavors. These considerations can be viewed as a
contemporary way of introducing the time-honored NJL model \cite{NJL61}, which has
been applied \cite{VW91,HK94} to a variety of problems in hadronic physics. The virtue of
the model is its simplicity in describing the basic mechanism behind spontaneous chiral
symmetry breaking, as we now outline.

In mean-field approximation, the equation of motion derived from the Lagrangian
(\ref{eq:NJL}, \ref{eq:Lint2}) leads to a gap equation
\begin{equation}
M_q = m_q - G\langle0|\bar{q}q|0\rangle\,\, ,
\end{equation}
which connects the dynamically generated constituent quark mass, $M_q$,
to the appearance of the chiral quark condensate
\begin{equation}
\langle 0|\bar{q}q|0\rangle= -{\rm tr}\lim_{\,x\rightarrow 0}\langle0| {\cal T}q(0)
\bar{q}(x)|0\rangle = -2iN_fN_c\int {d^4p\over (2\pi)^4}{M_q\,\theta(\Lambda -
|\vec{p}\,|)\over p^2 - M_q^2 + i\epsilon}\,\, .
\label{gapeq}
\end{equation}
A non-zero chiral condensate signals the spontaneous breaking
of chiral symmetry. In the chiral limit, starting from $m_q = 0$, a non-zero constituent quark
mass $M_q$ develops dynamically together with
$\langle 0|\bar{q}q|0\rangle \neq 0$, provided that $G$ exceeds a critical value on the order of
$G_{\rm crit} \sim 10$ GeV$^{-2}$. The integral in Eq.\ (\ref{gapeq}) requires a
momentum-space cutoff $\Lambda \simeq 2M_q$ beyond which the interaction is
"turned off". The strong non-perturbative interactions polarize the
vacuum and generate a condensate of quark-antiquark pairs, thereby turning an
initially point-like quark with its small bare mass $m_q$ into a dressed quasi-particle
with a size on the order of $(2M_q)^{-1}$ \cite{VLKW90}.

Solving Bethe-Salpeter equations in the color-singlet quark-antiquark channels,
the lightest mesons are generated as quark-antiquark excitations of the correlated QCD
ground state with its condensate structure. Many instructive calculations have been
performed previously in the three-flavor NJL model \cite{VW91,HK94,KLVW90}. This
model has an undesired $U(3)_L\times U(3)_R$ symmetry. The axial
$U(1)_A$ anomaly of QCD reduces this symmetry to 
$SU(3)_L\times SU(3)_R\times U(1)_V$, i.e., chiral symmetry in conjunction with baryon number conservation. In the three-flavor NJL model, the $U(1)_A$ symmetry is broken by instanton-driven interactions proportional to
a flavor determinant \cite{tH76} $\det[\bar{q}_i(1 \pm \gamma_5)q_j]$. This
interaction necessarily involves all three quark flavors $u, d, s$ simultaneously in a genuine
three-body contact term.

 \begin{figure}
       \centerline{\includegraphics[width=7cm] {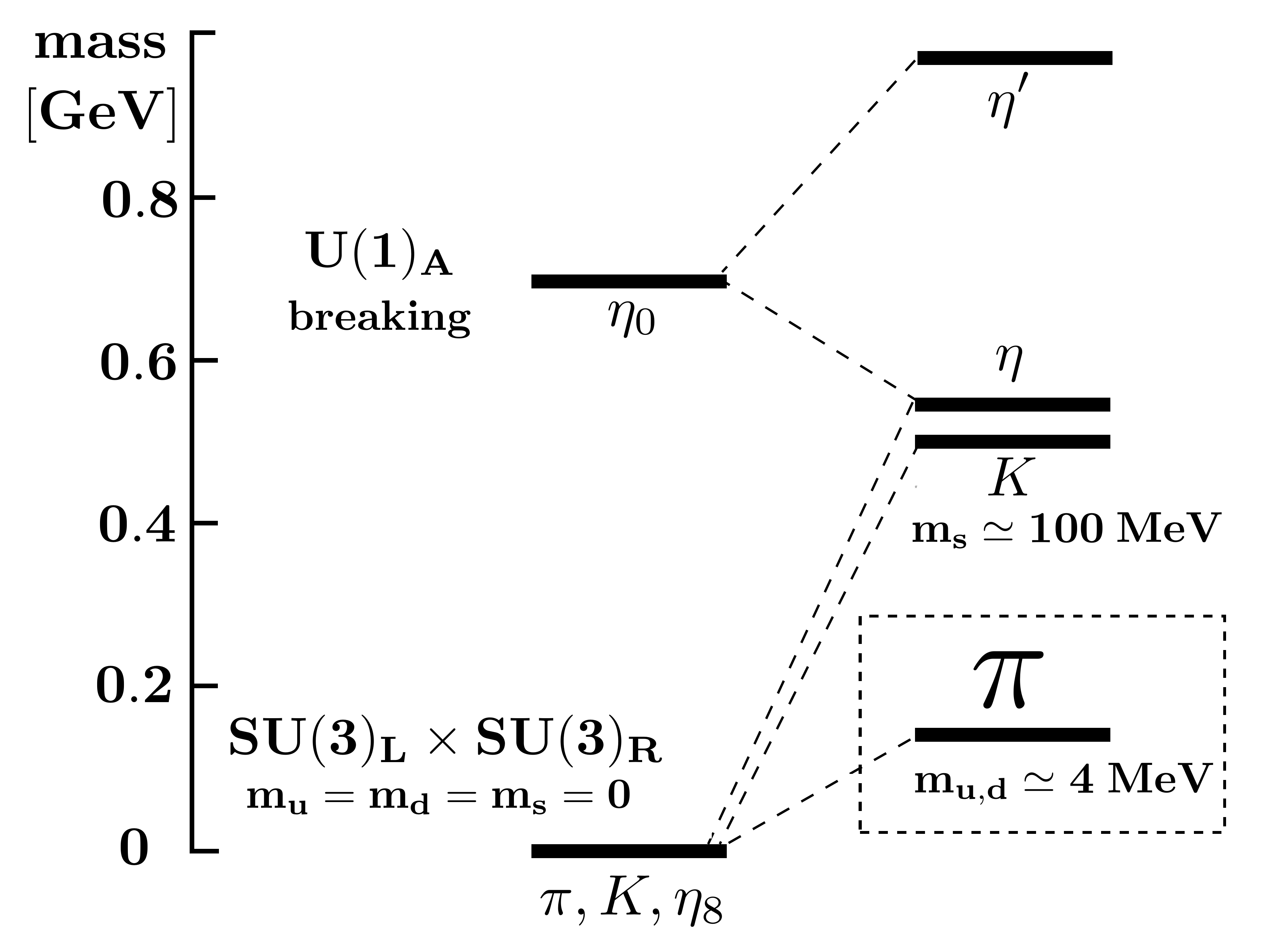}}
\caption{Symmetry breaking pattern in the pseudoscalar meson nonet calculated in a three-flavor NJL model \cite{KLVW90}.}
   \label{fig:1}
 \end{figure}

Fig.\,\ref{fig:1} shows the symmetry breaking pattern resulting from such a calculation
of the pseudoscalar meson spectrum. Assuming massless $u, d$ and
$s$ quarks, the pseudoscalar octet emerges as a set of massless Nambu-Goldstone bosons of the
spontaneously broken chiral $SU(3)_L\times SU(3)_R$ symmetry, while the anomalously broken
$U(1)_A$ symmetry drives the flavor-singlet $\eta_0$ away from the NG boson sector.
The inclusion of finite quark masses that explicitly break chiral symmetry shifts the
pseudoscalar ($J^\pi = 0^-$) nonet mesons into their empirically observed positions, including
$\eta_0$-$\eta_8$ mixing.

The pion mass is related to the scalar quark condensate and bare quark masses through
the famous Gell-Mann--Oakes--Renner relation \cite{GOR68}:
\begin{equation}
m^2_{\pi}\,f_{\pi}^2 = - {1\over 2}(m_u + m_d) \langle 0| \bar{q} q |0\rangle +
{\cal O}(m^2_{u,d},\,m^2_{u,d}\ln (m_{u,d})).
\label{eq:GOR}
\end{equation}
derived from current algebra and PCAC.
It involves additionally the pion decay constant\footnote{More precisely, the empirical $\pi^-$ decay constant is $f_{\pi^-} = (92.21\pm 0.16)$ MeV \cite{pdg2012}.}, $f_\pi = 92.2$\,MeV, which is defined
by the matrix element connecting the pion state with the QCD vacuum via the isovector
axial vector current, $A^\mu_a = \bar{q}\gamma^\mu\gamma_5 {\tau_a \over 2} q$:
\begin{equation}
\langle 0 | A^{\mu}_a (0) | \pi_b (p) \rangle = i \delta_{ab}\, p^{\mu} f _\pi\, .
\label{eq:fpi}
\end{equation}
The pion decay constant, like the chiral condensate $\langle 0| \bar{q} q |0\rangle$,
is a measure of spontaneous chiral symmetry breaking, with its associated characteristic
scale, $\Lambda_\chi \sim 4\pi f_\pi \simeq 1.2$ GeV. The non-zero pion mass,
$m_\pi \ll \Lambda_\chi$, is a reflection of the explicit chiral symmetry
breaking by the small quark masses, with $m^2_{\pi} \sim m_q$. Unlike the pion mass and decay constant, the quark masses $m_{u,d}$ as well as the quark condensate $\langle 0|\bar{q} q |0\rangle$
are scale-dependent quantities and therefore not separately observable. Only their product is renormalization group
invariant. At a renormalization scale of 2 GeV, a typical average quark mass
${1\over 2}(m_u + m_d) \simeq 3.5$ MeV goes together with a condensate 
\begin{equation}
\langle 0 |\bar{q} q|0 \rangle = \langle 0 |\bar{u} u + \bar{d}d|0 \rangle\simeq -(0.36\,\textrm{GeV})^3 \simeq -6\,\textrm{fm}^{-3}~~.
\nonumber
\end{equation}

The quark masses set primary scales in QCD, and their
classification into "light" and "heavy" quarks determines very different
physics phenomena. The heavy quarks (the $t$, $b$ and -- within limits --
the $c$ quarks), whose reciprocal masses offer a natural small parameter, can be
treated in non-relativistic approximations (that is, expansions of observables in powers of
$1/m_{t,b,c}$). The sector of the light quarks (the $u$, $d$ quarks and -- to some extent -- the
$s$ quark), on the other hand, is governed by quite different principles and rules. Evidently, the light quark
masses themselves are now small parameters to be compared with a
"large" scale of dynamical origin. In the hadronic sector this large scale is characterized by a
mass gap of about 1 GeV separating the QCD vacuum from nearly all of its
excitations, with the exception of the pseudoscalar meson octet shown in Fig.\,\ref{fig:1}.
This mass gap is indeed comparable to the scale, $4\pi f_\pi$, associated with
spontaneous chiral symmetry breaking in QCD\footnote{Note the close correspondence: 
${m_q\over\Lambda_{\rm QCD}}\sim \left({m_\pi\over 4\pi f_\pi}\right)^2 \simeq 0.015$ using $\Lambda_{\rm QCD} \sim 0.25$ GeV.}

\subsection{Notes on thermodynamics and dense baryonic matter}

The NJL model and its extended version incorporating the Polyakov loop (the PNJL model \cite{Fukushima2004,Fukushima2008,RTW2006,RRW2007}) has been widely used to explore the mean-field thermodynamics of strongly-coupled quark quasiparticles under the aspect of the chiral symmetry breaking scenario. In the chiral limit the pion decay constant acts as an order parameter. Its temperature dependence, $f_\pi(T)$, reflects a second-order chiral phase transition for $N_f = 2$ massless quarks, with $f_\pi(T\rightarrow T_c) \rightarrow 0$ at a critical temperature $T_c \simeq 0.2$ GeV. This transition becomes a crossover when chiral symmetry is explicitly broken by non-zero quark masses $m_q$.

While NJL type models (and their bosonized analogues, chiral quark-meson models) provide an instructive picture of the dynamics behind spontaneous chiral symmetry breaking, they have the principal deficiency of missing confinement. Introducing the Polyakov loop in PNJL models improves the situation by suppressing color-non-singlet degrees of freedom in the "hadronic" phase at temperatures below $T_c$. But confinement also implies clustering in space, such as the localization of three valence quarks in a small volume to form a baryon, for which there is no primary mechanism in standard PNJL models. This is obviously of some significance when studying the phase diagram of strong-interaction matter at finite baryon densities. Many (P)NJL and related mean-field calculations of isospin-symmetric matter predict the existence of a first-order chiral phase transition\footnote{For reviews see \cite{Buballa2005,CBMbook2011,FS2013} and references therein.}. In a $T-\mu_B$ phase diagram (temperature $T$ vs. baryon chemical potential $\mu_B$), this 1st order transition has a critical end point from where it extends downward to low temperatures and meets the $T = 0$ axis around $\mu_B \sim 0.9 - 1$ GeV. This, however, is the well established terrain of nuclear physics\footnote{The baryon chemical potential of ($T = 0$) nuclear matter at equilibrium is $\mu_B \simeq 923$ MeV, the nucleon mass minus the binding energy per nucleon.} where there is no indication of a chiral phase transition. The only empirically founded phase transition discussed in this region is the one from a nuclear Fermi liquid to an interacting gas of nucleons \cite{Po1995, Chomaz2001, Ago2005, Kar2008, ELMP2013}.

Motivated by such considerations, the focus of the present report will be on nuclear thermodynamics as it sets important constraints for exploring the QCD phase diagram in the region of baryon densities around and well above the density of equilibrium nuclear matter, and at temperatures $T\lesssim 100$ MeV within the hadronic phase. Apart from the lightest mesons, the relevant fermionic degrees of freedom in this domain are obviously nucleons (rather than quarks) which act as  "heavy" sources of the pion field. 

\subsection{The sigma model}

Upon bosonization, the NJL model with $N_f = 2$ flavors is demonstrated to be equivalent to a linear sigma model \cite{Ripka1997}. The scalar-isoscalar and pseudoscalar-isovector quark bilinears are associated with corresponding bosonic auxiliary fields:
\begin{equation}
s(x) = -G\,\bar{q}(x)\,q(x)~,~~~~p_a(x) =  -G\,\bar{q}(x)\,i\gamma_5\tau_a\,q(x)~~(a = 1,2,3)~.
\nonumber
\end{equation}
The bosonization is conveniently performed in Euclidean space, with $\int d_4 x \equiv\int_0^\beta d\tau \int d^3r$ and $\beta = 1/T$ the inverse temperature. In a first step the (Euclidean) NJL action is rewritten in terms of the auxiliary boson fields:
\begin{equation}
{\cal S}^E(q,q^\dagger ; s, p_a) = - {\rm Tr}    \ln (\partial_\tau + h) + {1\over 2G}\int d_4x \left[s^2(x) + p_a^2(x)\right]~.
\end{equation}
The Fermion determinant, $-{\rm Tr} \ln (\partial_\tau + h)$, involves quark loops generated by the Dirac Hamiltonian density
\begin{equation}
 h =  -i\boldsymbol{\alpha\cdot\nabla}+ \boldsymbol{\beta}m_q + \boldsymbol{\beta}(s + i\gamma_5\,p_a\tau_a)~~.
\end{equation}
When integrating out the quarks, the bosonic fields $s(x)$ and $p_a(x)$ become dynamical variables. The Fermion determinant generates time derivatives of the fields $s$ and $p_a$ appearing to all orders in the action. Higher time derivatives reflect the fact that $s(x), p_a(x)$ represent quark-antiquark composites. Keeping the leading second-order derivatives one arrives at the bosonic Euclidean action, ${\cal S}^E_\sigma(s, p_a) = \int d_4x\,{\cal L}^E_\sigma(s(x), p_a(x))$, with
\begin{eqnarray}
{\cal L}^E_\sigma(s, p_a)&=&{Z_\pi\over 2}\left[(\partial_\mu s)^2 + (\partial_\mu p_a)^2 + \left(s^2 + p_a^2 - M_q^2\right)^2\right]\nonumber \\ &+& {m_q\over 2GM_q}\left[(s-M_q)^2 + p_a^2\right]~~.
\end{eqnarray}
As a reminder of the intrinsic $q\bar{q}$ structure of the pion, the factor 
\begin{eqnarray}
Z_\pi = {N_c\over 2\pi^2}\int_0^\Lambda dk{k^3\over (k^2 + M_q^2)^2} = {f_\pi^2\over M_q^2}
\end{eqnarray}
represents the pertinent quark-antiquark loop integral with momentum space cutoff $\Lambda$. Introducing normalized $\pi$ and $\sigma$ fields,
\begin{eqnarray}
\pi_a(x) = \sqrt{Z_\pi}\,p_a(x)~~~~, ~~~~\sigma(x) = \sqrt{Z_\pi}\,s(x) ~~,
\end{eqnarray}
one arrives at the Euclidean form of the linear sigma model \cite{Lee72, Ripka1997}:
\begin{eqnarray}
{\cal L}^E_\sigma(\sigma,\pi_a)&=&{1\over 2}(\partial_\mu \sigma)^2 + {1\over 2}(\partial_\mu \pi_a)^2 + {M_q^2\over 2 f_\pi^2}\left(\sigma^2 + \pi_a^2 - f_\pi^2\right)^2\nonumber \\ &+&{m_\pi^2\over 2}\left[(\sigma - f_\pi)^2 + \pi_a^2\right]~~,
\label{eq:sigma}
\end{eqnarray}
or its familiar representation using the Minkowskian action ${\cal S}_\sigma = \int d^4x\,{\cal L}_\sigma$:
\begin{eqnarray}
{\cal L}_\sigma(\sigma,\boldsymbol{\pi})&=&{1\over 2}(\partial_\mu \sigma)(\partial^\mu \sigma) + {1\over 2}(\partial_\mu \boldsymbol{\pi})\cdot(\partial^\mu \boldsymbol{\pi}) - {\cal U}(\sigma, \boldsymbol{\pi}) ~~,
\label{eq:sigma2}
\end{eqnarray}
with the potential
\begin{equation}
{\cal U}(\sigma, \boldsymbol{\pi}) = {M_q^2\over 2 f_\pi^2}\left(\sigma^2 + \boldsymbol{\pi}^2 - f_\pi^2\right)^2 + {m_\pi^2\over 2}\left[(\sigma - f_\pi)^2 + \boldsymbol{\pi}^2\right]~~.
\end{equation}
This potential includes a term $\delta{\cal U}_{sb} = -m_\pi^2 f_\pi \sigma$ that represents the quark mass term of the QCD Lagrangian and breaks chiral symmetry explicitly. 
We express $\sigma = \sigma_0 + \delta\sigma$ in terms of its expectation value, $\sigma_0 \equiv \langle\sigma\rangle$, and a fluctuating part $\delta\sigma$. The minimum of ${\cal U}$ is located at $\sigma = \sigma_0 = f_\pi$ for $\boldsymbol{\pi} = 0$. The invariant chiral field $\sigma^2 + \boldsymbol{\pi}^2 = f_\pi^2$ at minimum defines the surface of a hypersphere with radius $f_\pi$ in four dimensions.

The pion mass is given by 
\begin{equation}
m_\pi^2 = {m_q\over Z_\pi G M_q} = {m_q M_q\over G f_\pi^2} \simeq -{m_q\over f_\pi^2}\langle\bar{q}q\rangle ~~.
\end{equation}
The last part is again the Gell-Mann--Oakes--Renner relation. Here it follows to leading order in $m_q$ when introducing the quark condensate $\langle\bar{q}q\rangle$ through the derivative of the action, $\delta{\cal S}^E/\delta m_q$, with respect to the quark mass.

The squared mass of the $\sigma$ boson is identified with the coefficient of the term proportional to $(\sigma - f_\pi)^2/2$ in Eq.\,(\ref{eq:sigma}):
\begin{equation}
m_\sigma^2 = 4M_q^2 + m_\pi^2 ~~.
\end{equation}
With a dynamical quark mass $M_q = 0.4 - 0.5$ GeV as discussed in \cite{Ripka1997}, this implies a large $\sigma$ mass, typically of a magnitude comparable to the mass of the $J^{PC} = 0^{++}$ meson $f_0(980)$. It is important to note that this $m_\sigma$ should in fact {\it not} be identified with the broad maximum around $400 - 500$ MeV seen in the spectral function of the isoscalar s-wave $\pi\pi$ scattering amplitude \cite{CCL2006}. This spectral function is in fact generated by the coupling of the "heavy" $\sigma$ to the interacting $\pi\pi$ continuum when solving the scattering equation in the respective channel \cite{WTSvS2014} . 

\subsection{Chiral nucleon-meson model}

The linear sigma model (\ref{eq:sigma2}) defines a framework for treating the chiral meson sector. At this stage nucleons can be added as sources of the chiral fields, $\phi = (\sigma, \boldsymbol\pi)$. A nucleon field $N(x) = \left(\psi_p(x), \psi_n(x)\right)^\top$ is introduced as an isospin doublet of proton and neutron Dirac fields, $\psi_{p,n}$, with minimal chiral couplings to $\sigma$ and $\boldsymbol{\pi}$. At the same time there is the option to generalize the potential ${\cal U(\sigma,\boldsymbol{\pi}})$ by including higher powers of the chiral invariant, $\phi^2 = \sigma^2 + \boldsymbol\pi^2$. This is the starting point for the construction of a chiral model of interacting nucleons and mesons. Its original form dates back to Ref.\,\cite{GML1960}:
\begin{eqnarray}
{\cal L}_0(N, \sigma,\boldsymbol{\pi}) &=& \bar{N}\left[i\gamma_\mu\partial^\mu - g(\sigma + i\gamma_5\,\boldsymbol{\tau\cdot\pi})\right]N\nonumber  \\ &+& {1\over 2}(\partial_\mu \sigma)(\partial^\mu \sigma) + {1\over 2}(\partial_\mu \boldsymbol{\pi})\cdot(\partial^\mu \boldsymbol{\pi}) - {\cal U}(\sigma, \boldsymbol{\pi}) ~~.
\label{eq:ChMN1}
\end{eqnarray}
The nucleon mass in vacuum is identified with
\begin{equation}
M_N = g\,\sigma_0 = g\,f_\pi ~~.
\label{eq:Nmass}
\end{equation}
Using physical values for the nucleon mass and pion decay constant, the Yukawa coupling constant is determined as $g = M_N/f_\pi \simeq 10.2$ or $g^2/4\pi \simeq 8.3$. Beyond mean-field level, the mass $M_N$ is renormalized when fluctuations of the chiral fields generate the meson cloud of the nucleon \cite{TW2001}.

The Lagrangian (\ref{eq:ChMN1}) is supposed to operate in the hadronic phase of QCD with spontaneously broken chiral symmetry, i.e., at energy and momentum scales well below $\Lambda_\chi \sim 4\pi f_\pi$. This Lagrangian generates single and multiple pion-exchange mechanisms in the nucleon-nucleon ($NN$) interaction at long and intermediate distances. However, a realistic treatment of the $NN$ force requires also to incorporate the effects  of the strong repulsion between nucleons at short distances, $r \lesssim 0.5$ fm. This repulsive core does not originate from chiral dynamics. Its origin in QCD presumably involves strongly interacting gluonic degrees of freedom, but the details of the underlying mechanisms need not be further specified as they are not resolved at the momentum scales characteristic of nuclear physics that we are going to focus on. Such short-distance effects can conveniently be included and parametrized by adding nucleon-nucleon contact interactions, e.g., terms proportional to $(\bar{N}\gamma_\mu N)^2$ and $(\bar{N}\gamma_\mu\boldsymbol\tau N)^2$ with corresponding coupling strengths of dimension ${\it (length)}^2$. An alternative and equivalent choice is to introduce isoscalar and isovector vector fields, $v_\mu$ and $\boldsymbol{\rho}_\mu$, coupled to the nucleons and adding the following piece to the Lagrangian (\ref{eq:ChMN1}) while still maintaining its chiral invariance:
\begin{eqnarray}
\delta{\cal L}(N, v, \boldsymbol\rho) &=& - \bar{N}\gamma_\mu\left(g_v \,v^\mu + g_\tau \boldsymbol{\tau\cdot\rho}^\mu\right) N \nonumber  \\ &+& {1\over 2}m_V^2\left(v_\mu v^\mu + \boldsymbol{\rho}_\mu\cdot\boldsymbol{\rho}^\mu\right) + {1\over 4}\left(v_{\mu\nu}v^{\mu\nu}+ \boldsymbol{\rho}_{\mu\nu}\cdot\boldsymbol{\rho}^{\mu\nu} \right)~~.
\label{eq:ChMN2}
\end{eqnarray}
Note that the vector bosons, $v_\mu$ and $\boldsymbol{\rho}_\mu$, are {\it not} to be identified with the physical $\omega$ and $\rho$ mesons. They simply act as phenomenological generators of the short-range $NN$ interactions. Their large mass parameter $m_V$ stands for the inverse correlation length characteristic of the short-distance core. In mean-field approximation the resulting low-momentum forces reduce to the contact interactions mentioned previously, with coupling strengths,
\begin{equation}
G_v = {g_v^2\over m_V^2} ~,~~~G_\tau = {g_\tau^2\over m_V^2}~,
\label{eq:vectorcouplings}
\end{equation}
where coupling constants and the vector mass do not appear individually but only their ratios enter. In this limit $v_\mu$ and $\boldsymbol{\rho}_\mu$ reduce to constant background fields, and their field tensors $v_{\mu\nu}$ and $\boldsymbol{\rho}_{\mu\nu}$ vanish. 

The combined Lagrangian,
\begin{equation}
{\cal L} = {\cal L}_0(N, \sigma,\boldsymbol{\pi}) + \delta{\cal L}(N, v, \boldsymbol\rho)~,
\label{eq:ChNM}
\end{equation}
of Eqs.\,(\ref{eq:ChMN1}) and (\ref{eq:ChMN2}) is referred to as the chiral nucleon-meson (ChNM) model in the present context. It is going to be used as the starting point for a functional renormalization group (FRG) approach to nuclear many-body systems and their thermodynamics \cite{DHKW2013, DW2014, DW2015}. A model of similar type has previously been studied in Refs. \cite{BJW2003,FW2012}. The non-perturbative treatment of fluctuations beyond mean-field approximation, a central feature of the FRG, includes the prominent low-energy modes: primarily pions and nucleon-hole excitations.
For the "heavy" degrees of freedom, such as the vector ($v$ and $\boldsymbol\rho$) bosons, fluctuations are suppressed by their large mass $m_V$.  

At low energy and momentum scales, $Q \ll m_\sigma$, the scalar $\sigma$ field is frozen and can be eliminated in favour of a non-linear realisation with only pions as active degrees of freedom.  In this limit, with the large sigma mass representing the steep curvature of the chiral potential ${\cal U}(\sigma,\boldsymbol{\pi})$, the system stays at the minimum of this potential and the fields are constrained by $\sigma^2 + \boldsymbol{\pi}^2 = f_\pi^2$. A new field  $\boldsymbol{\varphi}(x)$ satisfying this constraint can be defined as:
\begin{equation}
\sigma(x) = f_\pi\cos{\varphi(x)\over f_\pi}~~, ~~\boldsymbol\pi(x) = f_\pi\,\hat{\boldsymbol{\varphi}}\sin{\varphi(x)\over f_\pi}~~,
\label{eq:chiralfields}
\end{equation}
with $\varphi = |\boldsymbol{\varphi}| = \sqrt{\boldsymbol{\varphi\cdot\varphi}}$ and $\hat{\boldsymbol{\varphi}} = \boldsymbol{\varphi}/\varphi$. Next, a unitary transformation
\begin{equation}
u(x) = \exp\left[{i\over 2 f_\pi}\gamma_5\,\boldsymbol{\tau\cdot\varphi}(x)\right]
\end{equation}
acting on the nucleon field,
\begin{equation}
N(x) \rightarrow \Psi(x) = u(x)N(x)~~, ~~\bar{N}(x)\rightarrow \bar{\Psi}(x) = \bar{N}(x)\,u(x)~~,
\end{equation}
transforms the "bare" nucleon into a "dressed" one  with a multi-pion cloud attached. Introducing a chiral-covariant derivative: 
\begin{equation}
D_\mu\boldsymbol{\varphi} = (\partial_\mu\varphi)\hat{\boldsymbol{\varphi}} + \left(f_\pi\sin{\varphi(x)\over f_\pi}\right)\partial_\mu\hat{\boldsymbol{\varphi}}~~,
\end{equation}
the original linear sigma model Lagrangian (\ref{eq:ChMN1}) at the minimum of the potential, ${\cal U} = 0$, turns into a non-linear representation expressed in terms of only pionic fields coupled to isovector vector and axial currents of the nucleon:
\begin{eqnarray}
{\cal L}_0(N,\boldsymbol{\varphi}) &=& \bar{\Psi}\left[i\gamma_\mu\partial^\mu - M_N\right]\Psi + {1\over 2f_\pi}\bar{\Psi}\gamma_\mu\gamma_5\boldsymbol{\tau}\Psi\cdot D^\mu\boldsymbol{\varphi}\nonumber \\
&-&{1\over 2}\left[1-\cos{\varphi\over f_\pi}\right]\bar{\Psi}\gamma_\mu\boldsymbol{\tau}\Psi\cdot\left(\hat{\boldsymbol{\varphi}}\times\partial^\mu\hat{\boldsymbol{\varphi}}\right) + {1\over 2}D_\mu\boldsymbol{\varphi}\cdot D^\mu\boldsymbol{\varphi}~.
\label{eq:nonlinear1}
\end{eqnarray}
The nucleon mass is again given as $M_N = g\,f_\pi$. In the weak-field limit to leading order in $\boldsymbol{\varphi}$ (with $ \boldsymbol{\varphi}(x) = \boldsymbol\pi(x)$), one arrives at the familiar form:
\begin{eqnarray}
{\cal L}_0(N,\boldsymbol{\pi}) &=& \bar{\Psi}\left[i\gamma_\mu\partial^\mu - M_N\right]\Psi + {1\over 2f_\pi}\bar{\Psi}\gamma_\mu\gamma_5\boldsymbol{\tau}\Psi\cdot \partial^\mu\boldsymbol{\pi}\nonumber \\
&-&{1\over 4f_\pi^2}\bar{\Psi}\gamma_\mu\boldsymbol{\tau}\Psi\cdot\left(\boldsymbol{\pi}\times\partial^\mu\boldsymbol{\pi}\right) + {1\over 2}\partial_\mu\boldsymbol{\pi}\cdot \partial^\mu\boldsymbol{\pi}~,
\label{nonlinear2}
\end{eqnarray}
which now features the pseudovector pion-nucleon derivative coupling and the interaction of the isovector-vector currents of nucleon and pion. 

\subsection{Chiral effective field theory} 

While the linear sigma model is indeed a {\it model} and not an effective field theory, the transcription of a linear to a non-linear form of the chiral Lagrangian under the constraint $\sigma^2 + \boldsymbol{\pi}^2 = f_\pi^2$ can nonetheless be viewed as a basic step towards Chiral Effective Field Theory (ChEFT). As pointed out in \cite{weinbergbook} and references therein, it is however not necessary to start from a linear sigma model in order to construct ChEFT. It is sufficient to write down the most general Lagrangian that incorporates the basic properties of low-energy QCD, namely confinement and spontaneous chiral symmetry breaking, together with all other relevant symmetries of the underlying fundamental theory. This ChEFT is then the low-energy equivalent of QCD.

Chiral EFT with $N_f = 2$ is commonly phrased in terms of a chiral field $U(x)\in SU(2)$ that relates back to Eq.\,(\ref{eq:chiralfields}) of the previous subsection:
\begin{equation}
U(x) = \exp\left({i \over f_\pi}\,\boldsymbol{\tau\cdot\varphi}(x) \right) ~.
\end{equation}
The pion decay constant $f_\pi$ is first to be taken in the chiral limit ($m_\pi \rightarrow 0$).
The low-energy physics of QCD in the Nambu-Goldstone boson sector and in the absence of baryons is expressed in terms of an
effective Lagrangian involving $U(x)$ and its derivatives, ${\cal L}_{QCD} \to {\cal L}_{\rm eff} (U, \partial^\mu U, ...)$ .
Nambu-Goldstone bosons do not interact unless they carry non-zero four-momentum, so the low-energy
expansion of ${\cal L}_{\rm eff}$ allows for an ordering in powers of $\partial^{\mu}U$ and a perturbative treatment. Lorentz invariance implies that even numbers of derivatives in overall scalar combinations are permitted\footnote{A notable exception is the $N_f=3$ case in which the Wess-Zumino term appears with three derivatives.}. One writes for the first few terms of the
chiral Lagrangian:
\begin{equation}
{\cal L}_{\rm eff} = {\cal L}_{\pi}^{(2)} + {\cal L}_{\pi}^{(4)} + ...\,,
\end{equation}
where the first term (leading order in the non-linear sigma model) involves two derivatives:
\begin{equation}
{\cal L}^{(2)}_{\pi} = {f_\pi^2 \over 4} {\rm tr} ( \partial^{\mu} U \partial_{\mu}U^{\dagger} ) + \delta{\cal L}^{(2)}_\pi ~.
\end{equation}
The added symmetry-breaking mass term $\delta{\cal L}^{(2)}_\pi$ is small, so that it too can be handled perturbatively.
It introduces a term linear in the quark mass matrix $m_q$:
\begin{equation}
\delta{\cal L}^{(2)}_{\pi} = {f_\pi^2 \over 4} m_\pi^2 \, {\rm tr}(U + U^{\dagger}) \,,
\end{equation}
with $m_\pi^2 \sim (m_u+m_d)$.
At fourth order the additional terms permitted by symmetries are
\begin{equation}
{\cal L}^{(4)}_{\pi} = {\ell_1 \over 4} \Big[ {\rm tr}(\partial^{\mu} U \partial_{\mu}
U^{\dagger})\Big]^2 + {\ell_2 \over 4}  {\rm tr} (\partial_{\mu} U \partial_{\nu} U^{\dagger})
\, {\rm tr} (\partial^{\mu} U \partial^{\nu} U^{\dagger}) +\dots \, ,\end{equation}
where further contributions involving the light quark mass $m_q$ and external fields are not shown.
The constants $\ell_1, \ell_2$ absorb loop
divergences and their finite scale-dependent parts must be fixed by matching to experiment.

The framework for systematic perturbative calculations of (on-shell) $S$-matrix elements involving
Nambu-Goldstone bosons, chiral perturbation theory (ChPT), is then defined by the following
rules: Collect all Feynman diagrams generated by ${\cal L}_{\rm eff}$. Classify
individual terms according to powers of the small quantity $Q/(4\pi f_{\pi})$,
where $Q$ stands generically for three-momenta of NG bosons or for the pion mass $m_{\pi}$. Loops are evaluated in
dimensional regularization and are renormalized by appropriate chiral counter terms.

The prominent role played by the pion as a Nambu-Goldstone boson of spontaneously
broken chiral symmetry in QCD impacts as well the low-energy structure
and dynamics of nucleons \cite{TW2001}.  For example, when probing the nucleon with a
long-wavelength electroweak field, a substantial part of the response comes from
the pion cloud, comprising the "soft" surface of the nucleon.
The calculational framework for this, baryon chiral perturbation theory
\cite{EM96,BKM95} has been applied successfully to diverse low-energy
processes such as low-energy pion-nucleon scattering, threshold pion photo- and
electro-production and Compton scattering on the nucleon. 

Consider next the physics of the pion-nucleon system, the sector with baryon number $B = 1$.
Unlike the pion, the nucleon has a large mass of the same order as the chiral symmetry breaking
scale, $4\pi f_\pi$, even in the limit of vanishing quark masses.
The additional term in the chiral effective Lagrangian involving the nucleon, denoted
by ${\cal L}_{\pi N}$, is again dictated by chiral symmetry and expanded in powers of the quark masses and
derivatives of the Goldstone boson field:
\begin{equation}
{\cal L}_{\pi N} = {\cal L}_{\pi N}^{(1)} + {\cal L}_{\pi N}^{(2)} + \dots
\end{equation}
In the leading term, ${\cal L}_{\pi N}^{(1)}$, there is a vector current coupling between pions and
the nucleon (arising from the replacement of $\partial^{\mu}$
by a chiral covariant derivative) as well as an axial vector coupling:
\begin{equation}
{\cal L}_{\pi N}^{(1)} =  \bar{\Psi}\left[i\gamma_{\mu}(\partial^{\mu} +\Gamma^{\mu})- M_0 -
g_A \gamma_{\mu}\gamma_5\, a^{\mu}\right]\Psi \,\,,
\label{eq:LeffN}
\end{equation}
where $M_0$ is the nucleon mass in the chiral limit. The vector and axial vector quantities involve the pion fields via $\xi = \sqrt{U}$ in the form
\begin{eqnarray}
\Gamma^{\mu} & = & {1\over 2}[\xi^{\dagger},\partial^{\mu}\xi] = {i\over 4f_\pi^2}
 \, \boldsymbol\tau \cdot (\boldsymbol\pi \times \partial^{\mu}\boldsymbol\pi) + ...~~, \\
a^{\mu} & = & {i\over 2}\{\xi^{\dagger},\partial^{\mu}\xi\}= - {1\over 2f_\pi}\, \boldsymbol\tau
\cdot \partial^{\mu}\boldsymbol\pi + ...~~,
\end{eqnarray}
where the last steps result from expanding $\Gamma^{\mu}$ and $a^{\mu}$ to
leading order in the pion fields. Up to this point the only parameters that enter are the
nucleon mass $M_0$, the pion decay constant $f_\pi$, and the nucleon axial vector
coupling constant $g_A$. 
The next-to-leading order pion-nucleon Lagrangian, ${\cal L}_{\pi N}^{(2)}$, includes
the chiral symmetry breaking
quark mass term, which shifts the nucleon mass to its physical value, $M_N$. The nucleon sigma term
\begin{equation}
\sigma_N = m_q\frac{\partial M_N}{\partial m_q} =
\langle N | m_q(\bar{u}u + \bar{d}d) |N\rangle
\end{equation}
measures the contribution of the non-vanishing quark mass to the nucleon mass.
Its empirical number, deduced from low-energy pion-nucleon data, has for a long time been in the range
$\sigma_N \simeq (45 \pm 8)$ MeV \cite{GLS91}. A recent advanced analysis \cite{HEKM2016} arrives at a significantly larger value: $\sigma_N \simeq (59.1 \pm 3.5)$ MeV, while lattice QCD \cite{Durr2016} suggests a smaller value, $\sigma_N = 38\pm 3(stat.)\pm 3(syst.)$ MeV. The difference between these two results remains a challenging issue.

Up to next-to-leading order, the $\pi N$ effective Lagrangian has the form
\begin{eqnarray}
{\cal L}_{eff}^{N} & = & \bar{\Psi}(i\gamma_{\mu}\partial^{\mu} - M_N)\Psi +
{g_A \over 2f_{\pi}} \bar{\Psi}\gamma_{\mu}\gamma_5\boldsymbol\tau\Psi \cdot\partial^{\mu}
\boldsymbol\pi  \nonumber \\ && -{1 \over 4f_{\pi}^2} \bar{\Psi}\gamma_{\mu}\boldsymbol\tau\Psi
\cdot (\boldsymbol\pi\times \partial^{\mu}\boldsymbol\pi) +{\sigma_N\over  2f_\pi^2}\,\bar{\Psi}
\Psi\,\boldsymbol\pi^{2} + ...~~,
\label{eq:LeffPiN}
\end{eqnarray}
where the pieces not shown in Eq.\,(\ref{eq:LeffPiN}) include additional terms involving $(\partial^{\mu}\boldsymbol\pi)^2$ that arise from the complete Lagrangian ${\cal L}_{\pi N}^{(2)}$, higher powers of the pion fields and isospin-breaking corrections. These terms come with further low-energy constants that encode physics at smaller distance scales and that need to be fitted to experimental data, e.g., from pion-nucleon scattering. 
The nucleon's axial vector coupling constant $g_A$, relevant for the Born terms of the p-wave pion-nucleon scattering amplitude, is determined empirically from neutron beta decay as $g_A = 1.272\pm 0.002$ \cite{PDG2014}.  The connection with the coupling $g$ of Eqs.\,(\ref{eq:ChMN1},\ref{eq:Nmass}) is now modified such that the pion-nucleon Yukawa coupling constant becomes $g_{\pi N} = g_A M_N/f_\pi \simeq 13$. This is the Goldberger-Treiman relation \cite{GT1958}. 

\subsection{Nuclear forces and in-medium chiral perturbation theory}

\begin{figure}
\centerline{\includegraphics[width=8cm] {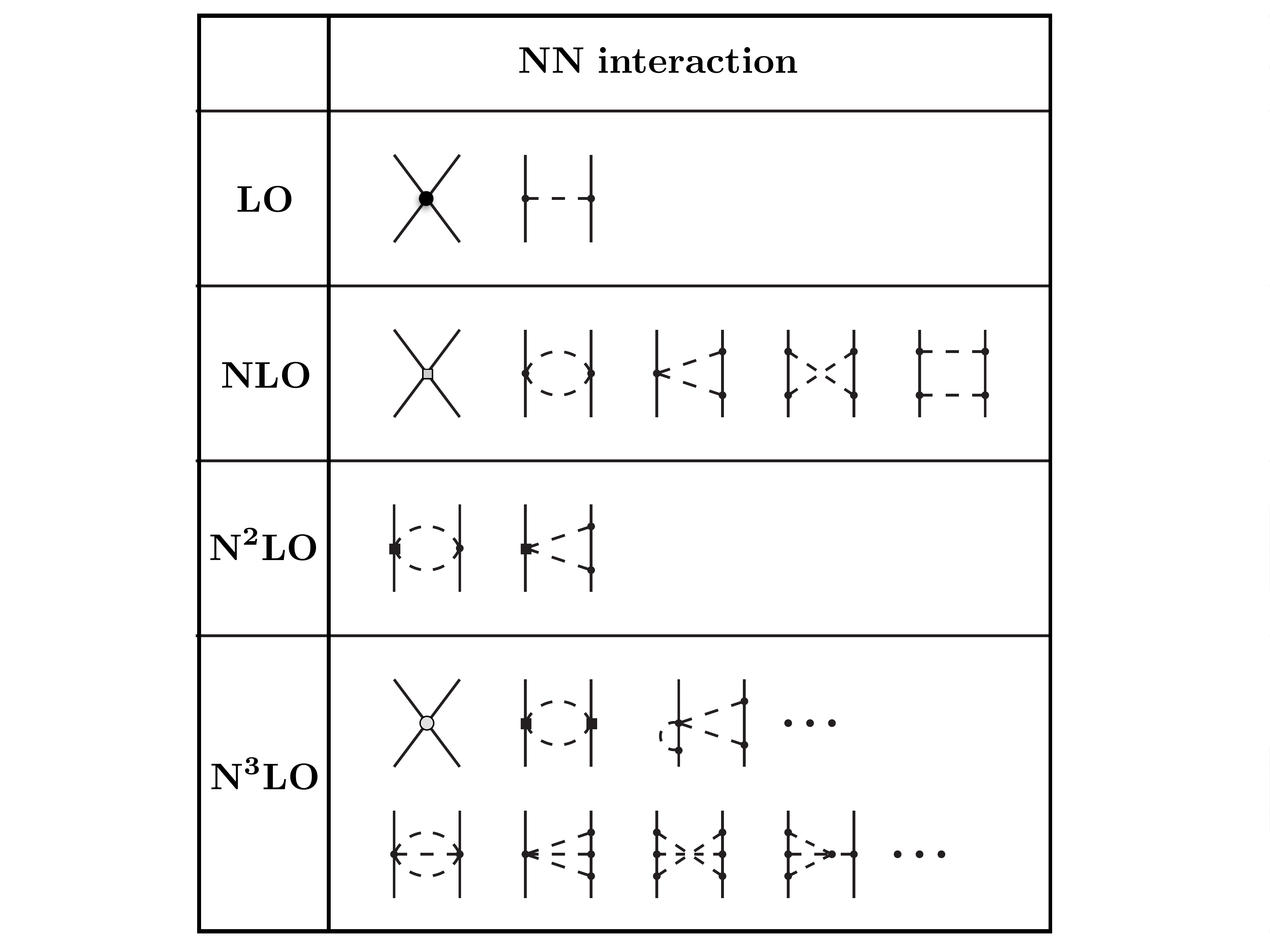}}
\caption{Diagrammatic hierarchy of chiral nuclear forces (single and multiple pion exchange) plus contact terms up to fourth order in powers of $Q/\Lambda$, the "light" scale divided by the "heavy" scale.}
\label{fig:2}
\end{figure}

In chiral effective field theory the nuclear forces are constructed in terms of explicit one-pion and
multi-pion exchange processes, constrained by chiral symmetry, plus a complete set of contact-terms
encoding unresolved short-distance dynamics \cite{evgenireview,hammerreview,machleidtreview}. A systematic hierarchy of diagrammatic contributions (see Fig.\,\ref{fig:2}) is organized in powers of $Q/\Lambda$, where $Q$ stands generically for small momenta or the pion mass, and $\Lambda < \Lambda_\chi = 4\pi f_\pi \sim 1$\,GeV is a conveniently chosen cutoff with reference to the chiral symmetry breaking scale. Two-body forces enter first at leading-order (LO), corresponding to $(Q/\Lambda)^0$, and include the
one-pion exchange contribution together with two contact terms acting in relative $S$-waves:
\begin{equation}
V_{NN}^{(\rm LO)} = -{g_A^2\over 4 f_\pi^2} \, {\boldsymbol\sigma_1 \cdot {\bf q}\,\,
\boldsymbol\sigma_2 \cdot {\bf q} \over m_\pi^2+{\bf q}^2}\,\boldsymbol\tau_1\cdot \boldsymbol\tau_2 + C_S + C_T \,
\boldsymbol\sigma_1 \cdot\boldsymbol\sigma_2 \,,
\end{equation}
where $\boldsymbol\sigma_{1,2}$ and $\boldsymbol\tau_{1,2}$ denote the spin- and isospin operators of
the two nucleons, and ${\bf q}$ is the three-momentum transfer carried by the exchanged pion.
Next-to-leading order (NLO) terms proportional to $(Q/\Lambda)^2$ include two-pion exchange diagrams
generated by the vertices of the chiral $\pi N$-Lagrangian ${\cal L}_{\pi N}^{(1)}$.
All relevant loop integrals can be performed analytically. Detailed expressions are found in the previously mentioned literature.
These NLO contributions generate isovector central, isoscalar
spin-spin and isoscalar tensor forces. Additional polynomial pieces produced
by the pion-loops can be absorbed into the set of contact terms at NLO which features seven low-energy constants associated with all combinations of spin, isospin and momentum operators appearing at that order. These low-energy constants are adjusted to fit NN scattering phase shifts.

At order N$^2$LO the most important pieces from chiral two-pion exchange arise, those that generate the prominent intermediate-range attraction in the
isoscalar central NN channel and reduce the overly strong one-pion-exchange isovector tensor force.
These terms \cite{nnpap1} come from $2\pi$-exchange
through chiral $\pi\pi NN$ contact couplings. They include the prominent effects of $\Delta(1232)$-isobar degrees of freedom in intermediate steps of the two-pion exchange process\footnote{In fact, treating the $\Delta(1232)$ as an additional baryonic degree of freedom promotes these mechanisms from N$^2$LO to NLO.}.
In addition there are relativistic $1/M_N$-corrections to $2\pi$-exchange \cite{nnpap1,evgenireview}. 

Chiral NN potentials constructed up to order N$^3$LO (i.e., fourth order in $Q/\Lambda$) include two-loop $2\pi$-exchange processes, $3\pi$-exchange terms plus contact forces quartic in the momenta and parameterized by 15 additional low-energy constants. When solving the Lippmann-Schwinger equation the potential is multiplied by an exponential regulator function with a cutoff scale $\Lambda = 400 - 700\,$MeV in order to
restrict integrations to the low-momentum region where chiral effective field theory is applicable.
At order N$^3$LO the chiral NN potential reaches the quality of a
"high-precision" ($\chi^2/d.o.f.\sim 1$) potential in reproducing empirical NN scattering
phase shifts and deuteron properties \cite{hammerreview,CHIMS2013}.  It simultaneously provides the foundation for systematic nuclear structure studies of few- and many-body systems. Current state-of-the-art potentials\footnote{For a recent review see Ref. \cite{MS2016}} have progressed to chiral order N$^4$LO (fifth order in $Q/\Lambda$) and achieved still further substantial improvements in comparison with empirical NN two-body data, including peripheral phase shifts and $np$ polarization observables \cite{EKMN2015, EKM2015}. Convergence has been successfully demonstrated by investigating dominant contributions of order N$^5$LO \cite{EKMN2015b}. 

\begin{figure}
\centerline{\includegraphics[width=6cm] {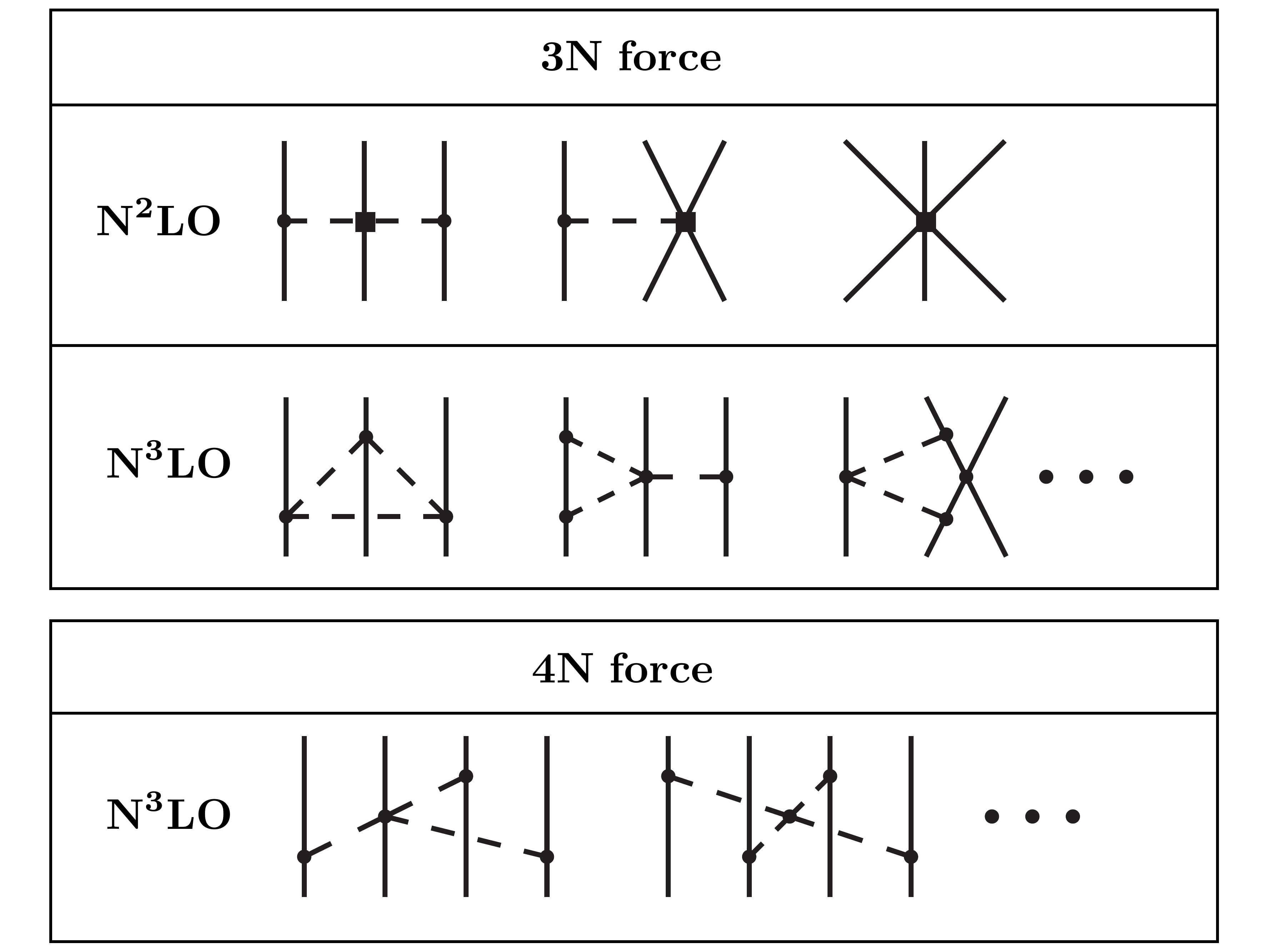}}
\caption{Three- and four-body nuclear forces generated in chiral EFT at orders N$^2$LO and N$^3$LO.}
\label{fig:3}
\end{figure}

Recent years have seen many applications of chiral NN and NNN interactions in various nuclear structure computations and calculations of nuclear and neutron matter (e.g., \cite{HNS2013,LENPIC2016,Coraggio2016} and refs.\,therein). The strength of the ChEFT approach is its well-defined perturbative hierarchy, increasing order by order in the complexity of two- and many-body interactions. Three-body forces first enter at order N$^2$LO as illustrated in Fig.\,\ref{fig:3}, with pion-loop effects added at order N$^3$LO. Four-body forces start to appear at this order as well, actually with no additional low-energy constants. 

ChEFT is basically a perturbative concept. In a nuclear medium, the new "small" scale that enters in addition to three-momenta and pion mass is the Fermi momentum, $p_F$, of the nucleons. Its value at the equilibrium density of $N = Z$ nuclear matter, $n_0 = 2 p_F^3/3\pi^2 = 0.17$ fm$^{-3}$, is $p_F = 1.36$ fm$^{-1} \sim 2 m_\pi$, small compared to the chiral scale $\Lambda_\chi = 4\pi f_\pi \sim 1$ GeV. Expansions in powers of $p_F/\Lambda_\chi < 0.3$ are thus likely to converge. Even at densities $n = 3 n_0$ the Fermi momentum still satisfies $p_F/\Lambda_\chi < 0.4$. These observations suggest the applicability, within limits, of perturbative approaches to the nuclear many-body problem. 

Efforts to understand the properties of nuclear matter from ChEFT generally fall
into two classes. In one approach free-space two- and three-body nuclear potentials, with low-energy
constants fitted to NN scattering phase shifts and properties of bound two- and three-body systems,
are combined with a many-body method of choice to compute the energy per particle. The unresolved
short-distance dynamics is therefore fixed at the few-body level. In the second approach, {\it in-medium chiral perturbation theory}, the energy per particle is constructed as a diagrammatic expansion in the
number of loops involving explicit pion-exchange processes. There are two small scales, $p_F$
and the pion mass. The nuclear matter equation-of-state is given by an expansion in powers of the Fermi momentum.
The expansion coefficients are non-trivial functions of the dimensionless ratio, $p_F/m_\pi$, of the
two relevant low-energy scales in the problem. 

The new feature in nuclear many-body calculations (compared to
scattering processes in the vacuum) is the in-medium nucleon propagator. For a
non-relativistic nucleon with four-momentum $p^\mu =(p_0, \boldsymbol{p})$ it reads:
\begin{equation}
G(p_0, \boldsymbol{p}) = { i \over p_0 - \boldsymbol{p}^{2}/2M_N + i \epsilon} -
2\pi \delta(p_0 - \boldsymbol{p}^{2}/2M_N) \, \theta(p_F -| \boldsymbol{p}|)\, .
\label{imp}
\end{equation}
The first term is the free-space propagator, while the second term accounts for the filled
Fermi sea of nucleons. We note that this expression for $G(p_0, \boldsymbol{p})$ can be rewritten as a sum of
particle and hole propagators, a form more commonly used in non-relativistic many-body perturbation
theory. With the decomposition in Eq.\,(\ref{imp}), closed
multi-loop diagrams representing the energy density can be organized
systematically in the number of medium insertions. Thermodynamics proceeds in an analogous way for the free energy density, with $\theta(p_F -| \boldsymbol{p}|)$ replaced by corresponding thermal distributions \cite{hkwreview}.

Various applications of {\it in-medium chiral perturbation theory} have been reported in recent years (see \cite{hkwreview,hrw2016} and refs. therein). In particular, the equations-of-state for nuclear and neutron matter have been computed, and we will refer to such results for comparison with calculations using non-perturbative functional renormalization group methods in subsequent sections. 

\subsection{Renormalization group strategies} 

In the preceding subsections we have outlined relevant scales and frameworks at the interface between QCD and nuclear physics. 
The focus will now be on functional renormalization group (FRG) methods applied to nuclear many-body systems. The logic is the following: in the domain of spontaneously broken chiral symmetry, the active degrees of freedom are pions and nucleons. A corresponding hierarchy of scales starts off at the chiral symmetry breaking scale, $\Lambda_\chi = 4\pi f_\pi \sim \Lambda_{UV}$. This is taken as the "ultraviolet" (UV) initialization of a RG flow equation that 
describes the evolution of the action down to low-energy "infrared" (IR) scales that characterize the nuclear many-body problem at Fermi momenta $p_F\ll  \Lambda_\chi$. At the UV scale a chiral nucleon-meson Lagrangian based on the linear sigma model with an appropriate potential is chosen as a starting point. The effective action in the IR limit then emerges by solving the non-perturbative RG flow equations. In principle, the physics results in this long-wavelength limit should match those from chiral effective field theory if the in-medium ChPT loop expansion is taken to sufficiently high order. The FRG treatment of fluctuations involving pions has a correspondence (although not one-to-one\footnote{See the instructive discussion in Appendix B of Ref.\,\cite{GL84}.}) in the loop expansion of ChEFT.   
It is then an interesting point to compare (non-perturbative) FRG results with (perturbative) in-medium ChEFT calculations, in particular with reference to convergence issues in the latter as one progresses to higher baryon densities. 

\begin{figure}
\centerline{\includegraphics[width=7cm] {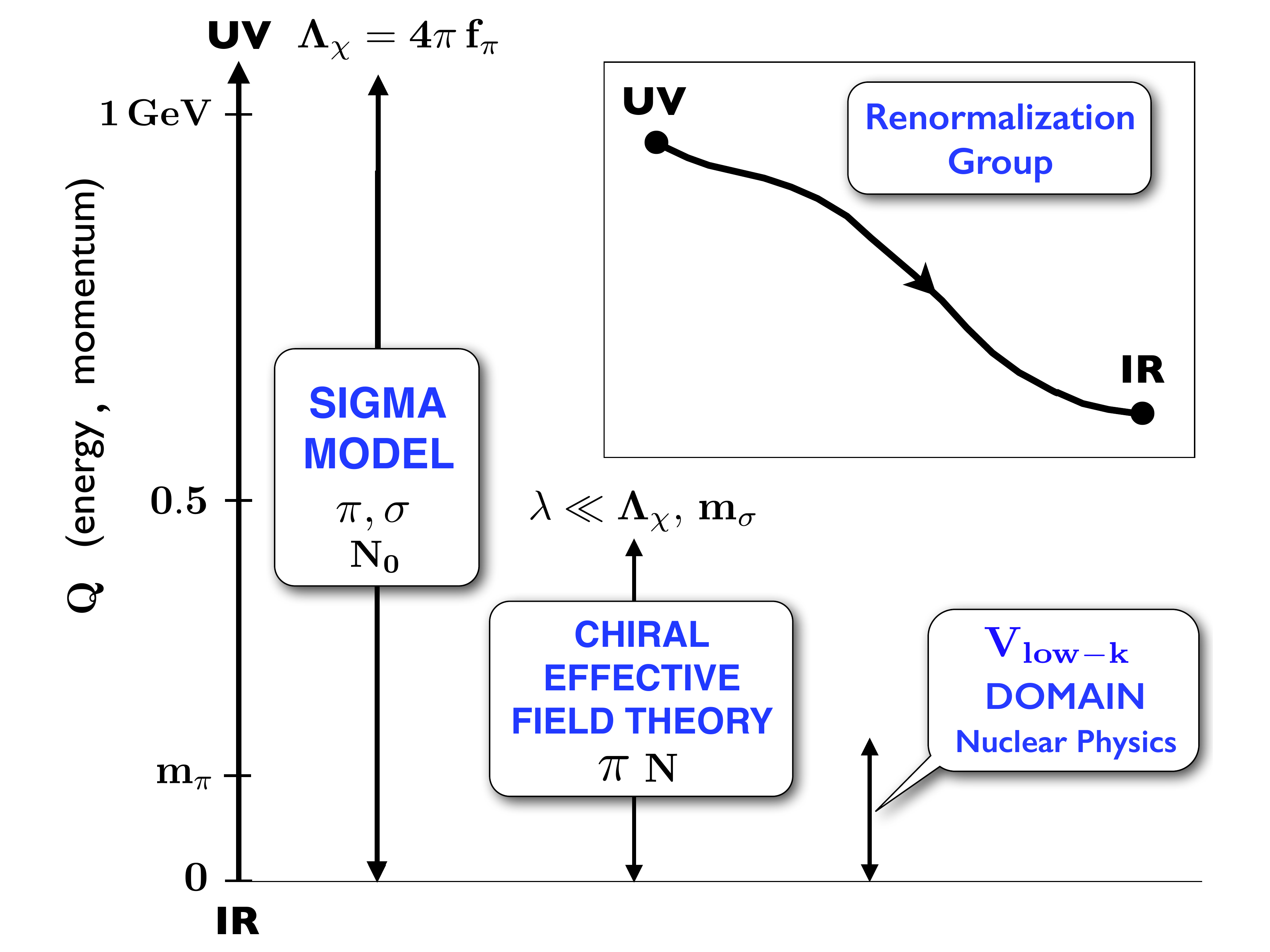}}
\caption{Illustration of renormalization group evolution concepts: from the (UV) scale of spontaneous chiral symmetry breaking in QCD, $\Lambda_\chi = 4\pi f_\pi \sim 1$ GeV, to the low-momentum (IR) scale relevant for nuclear physics.}
\label{fig:4}
\end{figure}

Fig.\,\ref{fig:4} illustrates what has just been described: a linear sigma model, treated non-perturbatively with inclusion of nucleons, undergoes RG evolution from the chiral UV scale, $\Lambda_\chi \sim 1$ GeV, all the way down to the effective action in the long-wavelength limit. At momentum scales below about $0.5$ GeV the heavy $\sigma$ boson decouples and the theory can be rephrased in terms of a non-linear sigma model (ChEFT) with pions as the only remaining "light" degrees of freedom, coupled to the "heavy" nucleons. Note that the linear and non-linear sigma models are not equivalent at a perturbative level: in confrontation with observables, resummations to high orders in the linear sigma model must be perfomed when comparing to leading orders in the non-linear sigma model. But such resummations are just what the FRG equations generate. 

In a broad context, renormalization group methods have previously been applied to a wide spectrum of topics such as pairing in many-fermion systems \cite{Krippa2005}, few-nucleon problems \cite{Birse2013} and the effective field theory treatment of two-body scattering with short-range interactions \cite{Birse2016}. 
At very low momentum scales, a successful strategy using renormalization group techniques is the "$V_{\rm low-k}$" approach, the construction of a universal low-momentum potential for nuclear many-body calculations \cite{Bogner2002,Schwenk2002,BKS2003}. These items will however not be covered in the present review.

We now proceed with an introductory overview of FRG concepts and then turn to the main theme of this review, the FRG approach to nuclear and neutron matter with a variety of applications.

\section{Functional renormalization group: concepts and principles}
\label{sec:FRG}

Quantitative descriptions and predictions for many-body systems close to critical phenomena require the proper inclusion of fluctuations beyond mean-field approximation \cite{Berges2002,Pawlowski2007,SchW2008,KBS2010,Braun2012,PS2012}. Renormalization group methods, widely used in many different fields of physics, are particularly useful to deal with these issues. The RG approach, pioneered by Wilson \cite{Wilson1974}, is based on the concept of the effective action, calculated by successive elimination of the high-energy degrees of freedom of a given theory. The FRG provides a mathematically elegant way for generating this effective action in terms of a formally exact functional differential equation. The present section gives a brief description of the FRG framework. 

\subsection{The effective action}
Consider a field theory of interacting bosons $(\varphi)$ and fermions $(\psi$ and $\bar{\psi})$. Each of the boson and fermion fields may have several components (such as the chiral field $\varphi = (\sigma,\boldsymbol\pi)$ and the Dirac field $\psi_N$ representing the iso-doublet of proton and neutron). For the Fourier-transformed fields, the momentum dependence will in the following be denoted by a subscript, 
\begin{align}\label{eq:field_index}
	\varphi_p\equiv \varphi(p)\,,\quad \psi_p\equiv\psi(p)\,,\quad \bar\psi_p\equiv\bar\psi(p)\,.
\end{align}
It is useful to combine the bosonic and fermionic fields into a multi-component vector
\begin{align}\label{eq:phi_components}
	\xi_p=\begin{pmatrix} \varphi_p \\ \psi_p \\ \bar\psi_{-p}^\top \end{pmatrix}\,,\quad \xi_p^\top:=(\xi^\top)_{-p}=\big(\varphi_{-p}\,,\; \psi_{-p}^\top\,,\; \bar\psi_p \big).
\end{align}
We adopt in the following a condensed (DeWitt) notation: a quantum field $\xi^a(x)$ is simply denoted by $\xi^I$, where now $I$ stands for space-time coordinates $x$ as well as external indices $a$. The product of two fields $\tilde\xi$ and $\xi$ is defined as
\begin{align}
	\begin{aligned}
		\tilde\xi\xi\equiv\tilde\xi_I\xi^I&=\sum_a\int d^4x\;\tilde\xi^a(x)\;\xi^a(x) = \sum_a\int \frac{d^4p}{(2\pi)^4}\;\tilde\xi^a(-p)\;\xi^a(p)\,.
	\end{aligned}
\end{align}
The partition function $Z[J]$ and the Schwinger functional $W[J]$ depend on the external sources $J$: 
\begin{align}\label{eq:partition_function_T0}
	Z[J]=\operatorname{e}^{-iW[J]}=\int \mathcal D\xi\;\operatorname{e}^{i\int d^4x\;\mathcal L+iJ\xi}\,\,,
\end{align}
given a Lagrangian $\mathcal L$. In the spirit of Feynman's path-integral formalism, the partition function is a sum over all possible field configurations, weighted with the respective measure. The expectation value of an operator $\mathcal O$ (generally composed of fields and their derivatives) in the presence of external sources $J$ is given by
\begin{align}
	\langle{\mathcal O}\rangle_J = \frac 1{Z[J]}\int {\mathcal D}\xi \; \mathcal O \exp\left[i\int d^4x\;\mathcal L+iJ\xi\right]\,.
\end{align}

For \mbox{$n\ge3$}, the Schwinger functional $W$ is the generating functional of all connected correlation functions:
\begin{align}
	{\delta^nW[J]\over i\delta J_{i_1}(x_1)\cdots i\delta J_{i_n}(x_n)}\bigg{|}_{J=0}=i\langle\xi_{i_1}(x_1)\cdots\xi_{i_n}(x_n)\rangle_{\rm{conn}}\,.
\end{align}
Moreover, the inverse of the propagator $D_i(x,y)$ associated with the field $\xi_i(x)$ is the connected two-point function given by the second derivative
\begin{align}\label{eq:propagator_1}
{\delta^2W[J]\over i\delta J_i(x_1)\;i\delta J^\top_i(x_2)}\bigg{|}_{J=0} = iD_i^{-1}(x_1,x_2)\,.
\end{align}
The Schwinger functional contains all information of the quantum field theory. In the presence of a small coupling, $W$ can be computed perturbatively with the help of Feynman diagrams. However, at strong coupling or in the vicinity of phase changes such as those in QCD at finite temperature and chemical potential, a non-perturbative treatment is mandatory.

Next, introduce {\it classical} fields $\Phi$ as expectation values of the quantum fields $\xi$ in the background source $J$:
\begin{align}\label{eq:classical_field}
	\Phi_i(x) = \langle\xi_i(x)\rangle_J=-\frac{\delta W[J]}{\delta J_i(x)}\,,
\end{align}
and perform a Legendre transformation of the Schwinger functional:
\begin{align}\label{eq:eff_action}
	\Gamma[\Phi]=-W[J]-J\Phi\,,
\end{align}
which defines the $\Phi$-dependent {\it(quantum) effective action}. From there the quantum equations of motion follow directly:
\begin{align}
	\frac{\delta\Gamma[\Phi]}{\delta\Phi_i(x)}=-J_i(x)\,.
\end{align}
In the absence of external sources, the right-hand side vanishes, so one important feature of the effective action is that it is minimized by the classical field configuration $\Phi(x)$. The minimizing field configuration is often homogeneously distributed\footnote{Exceptions are space-time dependent topological configurations such as solitons, monopoles and domain walls.}, i.e., \mbox{$\Phi(x)\equiv\Phi=\text{const.}$}, one example being the chiral condensate in nuclear matter. In this case, the effective action factorizes:
\begin{align}
	\Gamma[\Phi]=-{\cal VT}\cdot {\cal U}(\Phi)\,,
\end{align}
where ${\cal U}(\Phi)$ is the {\it (quantum) effective potential}, and ${\cal VT}$ is a four-dimensional volume factor, time ${\cal T}$ multiplied by the spatial volume ${\cal V}$. In this case $\Gamma$ is no longer a functional but reduces to a function of $\Phi$. 

While the Schwinger functional $W$ produces all connected diagrams, the effective action $\Gamma$ generates all one-particle irreducible (1PI) diagrams. These diagrams cannot be split into two by cutting a single line. For \mbox{$n\ge3$}, the 1PI diagrams are the $n$-th moments of $\Gamma$:
\begin{align}
	\frac{\delta^n\Gamma[\Phi]}{\delta\Phi_{i_1}(x_1)\cdots\delta\Phi_{i_n}(x_n)} = -i\langle\xi_{i_1}(x_1)\cdots\xi_{i_n}(x_n)\big\rangle_{\text{1PI}}\,.
\end{align}
Moreover, the second derivative of the effective action with respect to the fields is the inverse propagator,
\begin{align}\label{eq:propagator_2}
	\frac{\delta^2\Gamma[\Phi]}{\delta\Phi^\top_i(x_1)\;\delta\Phi_i(x_2)} = iD_i^{-1}(x_1,x_2)\,.
\end{align}
The quantum effective action incorporates all correlation functions already at tree-level. It is the low-energy action with all quantum fluctuations integrated out. If the effective action is known, the low-energy theory is completely determined. Hence it is of great interest to develop efficient methods for computing this effective action. In the weak coupling limit (or in the large-$N_c$ limit) the theory can be treated perturbatively. Then, up to one-loop order, the effective action can be computed (see, e.g., \cite{weinbergQFT1996}) as
\begin{align}
	\Gamma[\Phi]=S[\Phi] + \frac i2\operatorname{Tr}\ln\frac{\delta^2S}{\delta\Phi^2}+\text{higher orders} ~~,
\end{align}
with $S = \int d^4x\,{\cal L}$. In the next subsection these concepts will be extended to finite temperatures.

\subsection{Finite temperature}

The grand-canonical potential in statistical mechanics is formally similar to the partition function (\ref{eq:partition_function_T0}) of a quantum field theory, with a close correspondence between time and inverse temperature.  
In the Matsubara formalism (see e.g., \cite{Bellac2000}) a thermal field theory is constructed by Wick-rotating the time dimension, $t=x^0\rightarrow-i\tau=-ix^4_{\rm E}$. The time integral $\int dt$ is converted into an integral over Eucidean time $\tau$. The $\tau$-dimension is compactified on a circle with radius $\beta= 1/T$, the inverse temperature. The commutation and anticommutation relations imply that bosonic fields $\varphi$ are periodic while fermionic fields $\psi$ are antiperiodic, i.e.,
\begin{align}
	\varphi(\boldsymbol x,\beta) = \varphi(\boldsymbol x,0)\,,\quad \psi(\boldsymbol x,\beta) = -\psi(\boldsymbol x,0)\,.
\end{align}
From a Fourier analysis it follows that the corresponding momenta in the finite $\tau$-direction are discrete, i.e., $p_4\rightarrow -\omega_l$, where $\omega_l$ with integer $l$ are the {\it Matsubara frequencies},
\begin{align}\label{eq:matsubara}
	\begin{alignedat}{2}
		& \omega_l=2l\pi T\,, \qquad && \rm{for~ bosons,} \\
		& \omega_l=(2l+1)\pi T\,, \qquad && \rm{for ~fermions.}
	\end{alignedat}
\end{align}
The four-dimensional momentum space integral at zero temperature is thus replaced by a three-dimensional integral and a summation over Matsubara frequencies:
\begin{align}\label{eq:4d_to_3d}
	\int\frac{d^4p}{(2\pi)^4}\quad\rightarrow\quad T\sum_{\omega_l}\int \frac{d^3p}{(2\pi)^3}\,.
\end{align}
Altogether, the Minkowskian partition function at $T=0$ and $\mu=0$, given by
\begin{align}
	Z_{\rm M}[J_i]=\int\mathcal D\psi\,\mathcal D\psi^\dagger\,\mathcal D\varphi\;\exp\big[i \left(S_{\rm M}+J\xi\right)\big]~,
\end{align}
is replaced by a partition function in Euclidean space at finite temperature $T$ and chemical potential $\mu$:
\begin{align}
	Z_{\rm E}[J,T,\mu]=\int\mathcal D\psi\,\mathcal D\psi^\dagger\,\mathcal D\varphi\;\exp\bigg(-S_{\rm E}+\int_0^\beta d\tau\int d^3x \; \mu\,\psi^\dagger\psi+J\xi\bigg).
\end{align}
An effective action can now be derived in the presence of non-zero temperature and chemical potentials. The theory is  defined in Euclidean space, so the definitions differ by factors of $i$ and $-1$ as compared to the vacuum case. The Schwinger functional $W_{\rm E}[J,T,\mu]$ is the logarithm of the partition function, 
\begin{align}
	{\rm e}^{W_{\rm E}[J,T,\mu]}=Z_{\rm E}[J,T,\mu]\,.
\end{align}
The expectation value of $\xi$ in the background of the source $J$ defines the classical field
\begin{gather}\label{eq:expectation_value}
	\Phi_i(x)=\langle\xi_i(x)\rangle_{J,T,\mu}=\frac{\delta W_{\rm E}[J,T,\mu]}{\delta J_i(x)}\,.
\end{gather}
The effective action is the Legendre transform of the Schwinger functional,
\begin{align}
	\Gamma_{\rm E}[\Phi,T,\mu]=-W_{\rm E}[J,T,\mu]+J\,\Phi\,,
\end{align}
which implies the quantum equations of motion:
\begin{gather}\label{eq:EOM}
	\frac{\delta\Gamma_{\rm E}[\Phi,T,\mu]}{\delta\Phi_i(x)}=J_i(x)\,.
\end{gather}
The inverse propagator is given by
\begin{align}\label{eq:propagator_finite_T_2}
	\frac{\delta^2\Gamma_{\rm E}[\Phi,T,\mu]}{\delta\Phi^\top_i(x_1)\;\delta\Phi_i(x_2)} = D^{-1}_i(x_1,x_2)\,.
\end{align}
As in the vacuum case, for spatially constant solutions $\Phi$ of the quantum equations of motion (such as the chiral condensate), the effective action simplifies to
\begin{align}\label{eq:eff_action_finite_T}
	\Gamma_{\rm E}[\Phi,T,\mu] = \beta {\cal V}\cdot {\cal U}_{\rm E}(\Phi,T,\mu)\,,
\end{align}
where $\beta {\cal V} = {\cal V}/T$ is the Euclidean volume. The effective potential ${\cal U}_{\rm E}(\Phi,T,\mu)$ is again only a function of $\Phi$ and no longer a functional. Let $\Phi_{\rm {min}}(T,\mu)$ be the field at the minimum of the potential ${\cal U}$ for given temperature and chemical potential. If the potential is evaluated at its minimum and the external sources are set to zero, the partition function takes the simple form
\begin{align}\label{eq:partition_function}
	Z_{\rm E}(T,\mu) = {\rm e}^{-\beta {\cal V} \, \Omega(T,\mu)}\,,
\end{align}
where $\Omega$ is the {\it grand-canonical potential},
\begin{align}\label{eq:grandcanonical_potential}
	\Omega(T,\mu) = {\cal U}\big(\Phi_{\rm{min}}(T,\mu),T,\mu\big)\,.
\end{align}
All relevant quantities in thermodynamic equilibrium can be computed from the grand-canonical potential $\Omega$. In particular, the pressure $P$, the fermion density $n$, the entropy density $s$, and the energy density $\varepsilon$, are given by:
\begin{align}\label{eq:thermodynamics}
	\begin{alignedat}{3}
		P&=-\Omega(T,\mu)\,, \qquad &n&=-\frac{\partial \Omega(T,\mu)}{\partial\mu}\,,\\
		s&=-\frac{\partial \Omega(T,\mu)}{\partial T}\,, &\varepsilon&=-P+\mu n +Ts\,.
	\end{alignedat}
\end{align}
The next step is now to introduce a framework and tools for computing the effective action efficiently. 

\subsection{The functional renormalization group}

Wilson's original idea was to study how a theory changes as the momentum scale is lowered to a new scale $c\Lambda$, with \mbox{$c<1$}. His proposal was to integrate out momentum shells, i.e., all contributions to the path integral with $c\Lambda<|p|
<\Lambda$. After a rescaling procedure, a new effective theory results, now at a scale $c\Lambda$. All operators allowed by the underlying symmetries have to be included in this procedure. Their respective couplings are changing according to the renormalization group equations. In this way, new operators naturally appear in the effective Lagrangian. This is an entirely non-perturbative approach that does not depend on any small couplings.

There are several different implementations of Wilson's idea to compute the effective action, sometimes collectively called {\it exact renormalization group equations} (ERGE). The first one, Wilson's ERGE, employs the idea of momentum shells in a direct way. One rewrites the partition function in Euclidean space as
\begin{align}\label{eq:wilson}
	Z=\int[{\cal D}\xi]_{|p|<\Lambda}\;{\rm e}^{-S_\Lambda[\xi]} = \int[{\cal D}\xi]_{|p|<c\Lambda}\int [{\cal D}\xi]_{c\Lambda< |p|<\Lambda}\;{\rm e}^{-S_\Lambda[\xi]}\,,
\end{align}
where the path integral measure over the fields is split according to the contributing momentum scales. If one now defines
\begin{align}
	{\rm e}^{-S_{c\Lambda}[\xi]}\equiv \int[{\cal D}\xi]_{c\Lambda<|p|<\Lambda}\;{\rm e}^{-S_\Lambda[\xi]}\,,
\end{align}
then the partition function reads
\begin{align}
	Z=\int[{\cal D}\xi]_{|p|<c\Lambda}\;{\rm e}^{-S_{c\Lambda}[\xi]}\,.
\end{align}
The path integral has the same structure as in Eq.~\eqref{eq:wilson}. However, the momenta of the fields contributing to the partition function are now restricted to the range \mbox{$|p|<c\Lambda$}. 

Wilson's ERGE is straightforward to write down but almost impossible to tackle numerically. In general, non-local interactions are generated and a derivative expansion is not possible. An improvement proposed in Ref.~\cite{Polchinski1984} is to replace the sharp momentum cutoff by a softer one (see \cite{Rosten2012} for a review). A regulator term,
\begin{align}\label{eq:S_kR_k}
	\Delta S_k[\xi]=\frac 12\xi^\top R_k\, \xi\,,
\end{align}
is added to the action, where $R_k$ is a regulator function and $k$ is a renormalization scale with the dimension of a momentum. The renormalization scale plays the role of an intrinsic resolution length, $\sim k^{-1}$. The regulator function has to satisfy three conditions:
\begin{enumerate}
	\item \label{page:conditions} First, $R_k$ is supposed to regularize the theory in the infrared regime. Therefore, it must be positive for small momenta, i.e.,
		\begin{align}
			\lim_{p^2/k^2\rightarrow 0}R_k(p^2)>0\,.
		\end{align}
	\item Next, $R_k$ must vanish for $p^2\gg k^2$ and, in particular, for $k\rightarrow 0$, i.e.,
		\begin{align}
			\lim_{k^2/p^2\rightarrow 0}R_k(p^2)=0\,.
		\end{align}
		In this way, it is ensured that the partition function $Z_k$ reduces to the full low-energy partition function as $k$ goes to zero.
	\item Finally, for $k\rightarrow \Lambda$, the regulator function has to be large. In this way, one recovers the bare action as the starting point at a UV scale, $k=\Lambda$.
\end{enumerate}
\begin{figure}
	\centering
	\includegraphics[width=0.4\textwidth]{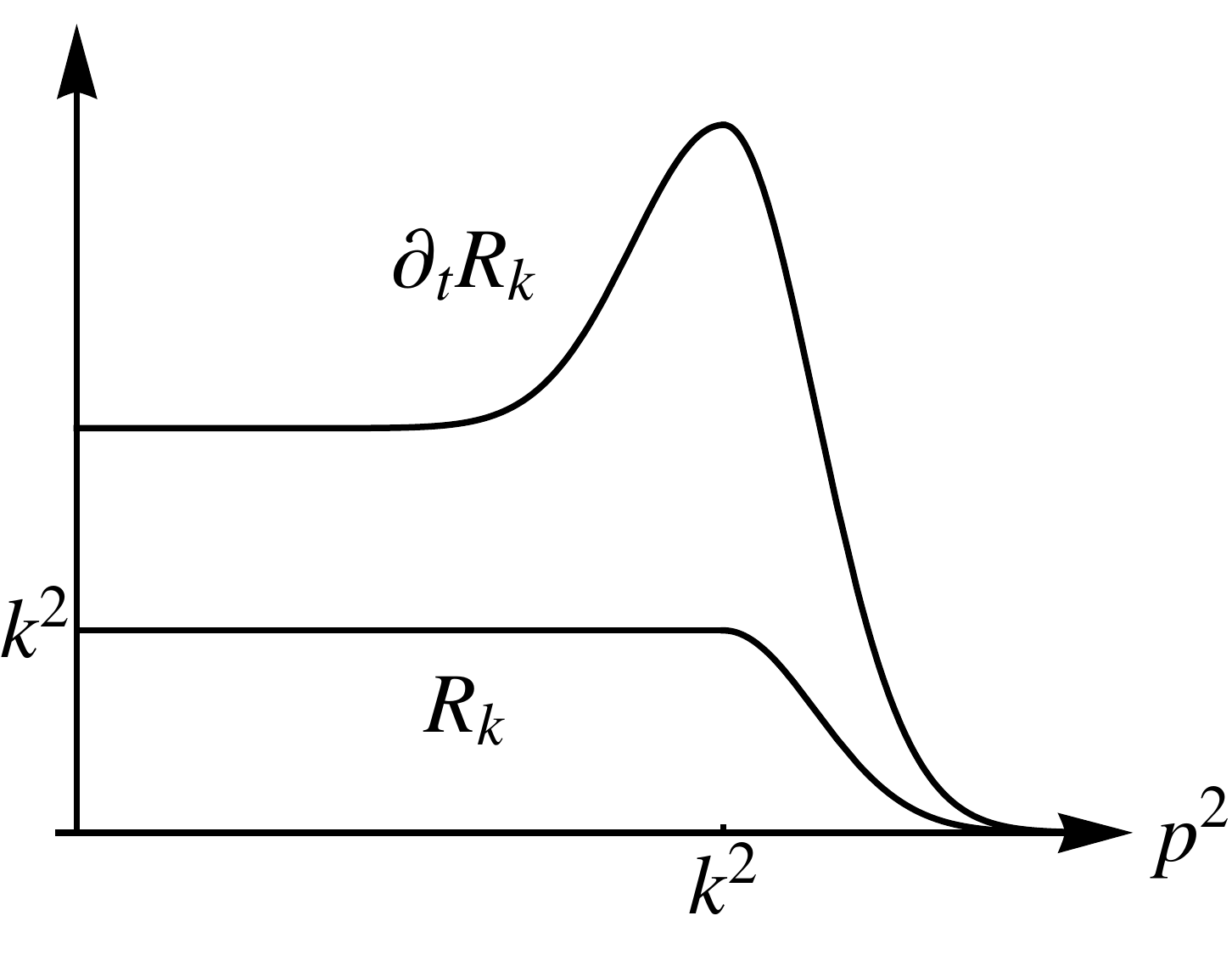}
	\caption{Typical behaviour of the regulator function $R_k$ and its derivative $\partial_tR_k=k\partial_k R_k$.}
	\label{fig:5}
\end{figure}
The regulator and its derivative typically behave as shown in Fig.\,\ref{fig:5}. Modes with squared momentum below $k^2$ are equipped with an effective squared mass $m^2\sim k^2$. Higher momentum modes are not altered but eliminated by the vanishing regulator. The exact form of the regulator function is in principle arbitrary. The flow pattern in the space of $k$-dependent actions will change, but the endpoint in the limit $k\rightarrow 0$ will be the same\footnote{In practice, of course, approximations must be made, and the endpoint will depend somewhat on the specific choice of the regulator function.}.

\begin{figure}
	\centering
	\begin{overpic}[width=0.5\textwidth]{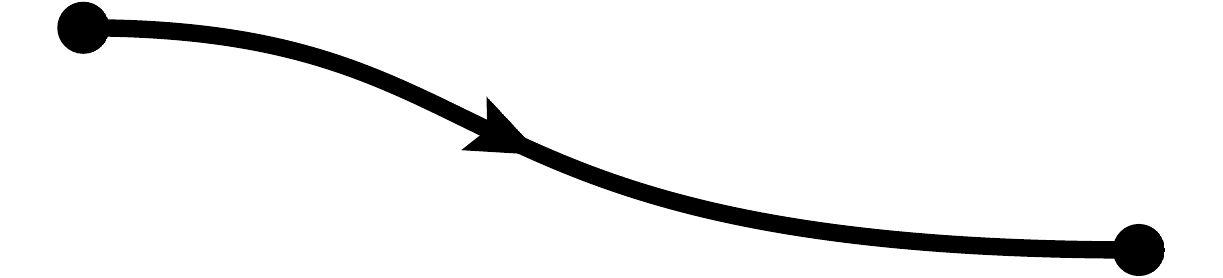}
		\put(-35,18.3){$\Gamma_{k=\Lambda}[\Phi]=S$}
		\put(53,10){$\Gamma_k[\Phi]$}
		\put(100,0){$\Gamma_{k=0}[\Phi]=\Gamma[\Phi]$}
	\end{overpic}
	\caption{The flow of the effective action $\Gamma_k$ in the infinite-dimensional space spanned by all operators allowed by the symmetries of the theory. The dynamics as a function of the renormalization scale $k$ is governed by the flow equation.\label{fig:6}}
\end{figure}

The starting point of the FRG is a $k$-dependent action,
\begin{align}
	S_k[\xi,J]=S[\xi]+\Delta S_k[\xi]-J\,\xi\,,
\end{align}
where $S[\xi]$ is the classical action and $\Delta S_k[\xi]$ is defined in Eq.~\eqref{eq:S_kR_k}. The regulator term $R_k$ has to satisfy the conditions as above on page~\pageref{page:conditions}. The construction of the $k$-dependent effective action parallels the discussion leading to Eq.~\eqref{eq:eff_action}. The partition function is
\begin{align}\label{eq:partition_function_k}
	Z_k[J]={\rm e}^{W_k[J]}=\int \mathcal D\xi\, {\rm e}^{-S_k[\xi,J]}\,.
\end{align}
The classical field,
\begin{align}\label{eq:classical_field_RG}
	\Phi=\frac{\delta W_k[J]}{\delta J} = \langle{\xi}\rangle_J\,,
\end{align}
is computed from the $k$-dependent Schwinger functional $W_k[J]$. The {\it $k$-dependent effective action} is defined as a slight modification of the Legendre transform\footnote{
	This modification establishes the correct connection between $\Gamma_k$ and the microscopic action $S$ in the limit $k\rightarrow\Lambda$.
}. If $\widetilde\Gamma_k[\Phi]$ denotes the Legendre transform of $W_k[J]$, then
\begin{align}\label{eq:effGammaLegendre}
	\Gamma_k[\Phi]=\widetilde\Gamma_k[\Phi]-\Delta S_k[\Phi]=- W_k[J]+\Phi^\top J-\frac 12\Phi^\top R_k\Phi.
\end{align}
The action $\Gamma_k$ is designed in such a way that it interpolates between the microscopic action (at the UV scale $k=\Lambda$) and the full quantum effective action (in the limit $k\rightarrow0$), as illustrated in Fig.\ref{fig:6}.

Derivatives of the effective action with respect to the fields (with the short-hand notation $\delta_{\Phi_p}\equiv\frac\delta{\delta\Phi_p}$) are defined as follows:
\begin{align}
	\Gamma^{(n,m)}_k(p_1,\ldots,p_n,q_1,\ldots,q_m) \equiv {\stackrel{\rightarrow}\delta}_{\Phi^\top_{p_1}} \cdots {\stackrel{\rightarrow}\delta}_{\Phi^\top_{p_n}} \Gamma_k {\stackrel{\leftarrow}\delta}_{\Phi_{q_1}} \cdots {\stackrel{\leftarrow}\delta}_{\Phi_{q_m}}\,.
\end{align}
For the flow equation the second derivative of the effective action is needed. This is the following matrix:
\begin{align}\label{eq:gamma11}
	\Gamma_k^{(1,1)}(p,p')&=\stackrel{\rightarrow}\delta_{\Phi^\top_p}\Gamma_k\stackrel{\leftarrow}\delta_{\Phi_{p'}}= \nonumber \\
	&=
		\begin{pmatrix} 
			\stackrel{\rightarrow}\delta_{\phi_{-p}}\Gamma_k\stackrel{\leftarrow}\delta_{\phi_{p'}} & \stackrel{\rightarrow}\delta_{\phi_{-p}}\Gamma_k\stackrel{\leftarrow}\delta_{\Psi_{p'}} & \stackrel{\rightarrow}\delta_{\phi_{-p}}\Gamma_k \stackrel{\leftarrow}\delta_{{\bar\Psi^\top}_{-p'}} \\
			\stackrel{\rightarrow}\delta_{\Psi^\top_{-p}}\Gamma_k\stackrel{\leftarrow}\delta_{\phi_{p'}} & \stackrel{\rightarrow}\delta_{\Psi^\top_{-p}}\Gamma_k\stackrel{\leftarrow}\delta_{\Psi_{p'}} & \stackrel{\rightarrow}\delta_{\Psi^\top_{-p}}\Gamma_k \stackrel{\leftarrow}\delta_{{\bar\Psi^\top}_{-p'}} \\
			\stackrel{\rightarrow}\delta_{\bar\Psi_{p}}\Gamma_k\stackrel{\leftarrow}\delta_{\phi_{p'}} & \stackrel{\rightarrow}\delta_{\bar\Psi_{p}}\Gamma_k\stackrel{\leftarrow}\delta_{\Psi_{p'}} & \stackrel{\rightarrow}\delta_{\bar\Psi_{p}}\Gamma_k \stackrel{\leftarrow}\delta_{{\bar\Psi^\top}_{-p'}}
		\end{pmatrix}~,
\end{align}
where $\phi$ and $\Psi$ are the bosonic and fermionic components of the classical field $\Phi$ in analogy to Eq.\,(\ref{eq:phi_components}). With these definitions we can now write down the {\it flow equation} for the effective action $\Gamma_k$ as it was first derived by Wetterich \cite{Wetterich1993}:
\begin{align}\label{eq:Wetterich}
	\begin{aligned}
		k\,\frac{\partial\Gamma_k[\Phi]}{\partial k}=
		\begin{aligned}
			\vspace{1cm}
			\includegraphics[width=0.12\textwidth]{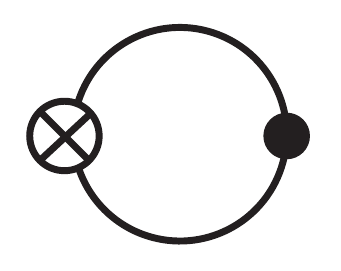}
		\end{aligned} \vspace{-1cm}=\frac 12 {\rm Tr}\left[k\frac{\partial R_k}{\partial k}\cdot\Big(\Gamma_k^{(1,1)}[\Phi]+R_k\Big)^{-1}\right]~~.
	\end{aligned}
\end{align}
Wetterich's flow equation relates the change of the $k$-dependent effective action to a one-loop diagram. In general, the flow equation is a functional differential equation because $\Phi$ depends on space-time coordinates.  The trace goes over internal and space-time indices. Moreover, it extends over the bosonic and fermionic subspaces as indicated in the structure of the matrix~\eqref{eq:gamma11}. It is understood that the fermionic subspace comes with an additional minus sign. In the pictorial version of the flow equation, the loop stands for the propagator of any of the active (bosonic and fermionic) degrees of freedom; the dot indicates that this is the {\it full} propagator. The cross symbolizes the insertion of the regulator function and its derivative, $k\frac{\partial R_k}{\partial k}$.

The $k$-dependent effective action interpolates by construction between the microscopic and quantum effective action, which can be seen as follows:
\begin{enumerate}
	\item By expanding the action around the classical (background) field $\Phi$, it is easy to derive from Eq.~\eqref{eq:effGammaLegendre} that
		\begin{align}
			{\rm e}^{-\Gamma_k[\Phi]}=\int\mathcal D\xi\,{\rm e}^{-S[\Phi+\xi]+\frac{\delta\Gamma_k}{\delta\Phi}\xi-\frac 12\xi^\top R_k\xi}.
		\end{align}
		For $k\simeq\Lambda$, the regulator function $R_k$ is required to be large. Consequently, $-\frac 12\xi^\top R_k\xi$ leads to a delta-functional $\delta[\xi]$. Only the classical field configuration contributes to the path integral and therefore $\Gamma_k[\Phi]\xrightarrow[]{k\rightarrow\Lambda}S[\Phi]$. The regulator equips all particles with large effective masses. Consequently, all fluctuations around the classical field configuration are highly suppressed and only the classical configuration contributes.
	\item In the opposite limit, when $k\rightarrow 0$, the regulator term vanishes. All modes are fully integrated out, so $\Gamma_k\rightarrow\widetilde\Gamma_k\rightarrow\Gamma$.
\end{enumerate}
In general, the effective action contains all possible operators that respect the symmetries of the underlying theory. The relative strength of these operators is then computed by Wetterich's flow equation. In this way, one gets an infinite tower of coupled equations. It is necessary to truncate the set of flow equations such that the most relevant operators are kept. In the {\it derivative expansion} \cite{Golner1986}, only powers of derivatives of the fields up to a certain order are kept. All higher derivative couplings are ignored. This approach is often combined with an expansion in powers of the fields. The derivative expansion provides a useful scheme that allows systematic improvement by going to higher orders. 

In practice, most calculations work at leading order in the derivative expansion. It is instructive at this point to recall an $O(N)$-symmetric model with scalars $\phi_i$, $i=1,\ldots,N$. The square of the fields is invariant under $O(N)$ transformations and denoted $\chi=\frac 12\sum_i\phi_i\phi_i$. To leading order in the derivative expansion, the effective action has the form \cite{Wetterich1993b}
\begin{align}\label{eq:gamma_k_LPA}
	\Gamma_k = \int d^4x\;\Big[\frac 12 Z_k(\chi)\sum_i\partial_\mu\phi_i\;\partial^\mu\phi_i+\frac 14 Y_k(\chi)\;\partial_\mu\chi\;\partial^\mu\chi+U_k(\chi)\Big].
\end{align}
A non-trivial wave-function renormalization $Z_k$ leads to an anomalous dimension. If $Z_k\equiv1$ and $Y_k\equiv0$ one speaks of the {\it local potential approximation} (LPA, \cite{Nicoll1974}). A study of an $O(N)$ model indeed indicates that the anomalous dimension is negligible \cite{Tetradis1994}. In the LPA, it is possible to derive a simple optimized regulator function \cite{Litim2001} for bosonic and fermionic fields:
\begin{align}\label{eq:regulator}
	\begin{aligned}
R_k^{\rm bos} & =(k^2-p^2)\,\theta(k^2-p^2)\,, \\
R_k^{\rm fer} &= \begin{pmatrix} 0 & ip_\mu(\gamma_\mu^{\rm E})^\top \\ i p_\mu\gamma_\mu^{\rm E} & 0 \end{pmatrix} \left(\sqrt{k^2\over p^2}-1\right)\,\theta(k^2-p^2)\,\,,
	\end{aligned}
\end{align}
where $\gamma_\mu^{\rm E}$ denotes the Euclidean representation of the Dirac gamma matrices. The role of this so-called Litim regulator can be understood quite intuitively in this case. On the one hand, its appearance in the full loop in the denominator acts as an IR cutoff: for small momentum modes with $p^2<k^2$, the squared momentum gets replaced by the larger regulator scale $k^2$. Even massless fluctuations acquire a finite effective mass and are therefore smoothed out. On the other hand, the insertion $~k\frac{dR_k}{dk}$ in the numerator regulates in the ultraviolet regime: modes with $p^2>k^2$ do not contribute to the flow. Therefore the flow equation is manifestly finite in both UV and IR regimes. 

The Litim regulator is optimal in the sense that it maximizes the denominator,
\begin{align}
	\min_{q^2\ge 0}\left[ {\stackrel{\rightarrow}\delta}_{\Phi^\top_{q}} \; \Gamma_k {\stackrel{\leftarrow}\delta}_{\Phi_{q}} + R_k(q^2)\right] = Ck^2>0\,.
\end{align}
The existence of a gap $C>0$ is necessary in order for the flow to be finite. The maximum is achieved for a regulator that renders the left-hand side momentum independent. Together with the constraints on $R_k$ stated on page~\pageref{page:conditions}, the regulator is then uniquely fixed. Moreover, it can be shown that this choice of regulator leads to the fastest decoupling of heavy modes~\cite{Litim2001}.

At finite temperatures, the time integral is converted into a sum over Matsubara frequencies according to Eq.~\eqref{eq:4d_to_3d}. The inverse temperature acts already as a UV regulator in the imaginary-time direction. The sum over Matsubara frequencies converges and the regulator function $R_k$ acts on the three-momenta only. The regulator takes the form \cite{Litim2006,Blaizot2007}:
\begin{align}\label{eq:regulator_finite_T}
	\begin{aligned}
		R_k^{\rm bos} &=(k^2- {\boldsymbol p}^2)\theta(k^2-{\boldsymbol p}^2)\,, \\
		R_k^{\rm fer} &=\begin{pmatrix} 0 & ip_i(\gamma_i^{\rm E})^\top \\ i p_i\gamma_i^{\rm E} & 0 \end{pmatrix} \left(\sqrt{k^2 \over {\boldsymbol p}^2}-1\right)\theta(k^2-{\boldsymbol p}^2)\,.
	\end{aligned}
\end{align}
Because of the structure of the regulators, the dependence on three-momenta is eliminated and only an integral over a theta-function remains. Then it is often possible to perform the Matsubara sum, and the flow equation can be written down analytically.

Numerical strategies used for solving the flow equation are briefly described in the Appendix.
\section{Functional renormalization group and nuclear many-body problem}\label{sec:nuclFRG}
Theoretical investigations of nuclear and neutron-rich matter have quite well converged in recent years. Different approaches, such as chiral effective field theory (ChEFT \cite{hkwreview, Fritsch2005, Fiorilla2012}), chiral Fermi liquid theory \cite{Holt2013}, as well as Quantum Monte Carlo (QMC) methods using either phenomenological interactions \cite{Gandolfi2012,Gandolfi2014} or ChEFT potentials \cite{Gezerlis2013,Roggero2014,Holt2014}, show satisfactory agreement in their common overlap regions of applicability at lower densities.

At higher densities, effects of three-body forces and higher-order pion-exchange processes become increasingly important \cite{Hebeler2010, Tews2016}, and it is crucial that any realistic approach takes into account fluctuations and correlations generated by those mechanisms. In Section \ref{sec:FRG} the functional renormalization group has been described as a powerful method to study the effects of fluctuations in a consistent and fully non-perturbative way. 
We now focus on the central theme of this review: developing the thermodynamics of symmetric and asymmetric nuclear matter based on a chiral nucleon-meson effective action with its scale evolution controlled by the functional renormalization group equations. Physics issues of high current interest will be investigated: the nuclear liquid-gas phase transition and its disappearance as one proceeds from isospin-symmetric to asymmetric nuclear matter and neutron matter; the quest for chiral symmetry restoration in dense matter and the crucial role played by fluctuations beyond mean-field approximation in shifting the chiral transition to extremely high baryon densities; the in-medium pion mass as a test case for pionic fluctuations; and a study of the mass-radius relation for neutron stars with emphasis on the constraints provided by the existence of two-solar-mass stars. Throughout these discussions, while primarily concentrating on the non-perturbative FRG approach, we will have an eye on results from in-medium chiral perturbation theory for comparison. In order to set a baseline we start from the mean-field limit, mainly for later comparisons with more complete FRG results including fluctuations.

\subsection{Chiral nucleon-meson model}

Our starting point is now the chiral nucleon-meson model of Eqs.\,(\ref{eq:ChMN1} - \ref{eq:ChMN2}). Here it first serves as a basis for mean-field studies. Later it will define the input at the UV scale for the FRG equations. From the beginning we shall assume that the short-distance nucleon-nucleon dynamics can be parametrized in terms of approximately constant, massive isoscalar- and isovector-vector fields. Assuming furthermore an isotropic medium, only the temporal components, $v_0$ and ${\boldsymbol \rho}_0$, of these vector fields survive. The starting Lagrangian (in Minkowski space) is therefore:
\begin{eqnarray}
{\cal L} &=& \bar{N}\left[i\gamma_\mu\partial^\mu - g(\sigma + i\gamma_5\,\boldsymbol{\tau\cdot\pi})\right]N - N^\dagger\left(g_v\,v_0 + g_\tau\,\boldsymbol{\tau\cdot\rho}_0\right)N\nonumber  \\ &+& {1\over 2}\partial_\mu \sigma \partial^\mu \sigma + {1\over 2}\partial_\mu \boldsymbol{\pi}\cdot\partial^\mu \boldsymbol{\pi} - {\cal U}(\sigma, \boldsymbol{\pi})+{1\over 2}m_V^2\left(v_0^2 + \boldsymbol{\rho}_0^2\right)~~.
\label{eq:ChMNmf}
\end{eqnarray}
As mentioned before, the vector couplings $g_v, g_\tau$ and the (large) vector boson mass $m_V$ are of no individual relevance; only their ratio plays a role. The potential ${\cal U}(\sigma, \boldsymbol{\pi})$ will be constructed such as to be consistent with pion-nucleon data and selected ground state properties of nuclear matter.

Finite temperatures and chemical potentials require the following modifications. In the Matsubara formalism, space-time is Wick-rotated to Euclidean space. Time components are rotated as $x^0\rightarrow-i\tau$. The $\tau$-dimension is compactifed on a circle, such that $\tau$ is restricted to $[0,\beta]$ with the inverse temperature $\beta=1/T$. Time-integrals are replaced by $-i\int_0^\beta d\tau$. Boson and fermion fields are periodic or anti-periodic, respectively, under $\tau\rightarrow \tau+\beta$. The Minkowski-space action $S=\int d^4x\;\mathcal L$ is transformed into the Euclidean action $S_{\rm E}=\int_0^\beta d\tau\int d^3x\;{\cal L}_{\rm E}$ with
\begin{align}
		{\cal L}_{\rm E}&=\bar{N}\left[\gamma_\mu\partial_\mu+g(\sigma+i\gamma_5\boldsymbol\tau\cdot\boldsymbol\pi)\right]N - i N^\dagger \left(g_v v_0+g_\tau \boldsymbol\tau\cdot\boldsymbol\rho_0\right) N \nonumber\\
		&\hspace{0.4cm}+\frac 12\big(\partial_\mu\sigma\;\partial_\mu\sigma+\partial_\mu\boldsymbol\pi\cdot\partial_\mu\boldsymbol\pi\big)+\mathcal U(\sigma, \boldsymbol\pi) - {1\over 2}m_V^2\left(v_0^2 + \boldsymbol{\rho}_0^2\right)~,
\end{align}
where the Dirac gamma matrices and four-gradients are now understood to be written in Euclidean form, dropping the index "$\rm E$" for convenience. For the temporal components of vector fields the Euclidean representation $v_{\rm E}^4 = -iv_0$ has been used. 

Baryon-number conservation is implemented by the baryon chemical potential, $\mu$, as a Lagrange parameter. In the path integral formalism the exponent picks up an additional term $-\mu\int_0^\beta d\tau\int d^3x\;N^\dagger N$ proportional to the baryon number. For isospin asymmetric matter, separate proton and neutron chemical potentials, $\mu_p$ and $\mu_n$, can be introduced. The partition function gets an additional factor $\exp(\beta\mu_pN_p+\beta\mu_nN_n)$, with proton and neutron numbers,
		\begin{align}
			N_p=\int d^3x\;\psi_p^\dagger(x)\psi_p(x)\,,\quad N_n=\int d^3x\;\psi_n^\dagger(x)\psi_n(x)~,
		\end{align}
related to the charge
\begin{align}
			Q_N=\int d^3x\;N^\dagger(x)\,\tau_3\,N(x) = N_p - N_n~.
		\end{align}
Exchange processes involving charged pions imply that $Q_N$ alone is not a conserved charge. With $\boldsymbol{\pi} = (\pi_1, \pi_2, \pi_3)$ and $\pi_\pm = \pi_1\pm i\pi_2$, the isospin Noether current is in fact
\begin{align}
J_\mu^3 &= \bar{N}\tau_3\,\gamma_\mu N + i(\pi_-\partial_\mu\pi_+-\pi_+\partial_\mu\pi_-)\nonumber\\&=\bar{N}\tau_3\gamma_\mu N + 2(\pi_2\partial_\mu\pi_1-\pi_1\partial_\mu\pi_2)~,
\end{align}
and the associated conserved charge is
\begin{align}\label{eq:charge_isospin}
			Q=\int d^3x\;\left(N^\dagger\tau_3\,N + 2\pi_2{\partial\pi_1\over\partial t} - 2\pi_1{\partial\pi_2\over\partial t}\right)\,.
		\end{align}
In the absence of a term which explicitly breaks isospin symmetry, the model depends on the chiral meson fields only through their square,
\begin{equation}
\chi = \frac 12(\sigma^2+\boldsymbol\pi^2)~.\nonumber
\end{equation}
The charge \eqref{eq:charge_isospin} corresponds to isospin-3 rotations that involve the charged pionic fields $\pi_{1,2}$. In matter with different proton and neutron chemical potentials, the partition function is no longer invariant under the full isospin group $\operatorname{SO}(4)$ but broken down to an $\operatorname{SO}(2)\times\operatorname{SO}(2)$ subgroup. The invariant squares transforming under this subgroup are
\begin{align}\label{eq:chi12}
	\chi_1 = \frac 12(\pi_1^2+\pi_2^2)\,,\quad \chi_2=\frac 12(\pi_3^2+\sigma^2)\,.
\end{align}
A non-vanishing expectation value of $\chi_1$ would be identified as a charged pion condensate. 

\subsection{Mean-field approximation}

As a first approximation to the full path integral, all bosonic fields are replaced by their (temperature and density-dependent) expectation values and are treated as background fields: $\phi = \langle\phi\rangle + \delta \phi$ with fluctuating parts set to zero, $\delta \phi = 0$. This is the mean-field approximation, applied to the ChNM model in the present context \cite{FW2012, DHKW2013, DW2015}. Only rotationally invariant solutions are considered, so the spatial components of the vector fields vanish. 
A possible pion condensate is assumed not to appear, i.e., $\boldsymbol{\pi} = 0$. This is consistent with phenomenology in the presence of repulsive short-range nuclear spin-isospin correlations. Isospin-symmetry breaking effects occur through different chemical potentials for protons and neutrons, and they can be induced by a non-vanishing expectation value of the isovector $\rho_0^3$ field in isospin-3 direction. In summary, the fields that can get non-zero expectation values are $\sigma(x)$, $v_0(x)$ and $\rho_0^3(x)$. In thermodynamic equilibrium of homogeneous matter, only spatially constant fields are considered. In order to simplify notations from here on, we drop the time-component index in the following and write $v_0\equiv v$ and $\rho_0^3\equiv \rho_3$. 

The mean fields of the vector bosons enter the partition function just by shifting the chemical potentials. We introduce effective proton and neutron chemical potentials,
\begin{align}
	\begin{aligned}
		\bar{\mu}_p&=\mu_p - g_v v - g_\tau\rho_3\,,\\
		\bar{\mu}_n&=\mu_n - g_v  v + g_\tau\rho_3\,,
	\end{aligned}
\end{align}
As already outlined, the $\sigma$ mean field equips the nucleons with a mass term:
\begin{align}
	M_N&=g\,\sigma\,.
\end{align}

The partition function evaluated in mean-field approximation and with vanishing sources takes the form
\begin{align}
		Z_{\rm E}&=\int{\cal D}\psi_p\,{\cal D}\psi_p^\dagger\,{\cal D}\psi_n\,{\cal D}\psi_n^\dagger\exp\Bigg\{-\int_0^\beta d\tau\int d^3x\,\hspace{3cm}\\
		\times\bigg[&\bar{N}\left(\gamma_\mu\partial_\mu+M_N\right)N+ \mu_p\,\psi_p^\dagger\psi_p +\mu_n\,\psi_n^\dagger\psi_n-\frac12m_V^2\left(v^2+\rho_3^2\right)+{\cal U}(\sigma)\bigg]\Bigg\}\,.
\end{align}
The potential ${\cal U}$ includes in its parametrization vacuum fluctuation effects of pions and the $\sigma$-field. For fixed, space-time independent bosonic fields, this potential and the mass terms factor out and give a contribution \mbox{$\exp\{-\beta {\cal V}\cdot {\cal U}_{\rm B}(\sigma, v, \rho_3)\}$} with the bosonic potential
\begin{align}
	{\cal U}_{\rm B}(\sigma, v, \rho_3) = -\frac12m_V^2\left(v^2+\rho_3^2\right)+{\cal U}(\sigma)\,.
\end{align}
The nucleon fields appear only quadratically. As a consequence, the path integral can be performed and the logarithm of the partition function becomes:
\begin{align}\label{logdettrlog}
	\ln Z_E = \ln {\rm det} \,D-\beta {\cal V}\cdot {\cal U}_{\rm B}(\sigma, v,\rho_3)\,.
\end{align}
The fermion determinant, with $\ln {\rm det} \,D = {\rm Tr} \ln D$, involves traces over Dirac and isospin indices, a summation over Matsubara frequencies and an integral over three-monenta. The intermediate result is 
\begin{align}\label{eq:trlogD}
	{\rm Tr}\ln D = 2\beta {\cal V}\sum_{i=p,n}\int \frac{d^3p}{(2\pi)^3}\bigg\{E_N+\frac{p^2}{3E_N}n_{\rm F}(E_N- \bar{\mu}_i) + \frac{p^2}{3E_N}n_{\rm F}(E_N+\bar{\mu}_i)\bigg\}~,
\end{align}
with $E_N = \sqrt{\boldsymbol{p}^2 + M_N^2}$ and Fermi distributions 
\begin{align}
n_{\rm F}(E-\mu) = {1\over{\rm e}^{\beta (E-\mu)} + 1} \nonumber
\end{align}
for protons and neutrons. The divergent first term on the r.h.s. of Eq.\,(\ref{eq:trlogD}) can be handled using dimensional regularisation. This introduces a term proportional to $M_N^4\,\ln{(M_N^2/\lambda^2)}$ where $\lambda$ is a renormalization scale. This term can in turn be absorbed in the bosonic potential by introducing a corresponding piece proportional to $(g\sigma)^4\,\ln{((g\sigma)^2/\lambda^2)}$. Altogether one arrives at the {\it mean-field potential}:
\begin{align}\label{eq:MF_potential}
		{\cal U}_{\rm MF}(T,\mu_p,\mu_n,\sigma,v,\rho_3)={\cal U}_{\rm F}(T,\mu_p,\mu_n,\sigma,v,\rho_3)+{\cal U}_{\rm B}(\sigma,\boldsymbol{\pi} = 0, v,\rho_3)\,,
\end{align}
with the fermionic potential
\begin{align}\label{eq:U_F}
{\cal U}_{\rm F}(T,\mu,\sigma,v,\rho_3)&=-\frac2\beta\sum_{i=p,n}\sum_{r=\pm 1}\int\frac{d^3p}{(2\pi)^3}\ln\left[1+{\rm e}^{-\beta(E_N-r\bar{\mu}_i)}\right]\nonumber \\
			&=-2\sum_{i=p,n}\sum_{r=\pm 1}\int\frac{d^3p}{(2\pi)^3}\frac{p^2}{3E_N}n_{\rm F}(E_N-r\bar{\mu}_i)\,.
\end{align}
For the bosonic potential we return for the moment to its general form, keeping the dependence on both $\sigma$ and pion fields: 
\begin{align}
{\cal U}_{\rm B}(\sigma,\boldsymbol{\pi},v,\rho_3) =  {\cal U}(\sigma,\boldsymbol\pi)  -\frac12m_V^2\left(v^2 + \rho_3^2\right) \,.
\end{align}
The potential ${\cal U}(\sigma,\boldsymbol\pi)$ has a chirally symmetric part that depends only on the invariant square  $\chi=\frac12(\sigma^2+\boldsymbol\pi^2)$ and a symmetry-breaking piece proportional to $m_\pi^2$. In the vacuum, at $T = 0$ and $\mu_p = \mu_n = 0$, this potential is parametrized in terms of an expansion around the vacuum expectation values, $\sigma = \sigma_0 = f_\pi$ and $\boldsymbol\pi = 0$. The corresponding invariant square is denoted as $\chi_0=\frac 12\sigma_0^2$.
\begin{align}
{\cal U}(\sigma,\boldsymbol{\pi}) =   \sum_{n=0}^{n_{\rm{max}}}\frac{a_n}{n!}(\chi-\chi_0)^n - \frac{g^4\chi^2}{2\pi^2}\ln\frac{2g^2\chi}{\lambda^2} - m_\pi^2\,f_\pi (\sigma - \sigma_0)\,\,.
\end{align}
The logarithmic piece replaces, as mentioned, the divergent part in Eq.\,(\ref{eq:trlogD}). 

For given temperature $T$ and chemical potentials $\mu_{n,p}$, the mean-field potential is minimized as a function of its parameters $\sigma$, $\omega_0$ and $\rho_0^3$. The corresponding mean field equations are
\begin{equation}\label{eq:MF}
\frac{\partial {\cal U}_{\rm B}}{\partial\sigma}=-g\,n_s\,,
\end{equation}
and 
\begin{eqnarray}
		g_v\,v_{\rm min}=G_v\,(n_p+n_n)\,,\quad
		g_\tau\,\rho_{3,\rm min}=G_\tau\,(n_p-n_n)\,,
\end{eqnarray}
with neutron-, proton- and scalar-densities:
\begin{align}\label{eq:densities}
          \begin{gathered}
		n_i=2\int\frac{d^3p}{(2\pi)^3}\Big[n_{\rm F}(E_N - \bar{\mu}_i)+n_{\rm F}(E_N +\bar{\mu}_i)\Big]\,,\;\; i=n,p\,, \\
		n_s=2\sum_{i=n,p}\int\frac{d^3p}{(2\pi)^3}\frac{M_N}{E_N}\Big[n_{\rm F}(E_N -\bar{\mu}_i)+n_{\rm F}(E_N +\bar{\mu}_i)\Big]\,,
           \end{gathered}
\end{align}
and coupling strengths $G_v, G_\tau$ given in Eq. (\ref{eq:vectorcouplings}).
The values of the fields at the minimum of the potential are denoted as $\bar\sigma(T,\mu_{n,p})$, $\bar{v}(T,\mu_{n,p})$ and $\bar{\rho}_3(T,\mu_{n,p})$, respectively. The potential ${\cal U}_{\rm {MF}}$ evaluated at this minimum equals the grand canonical potential,
\begin{align}
	\Omega(T,\mu_n,\mu_p)&\equiv {\cal U}_{\rm{MF}}(T,\mu_n,\mu_p;\bar\sigma,\,\bar{v},\,\bar{\rho}_3)\,.
\end{align}
Pressure $P$, proton densities and neutron densities $n_{i=n,p}$, entropy density $s$, and energy density $\varepsilon$ are then determined from the standard thermodynamical relations:
\begin{align}\label{eq:thermodynamics}
	\begin{gathered}
		P=-\Omega(T,\mu_n,\mu_p)~,~~~\; n_i=-\frac{\partial \Omega(T,\mu_n,\mu_p)}{\partial\mu_i}\,,\\
		s=-\frac{\partial \Omega(T,\mu_n,\mu_p)}{\partial T}~,~~~\;\varepsilon=-P+Ts+\sum_{i=n,p}\mu_i\, n_i\,.
	\end{gathered}
\end{align}
Parameter fixing involves vacuum constraints and ground state properties of nuclear matter. The scalar-pseudoscalar Yukawa coupling $g$ is determined  as $g=M_N/\sigma_0 = 939 \,{\rm MeV}/f_\pi =10.2$. At vanishing temperature and chemical potential, the minimum of the mean-field potential is located at $\sigma=f_\pi$ with vanishing pressure $P_{\rm vac}$, so ${\cal U}_{\rm MF}$ must satisfy
\begin{align}\label{eq:Uvac1}
	{\cal U}_{\rm MF}\Big|_{\sigma=f_\pi}=-P_{\rm vac}=0\,,\quad \left.\frac{\partial {\cal U}_{\rm MF}}{\partial\sigma}\right |_{\sigma=f_\pi}=0\,.
\end{align}
The masses of the physical pion and of the scalar field $\sigma$ are:
\begin{align}\label{eq:Uvac2}
	\left.\frac{\partial{\cal U}_{\rm MF}}{\partial\chi}\right|_{\chi_0}=m_\pi^2\,,~~ \left.\left(\frac{\partial{\cal U}_{\rm MF}}{\partial\chi}+2\chi\frac{\partial^2{\cal U}_{\rm MF}}{\partial\chi^2}\right)\right|_{\chi_0}=m_\sigma^2\,.
\end{align}
The physical pion mass will be used, while $m_\sigma$ is given a large value, typically $m_\sigma \sim 0.9$ GeV.
With these constraints the potential parameters $a_1, a_2$ are fixed and the mean-field potential \eqref{eq:MF_potential} is given by:
\begin{align}\label{eq:MF_potential_2}
	\begin{gathered}
		{\cal U}_{\rm {MF}}=-2\sum_{i=n,p}\int\frac{d^3p}{(2\pi)^3}\frac{p^2}{3E_N}\left[n_{\rm F}(E_N- \bar{\mu}_i)+n_{\rm F}(E_N + \bar{\mu}_i)\right] \\
		+ \left[m_\pi^2+\frac{g^4}{4\pi^2}f_\pi^2\left(1+2\ln\frac{f_\pi^2}{2\chi}\right)\right]\,(\chi-\chi_0) \\
		+ \frac 12\left[\frac{m_\sigma^2-m_\pi^2}{f_\pi^2}+\frac{g^4}{2\pi^2}\left(3+2\ln\frac{f_\pi^2}{2\chi}\right)\right]\,(\chi-\chi_0)^2 \\
		+ \frac{g^4}{8\pi^2}f_\pi^4\ln\frac{f_\pi^2}{2\chi} + \sum_{n=3}^{n_{\rm max}}\frac{a_n}{n!}(\chi-\chi_0)^n 
		- m_\pi^2f_\pi(\sigma-f_\pi)-\frac 12m_V^2\Big(v^2+\rho_3^2\Big)\,.
	\end{gathered}
\end{align}
Note that the dependence on the renormalization scale $\lambda$ has dropped out as it should, and that ${\cal U}_{\rm MF}$ is finite for $\chi\rightarrow 0$.

\subsection{Mean-field parameter settings}

In practical applications of Eq.\,(\ref{eq:MF_potential_2}), a typical and efficient choice of truncation is $n_{\rm max}=4$ which leaves five parameters to be determined: $a_3$, $a_4$, $G_v = g_v^2/m_V^2$, $G_\tau=g_\tau^2/m_V^2$ and $m_\sigma$. Some of these parameters are fixed by empirical properties of symmetric nuclear matter. In the isospin-symmetric case with $n_p=n_n$ the $\rho_3$ field vanishes identically, as seen from the mean-field equation~\eqref{eq:MF}. In this case $G_\tau$ does not enter the calculations and there is only a single chemical potential, $\mu\equiv\mu_p=\mu_n$. 

At zero temperature, the nuclear liquid-gas first-order phase transition sets in at a critical chemical potential equal to the difference between nucleon mass and binding energy per nucleon, $\mu_c=M_N-B = 923\,{\rm MeV}$ with $B = 16$ MeV. The potential has two degenerate minima. The one at $\sigma=f_\pi$ corresponds to the vacuum, while a second minimum at $\sigma_c < f_\pi$ corresponds to nuclear matter in its ground state. Let $v_c$ be the expectation value of the isoscalar-vector field for $T = 0$  and $\mu=\mu_c$ at $\sigma=\sigma_c$. The effective chemical potential at $\mu_c$ is given by $\bar{\mu}_c = \mu_c - g_v\,v_c$. It is equal to the Landau effective mass, $M_{\rm L}$, of a nucleon quasi-particle at the Fermi surface:
\begin{align}
M_{\rm L}=\sqrt{p_{\rm F}^2+ (g\,\sigma_c)^2}~, 
\end{align}
where $p_{\rm F}=1.36$ fm$^{-1}$ is the Fermi momentum at nuclear saturation density $n_0 = 0.17$ fm$^{-3}$, and $M_N(n_0) = g\sigma_c$ is the dynamical in-medium nucleon mass. The Landau mass equals the effective chemical potential at the second minimum located at $\sigma_c$: 
\begin{align}
	M_{\rm L}=\mu_c-g_v\,v_c = \mu_c-G_v\, n_0\,~.
\end{align}
With $M_{\rm L}\simeq 0.8 M_{\rm N}$ corresponding to $\sigma_c \simeq 70\,{\rm}MeV \simeq 0.76\,f_\pi$ one finds 
\begin{equation}
G_v= g_v^2/m_V^2 \simeq 5.12\,{\rm fm}^2~. \nonumber
\end{equation}
As already mentioned, at mean-field level this is the only relevant combination of $g_v$ and $m_V$. It acts as the coupling strength of a corresponding nucleon-nucleon contact interaction, $G_v (N^\dagger N)^2$. Note that in phenomenological relativistic mean-field (RMF) models which identify the isoscalar-vector boson with a physical $\omega$ meson, the corresponding coupling $G_\omega = 
(g_\omega/m_\omega)^2$  is typically about twice as strong as the $G_v$ deduced here. Even at the level of the mean-field approximation, the present approach differs significantly from the RMF model.

The first-order liquid-gas phase transition at \mbox{$\mu=\mu_c$} with two degenerate minima of the potential ${\cal U}_{\rm MF}$ implies, in addition to the vacuum constraints (\ref{eq:Uvac1}), the following conditions at the second minimum:
\begin{align}
	{\cal U}_{\rm MF}\big|_{T=0,\mu_c,\sigma_c, v_c} = 0~,~~~
\left.\frac{\partial {\cal U}_{\rm MF}}{\partial\sigma}\right|_{T=0,\mu_c,\sigma_c, v_c} = 0~. 
\end{align}
These conditions allow to solve for $a_3$ and $a_4$ as a function of $m_\sigma$, such that only $m_\sigma$ remains as a free parameter for symmetric nuclear matter. At this point it is important to note again that the sigma mass is {\it not} to be identified with the complex pole in the $I=0$ $s$-wave channel of the pion-pion scattering amplitude at $\sqrt s\simeq (500-i\,300)$ MeV \cite{Colangelo2001,Garcia-Martin2007}. The $\sigma$ boson in the ChNM model parametrizes part of the $NN$ and $\pi N$ s-wave interactions at short-distance. Its mass is chosen around $m_\sigma \sim 0.9$ GeV, close to the mass of the $f_0(980)$. 

A further constraint that depends on $m_\sigma$ is the surface tension of a large nuclear droplet. Such a droplet, or bubble, is described by a sigma field $\sigma(r)$ with $\sigma = \sigma_c$ inside the bubble and $\sigma = f_\pi$ in the vacuum outside. For a thin wall separating the in- and outside, the surface tension $\Sigma$ can be calculated following Coleman \cite{Coleman1977}:
\begin{align}
\Sigma = \int_{\sigma_c}^{f_\pi} d\sigma\sqrt{2\,{\cal U}_{\rm MF}(\sigma)}~~.
\end{align}
The empirical $\Sigma \simeq 1.1\,{\rm MeV}/{\rm fm}^2$ is well reproduced with $m_\sigma = 880$ MeV. Note that, given the large $\sigma$ mass, the equivalent coupling strength corresponding to a four-point interaction $G_\sigma(\bar{N}N)^2$ of scalar nucleon densities, $G_\sigma = g^2/m_\sigma^2\simeq 5.2$ fm$^2$, is again less than half of the coupling strength commonly used in phenomenological RMF models.

\begin{figure}
  \centerline{\includegraphics[width=7cm] {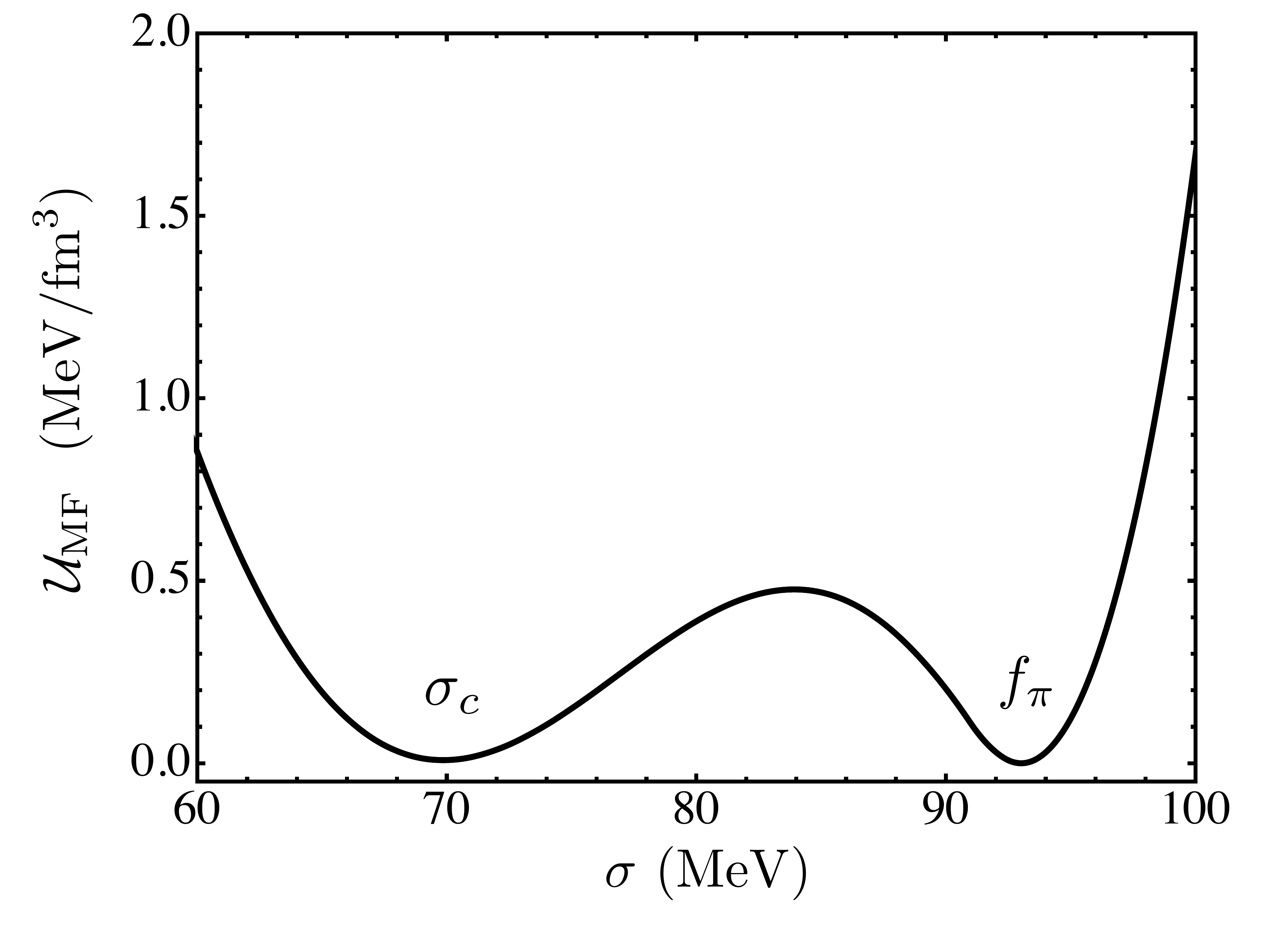}}
	\caption{Mean-field potential of symmetric nuclear matter at $T=0$ as function of $\sigma$ for $\mu=\mu_c = 923$ MeV. The minimum at $\sigma = f_\pi$ represents the vacuum, the one at $\sigma = \sigma_c = 69.8$ MeV corresponds to nuclear matter in its ground state.}
	\label{fig:MF_potential}
\end{figure}

Altogether, the resulting mean-field potential for symmetric nuclear matter at $T=0$ and $\mu = \mu_c = 923$ MeV (see Fig.~\ref{fig:MF_potential}) shows the two degenerate minima characteristic of a first-order phase transition with its liquid-gas coexistence domain.

The last remaining parameter still to be fixed is the isovector-vector coupling strength $G_\tau$. Once again, only the combination $G_\tau = g_\tau^2/m_V^2$ enters the equations, corresponding to the strength of an isovector-vector four-nucleon interaction $G_\tau(N^\dagger\boldsymbol\tau N)^2$. The parameter $G_\tau$ is chosen to reproduce the symmetry energy $E_{\rm sym}$ at nuclear saturation density, $n_0$. The symmetry energy $S(n)$ is defined as the difference between the energy per particle of pure neutron matter and symmetric nuclear matter at a given density $n$: 
\begin{align}\label{eq:SL}
	\begin{gathered}
		\frac EA(n,x)=\frac EA(n,0.5) + S(n)\,(1-2x)^2+\ldots\,,\\
		S(n)=E_{\rm sym}+\frac L3(n-n_0)+\ldots\,,
	\end{gathered}		
\end{align}
where the proton fraction $x=Z/A=n_p/(n_p+n_n)$ is a measure of the asymmetry. The $L$-parameter is related to the slope of the symmetry energy as a function of density $n$ around nuclear saturation density. The symmetry energy and the $L$-parameter can be inferred from measurements of neutron skin thickness, heavy ion collisions, dipole polarizabilities, giant and pygmy dipole resonance energies, as well as from fitting nuclear masses. A combined analysis gives values in the range $29\,{\rm MeV}\lesssim E_{\rm sym}\lesssim 33\,{\rm MeV}$ and $40\,{\rm MeV}\lesssim L \lesssim 62\,{\rm MeV}$ \cite{Tsang2012,Lattimer2013,Lattimer2014}.  Reproducing the symmetry energy value $E_{\rm sym}=32 \,{\rm MeV}$ fixes $G_\tau=1.07\,{\rm fm}^2$. All mean-field input parameters are now determined and summarized in Table \ref{tab:MF_parameters}.

\begin{table}
	\centering
	\begin{tabular}{cccccc}
\hline\\
\vspace{0.3cm}
	$a_3$\,(MeV${}^{-2}$) & $a_4$\,(MeV${}^{-4}$) & $m_\sigma$\,(MeV) & $g$ & $G_v$\,(fm${}^2$) & $G_\tau$\,(fm${}^2$) \\ 
\hline\\
\vspace{0.3cm}
$6.87\cdot 10^{-2}$ & $2.05\cdot 10^{-4}$ & 880 & 10.2 & $5.12$ & $1.07$ \\ 
\hline
	\end{tabular}
	\caption{List of all mean-field parameters entering the potential (\ref{eq:MF_potential_2}).}
	\label{tab:MF_parameters}
\end{table}  

\begin{figure}
  \centerline{\includegraphics[width=9cm] {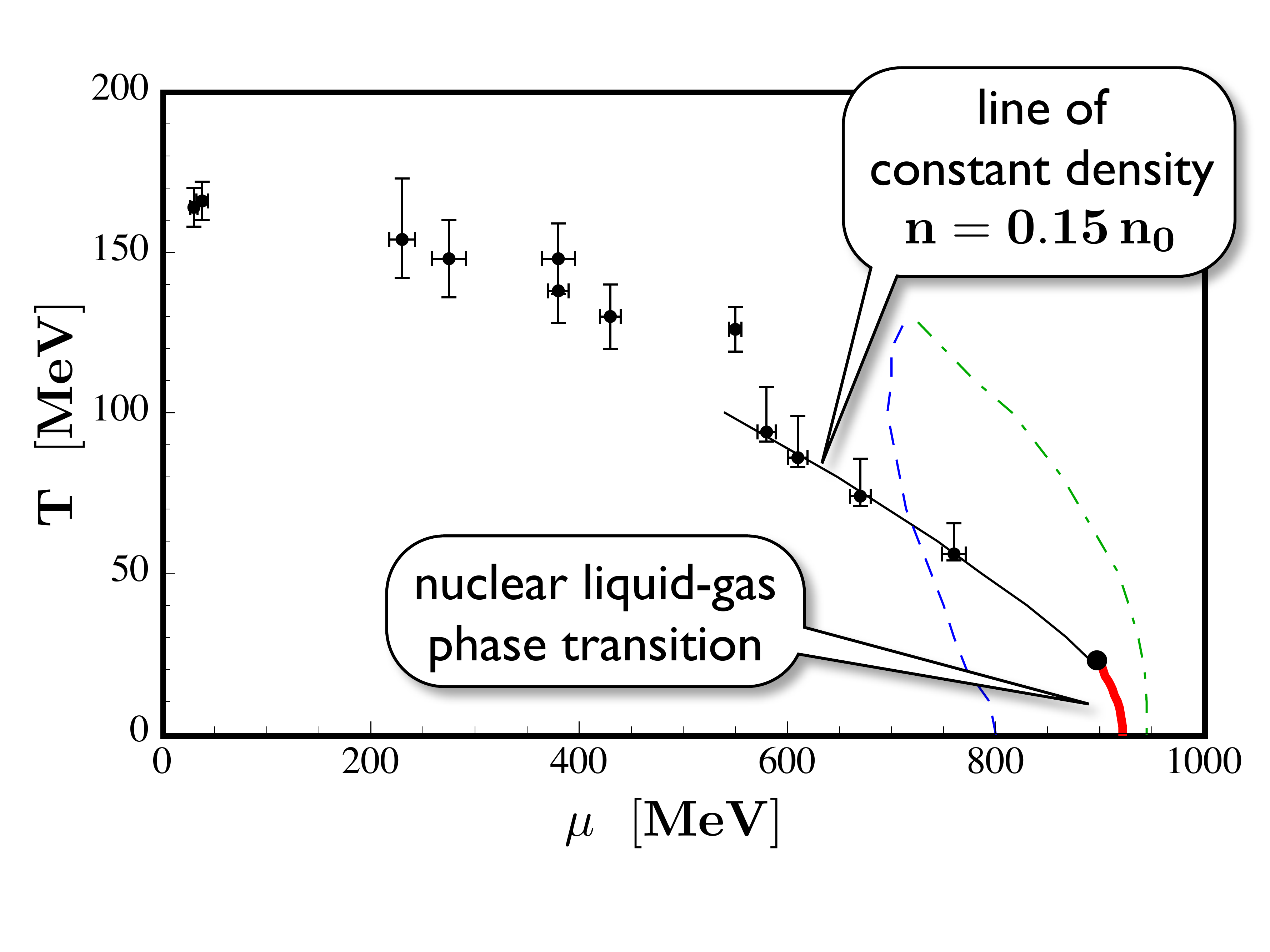}}
	\caption{Phase diagram displayed as temperature versus baryon chemical potential together with chemical freeze-out data from \cite{ABS2009}. The liquid-gas transition line and its extension has been adapted from \cite{FW2012}. }
	\label{fig:phasediagram}
\end{figure}

An instructive investigation using the ChMN model at mean-field level was performed in \cite{FW2012}. The nuclear $T-\mu$ phase diagram displaying the liquid-gas transition was calculated and extended in comparison with chemical freeze-out data from high-energy heavy-ion collisions \cite{ABS2009}. Whereas the crossover at high temperature and small $\mu$ features a close correspondence between chiral and deconfinement transitions, the behaviour at low temperature and larger $\mu$ is qualitatively different. An extension of the first-order liquid-gas transition line beyond its critical point demonstrates that the freeze-out data in this region correspond, as expected, to very low baryon densities of only a small fraction of $n_0$, far distant from a transition towards chiral symmetry restoration. This is shown in Fig. \ref{fig:phasediagram}.

\subsection{Beyond mean-field: fluctuations and functional renormalization group}\label{fluctuations}

Fluctuations beyond the mean-field approximation in the ChNM model are included \cite{DHKW2013, DW2015} using the framework of the functional renormalization group that has been outlined in Section \ref{sec:FRG}. The ChNM Lagrangian serves as input at a UV scale $\Lambda$ that is characteristically of the order of the chiral scale, $\Lambda_\chi = 4\pi f_\pi$.  Then the effective action $\Gamma_k$ is introduced which depends on a renormalization scale $k$. The flow of this effective action is determined in such a way that it interpolates between the ultraviolet action at the scale $\Lambda$ and the full quantum effective action $\Gamma_{\rm eff}=\Gamma_{k=0}$ in the infrared limit $k\rightarrow 0$. The evolution of $\Gamma_k$ as a function of $k$ is given by Wetterich's flow equation (\ref{eq:Wetterich}):
\begin{align}\label{eq:FRGflow}
	\begin{aligned}
		k\,\frac{\partial\Gamma_k}{\partial k}=
		\begin{aligned}
			\vspace{1cm}
			\includegraphics[width=0.08\textwidth]{wetterich_fermion}
		\end{aligned} \vspace{-1cm}=\frac 12 \operatorname{Tr}\frac{k\,\frac{\partial R_k}{\partial k}}{\Gamma_k^{(1,1)}+R_k}\,.
	\end{aligned}
\end{align}
As described in Section \ref{sec:FRG}, the "soft" degrees of freedom that contribute most prominently to fluctuations beyond mean-field approximation are the pion field with its relatively small mass, and low-energy nucleon-hole excitations. Both these types of excitations enter non-perturbatively through their full propagators in the flow equation (\ref{eq:FRGflow}). The regulator $R_k$ is chosen as in Eq.\,(\ref{eq:regulator}). 

The mass $m_V$ associated with the vector bosons $v_\mu$ and $\boldsymbol{\rho}_\mu$ is large compared to the relevant low-energy scales. Therefore, fluctuations of these vector fields are suppressed and their "frozen" degrees of freedom are treated as background fields in mean-field approximation as before. In contrast, the fluctuations of the pions and (in order to maintain chiral symmetry) also of the $\sigma$ are included, as well as important particle-hole excitations of the nucleons around the Fermi surface. 

For the treatment of the thermodynamics with inclusion of fluctuations it is useful to compute the flow of the difference between the effective action at given values of temperature and chemical potential, $\Gamma_k(T,\mu)$, as compared to the potential at the liquid-gas phase transition point at zero temperature, $\Gamma_k(0,\mu_c)$, at which nuclear matter is in equilibrium. In analogy to Ref.\,\cite{Litim2006}, we study the flow of the difference
\begin{align}
	\bar\Gamma_k=\Gamma_k(T,\mu)-\Gamma_k(0,\mu_c)\,.
\end{align}
The $k$-dependence of $\bar\Gamma_k$ is thus given by
\begin{align}
	\begin{aligned}
		\frac{k\,\partial\bar \Gamma_k}{\partial k}(T,\mu)&=
		\begin{aligned}
			\hspace{-.1cm}
			\vspace{1cm}
			\includegraphics[width=0.08\textwidth]{wetterich_fermion}
		\end{aligned} \vspace{-1cm}\Bigg|_{T,\mu}-
		\begin{aligned}
			\hspace{-.1cm}
			\vspace{1cm}
			\includegraphics[width=0.08\textwidth]{wetterich_fermion}
		\end{aligned} \Bigg|_{\begin{subarray}{l} T=0 \\ \mu=\mu_c \end{subarray}}.
	\end{aligned}
\end{align}
The effective action is treated in leading order of the derivative expansion and we work in the local potential approximation, as discussed in Section \ref{sec:FRG}: a possible anomalous dimension (a $Z$-factor) or the so-called $Y$-term, which includes a derivative coupling together with higher powers in the fields, $Y(\phi)\,(\partial\phi)^2$, are not considered. Moreover, the running of the Yukawa couplings is ignored. With these simplifications, the effective action is written as
\begin{align}
	\begin{aligned}
	&\Gamma_k=\int d^4x\;\bigg\{\bar{N} i\slashed\partial N +\frac 12\partial_\mu\sigma\,\partial^\mu\sigma+\frac 12\partial_\mu\boldsymbol\pi\cdot\partial^\mu\boldsymbol\pi \\
	&-\bar N \Big[g(\sigma+i\gamma_5\,\boldsymbol\tau\cdot\boldsymbol\pi)+\gamma_0(g_v\,v_0+g_\tau\,\rho_{0,3}\,\tau^3)\Big] N - {\cal U}_k\bigg\}\,.
	\end{aligned}
\end{align}
As mentioned, the vector fields $v_0$ and $\rho_{0,3}$ appear here only as mean fields. The complete $k$-dependence is in the effective potential ${\cal U}_k$. In analogy to the mean-field potential \eqref{eq:MF_potential_2}, the effective potential has a chirally symmetric piece, ${\cal U}^{(\chi)}$, the explicit chiral symmetry breaking term and the mass terms of the vector bosons\footnote{The subscripts indicating the temporal components of the vector fields are again suppressed for convenience.}:
\begin{align}\label{eq:decomposition}
	{\cal U}_k= {\cal U}_k^{(\chi)}-m_\pi^2f_\pi(\sigma-f_\pi)-\frac 12m_V^2\Big(v^2+\rho_3^2\Big)\,.
\end{align}
The second derivative $\Gamma^{(1,1)}$ is computed and the Dirac and isospin traces are performed.
With the choice of the optimized regulators \eqref{eq:regulator} and \eqref{eq:regulator_finite_T}, the momentum dependence comes in through a step function and the momentum integral can be performed trivially. The remaining flow equations depend only on the chirally invariant field $\chi = \frac12\left(\sigma^2 + \boldsymbol{\pi}^2\right)$.

The flow of the subtracted chirally symmetric potential $\bar{\cal U}_k^{(\chi)}={\cal U}_k^{(\chi)}(T,\mu)-{\cal U}^{(\chi)}_k(0,\mu_c)$ is computed from the equation
\begin{gather}\label{eq:flow_equation}
	\frac{\partial\bar{\cal U}_k^{(\chi)}(T,\mu)}{\partial k}=f_k(T,\mu)-f_k(0,\mu_c)\,,
\end{gather}
with
\begin{multline}
	f_k(T,\mu)=\frac {k^4}{12\pi^2} \bigg\{3\cdot\frac{1+2n_{\rm B}(E_\pi)}{E_\pi}+\frac {1+2n_{\rm B}(E_\sigma)}{E_\sigma} \\
	-4 \sum_{i=n,p}\frac{1-n_{\rm F}\left(E_N - \mu^{\rm eff}_i\right)-n_{\rm F}\left(E_N + \mu^{\rm eff}_i\right)}{E_N}\bigg\}\,.
\end{multline}
Here,
\begin{align}\label{eq:mpi}
	\begin{gathered}
		E_N^2=k^2+2g^2\chi\,, \\
		E_\pi^2=k^2+\frac{\partial {\cal U}_k}{\partial\chi}\,,\quad E_\sigma^2=k^2+\frac{\partial {\cal U}_k}{\partial\chi}+2\chi\frac{\partial^2{\cal U}_k}{\partial\chi^2}\,, \\
		n_{\rm B}(E)=\frac 1{{\rm e}^{E/T}-1}\,,~~\text{ and }~˝\, n_{\rm F}(E-\mu)=\frac 1{{\rm e}^{(E-\mu)/T}+1}\,.
	\end{gathered}
\end{align}
So far $v$ and $\rho_3$ are constant background fields, to be determined self-consistently in such a way that the effective potential at $k=0$ is minimized as a function of these fields. A further step is to introduce a $k$- and $\chi$-dependence for $v$ and $\rho_3$ such that the effective potential ${\cal U}_k$ is minimized at each scale $k$ for each $\chi$. From Eq.~\eqref{eq:decomposition} follow the two gap equations for $v(k,\chi)$ and $\rho_3(k,\chi)$:
\begin{align}
	\begin{gathered}
		\frac\partial{\partial y}\Big[{\cal U}_k^{(\chi)}\Big(y,\rho_3(k,\chi)\Big)-\frac 12m_V^2y^2\Big]\Big|_{y=v(k,\chi)}=0\,, \\
		\frac\partial{\partial y}\Big[{\cal U}_k^{(\chi)}\Big(v(k,\chi),y\Big)-\frac 12m_V^2y^2\Big]\Big|_{y=\rho_3(k,\chi)}=0\,.
	\end{gathered}
\end{align}
With the help of Eq.\,\eqref{eq:flow_equation} it is possible to rewrite these gap equations in the following form:
\begin{gather}\label{eq:flow_equation_2}
	\begin{gathered}
	g_v\,v(k,\chi) = 
\frac {G_v}{3\pi^2}\int_k^\Lambda dp \; \frac{p^4}{E_N} \hspace{3.3cm} \\
	\quad\cdot\sum_{r=\pm}\frac\partial{\partial \mu}\Big[n_{\rm F}\big(E_N -r\mu^{\rm eff}_p(k,\chi)\big)+n_{\rm F}\big(E_N -r\mu^{\rm{eff}}_n(k,\chi)\big)\Big] \,, \\
	g_\tau\,\rho_3(k,\chi) = \frac {G_\tau}{3\pi^2}\int_k^\Lambda dp \; \frac{p^4}{E_N}\hspace{3.3cm} \\
	\quad\cdot\sum_{r=\pm}\frac\partial{\partial \mu}\Big[n_{\rm F}\big(E_N-r\mu^{\rm{eff}}_p(k,\chi)\big)-n_{\rm F}\big(E_N -r \mu^{\rm{eff}}_n(k,\chi)\big)\Big] \,,
	\end{gathered}
\end{gather}
where the effective chemical potentials now depend on $k$ and $\chi$ according to
\begin{align}
	\mu^{\rm{eff}}_{n,p}(k,\chi)=\mu_{n,p}-g_v\,v(k,\chi) \pm g_\tau\,\rho_3(k,\chi)\,.
\end{align}
These equations can be considered as generalizations of the mean field equations \eqref{eq:MF} in the context of the functional renormalization group. After an integration by parts, the gap equations can be brought into a form similar to Eq.~\eqref{eq:densities}, where the momenta of the nucleons contributing to the mean fields $v(k,\chi)$ and $\rho_3(k,\chi)$ at a certain step are restricted to the range \mbox{$k\le p \le\Lambda$}, as is clear from the integral boundaries of Eq.~\eqref{eq:flow_equation_2}. In addition, boundary terms from the integration by parts appear which vanish in the limit $k\rightarrow 0$ and for large cutoffs $\Lambda$. In this way it is possible to show that the flow equations reproduce the mean field results if bosonic loops are ignored.

Finally, the ultraviolet potential is fixed in such a way that for $T=0$ and $\mu=\mu_c$ the mean field potential \eqref{eq:MF_potential} is reproduced. The pion mass and pion decay constant, both determined from the behavior of the potential at its minimum, are correctly kept. Nonetheless, as explained in Ref.~\cite{DHKW2013}, the input parameters need to be readjusted in order to reproduce the correct nuclear saturation density, because the dependence of the minimum of the effective potential ${\cal U}_k$ on the chemical potentials is influenced by the fluctuations. Again, the coupling $G_\tau$ is fixed  to reproduce a symmetry energy of 
$E_{\rm{sym}}=32\,\rm{MeV}$. The updated parameters of the model with inclusion of fluctuations are
\begin{align}\label{eq:parameters}
	\begin{gathered}
		G_v=\frac{g_v^2}{m_V^2}=4.04\text{ fm}^2\,,\quad G_\tau=\frac{g_\tau^2}{m_V^2}=1.12\text{ fm}^2\,, 
		\quad m_\sigma=770\text{ MeV}\,,\\~~a_3=5.55\cdot 10^{-3}\text{ MeV}^{-2}\,,~~
		a_4=8.38\cdot 10^{-5}\text{ MeV}^{-4}\,.
	\end{gathered}
\end{align}
The ultraviolet cutoff is chosen as $\Lambda=1.4\text{ GeV}$, slightly above the chiral scale $4\pi f_\pi$ below which the hadronic effective Lagrangian ${\cal L}_{\rm{ChNM}}$ is applicable. The parameter set \eqref{eq:parameters} is understood in conjunction with this given value of $\Lambda$. In general the UV scale is to be chosen large enough so that for large chemical potentials in the GeV range and high temperatures up to about 100 MeV, all relevant contributions to the partition function (in particular, multi-pion loops and multi-particle-hole excitations around the Fermi surface) are properly incorporated. Small changes of $\Lambda$ can be compensated by correspondingly small variations of the parameters, keeping the low-energy physics output unchanged.

It is instructive at this point to comment on the coefficients $a_3$ and $a_4$ of the terms in the expansion of the potential ${\cal U}$ that are of 3rd and 4th order in $\chi - \chi_0$, the shift of the chiral field in matter from its vacuum value, $\chi_0 = \frac12 \sigma_0^2 = \frac12 f_\pi^2$. In practice these are subleading terms. Around normal nuclear matter density they are typically one or two orders of magnitude smaller than those involving leading (linear and quadratic) terms in $\chi - \chi_0$. However, the relative importance of these higher-order terms increases with increasing baryon density through the multi-nucleon interactions generated by the coupling of the chiral fields to the nucleons. It is interesting to observe how $a_3$ and $a_4$ change as one proceeds from mean-field approximation to the full FRG scenario. At mean-field level these parameters encode indirectly, by their adjustment to the nuclear matter ground state, effects of fluctuations not treated explicitly. On the other hand, solving the FRG flow equations generates these fluctuations, i.e. loops involving the chiral fields and nucleon-hole excitations. Consequently the FRG-updated values of $a_3$ and $a_4$ in \eqref{eq:parameters} are reduced significantly in magnitude from their mean-field values in Table~\ref{tab:MF_parameters}.

For given temperature $T$ and chemical potentials $\mu_n$ and $\mu_p$, the full sets of flow equations \eqref{eq:flow_equation} and \eqref{eq:flow_equation_2} are then solved self-consistently using the grid method described in the Appendix. The potential is expanded as a function of $\chi$ around grid points and then matched continuously between any two adjacent grid points. In this way, the potential is not Taylor expanded but kept as a general function of $\chi$. Upon expansion of the effective potential around its absolute minimum, $n$-point interactions involving pions and sigma fields are generated in the effective action. In the context of a generalized sigma model, multi-pion and $\sigma$ exchange interactions are thus incorporated to all orders. The non-linear structure of the effective potential generates a hierarchy of many-body interactions between nucleons, including important three-body forces but also higher-order cluster interactions which become increasingly significant at higher baryon densities. In contrast to (perturbative) in-medium chiral effective field theory based on a non-linear sigma model, the $n$-point correlators in the effective action are now computed in a fully non-perturbative fashion. It is at the same time of great interest to compare the FRG results with those from in-medium chiral perturbation theory in order to examine convergence issues in perturbative approaches.

\section{From symmetric to asymmetric nuclear matter and neutron stars}

\subsection{Nuclear thermodynamics and the liquid-gas transition}

\begin{figure}
	\centerline{\includegraphics[width=9cm] {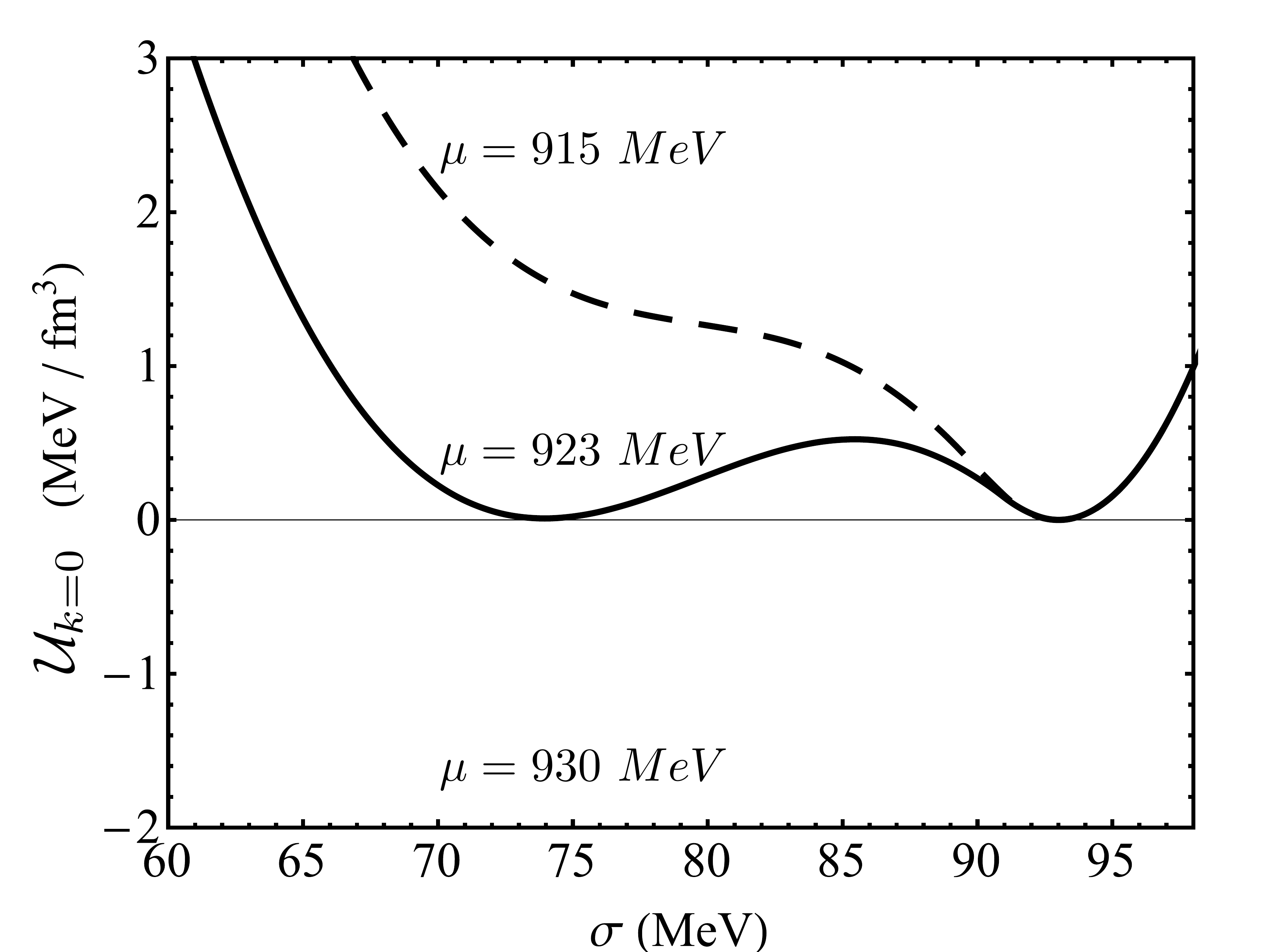}}
	\caption{The $\sigma$-dependent effective potential at zero temperature for three different baryon chemical potentials around the liquid-gas phase transition.}
	\label{fig:U_T0}
\end{figure}

Following these preparations we can now investigate the thermodynamics of symmetric nuclear matter, with special focus on the liquid-gas first-order phase transition. Fig.\,\ref{fig:U_T0} shows the effective potential at $T=0$ for several baryon chemical potentials in the vicinity of $\mu_c = 923$ MeV. Fluctuations are included around the liquid-gas transition as discussed, so the effective potential at $\mu=\mu_c$ looks similar to the mean-field potential of Fig.\,\ref{fig:MF_potential}. Slight differences originate from the re-adjustment of parameters. At smaller chemical potentials, the minimum of the effective potential is located at $\sigma=f_\pi$: the system is in its vacuum state. Therefore, in a $T - \mu$~phase-diagram, the $\mu$-axis at \mbox{$T=0$} up to \mbox{$\mu=\mu_c$} corresponds to a single physical state, the vacuum. At $\mu=\mu_c$, the two minima of the effective potential are degenerate. Vacuum and nuclear matter both have vanishing pressure and coexist. The Fermi sea is being filled, droplets of nuclear liquid form and the density increases up to saturation density, $n_0$. The whole coexistence region up to saturation density corresponds to a single chemical potential, $\mu_c$. Finally, for larger chemical potentials the system is characterized by a decreasing expectation value of the $\sigma$-field as the density increases.

\begin{figure}
	\centerline{\includegraphics[width=9cm] {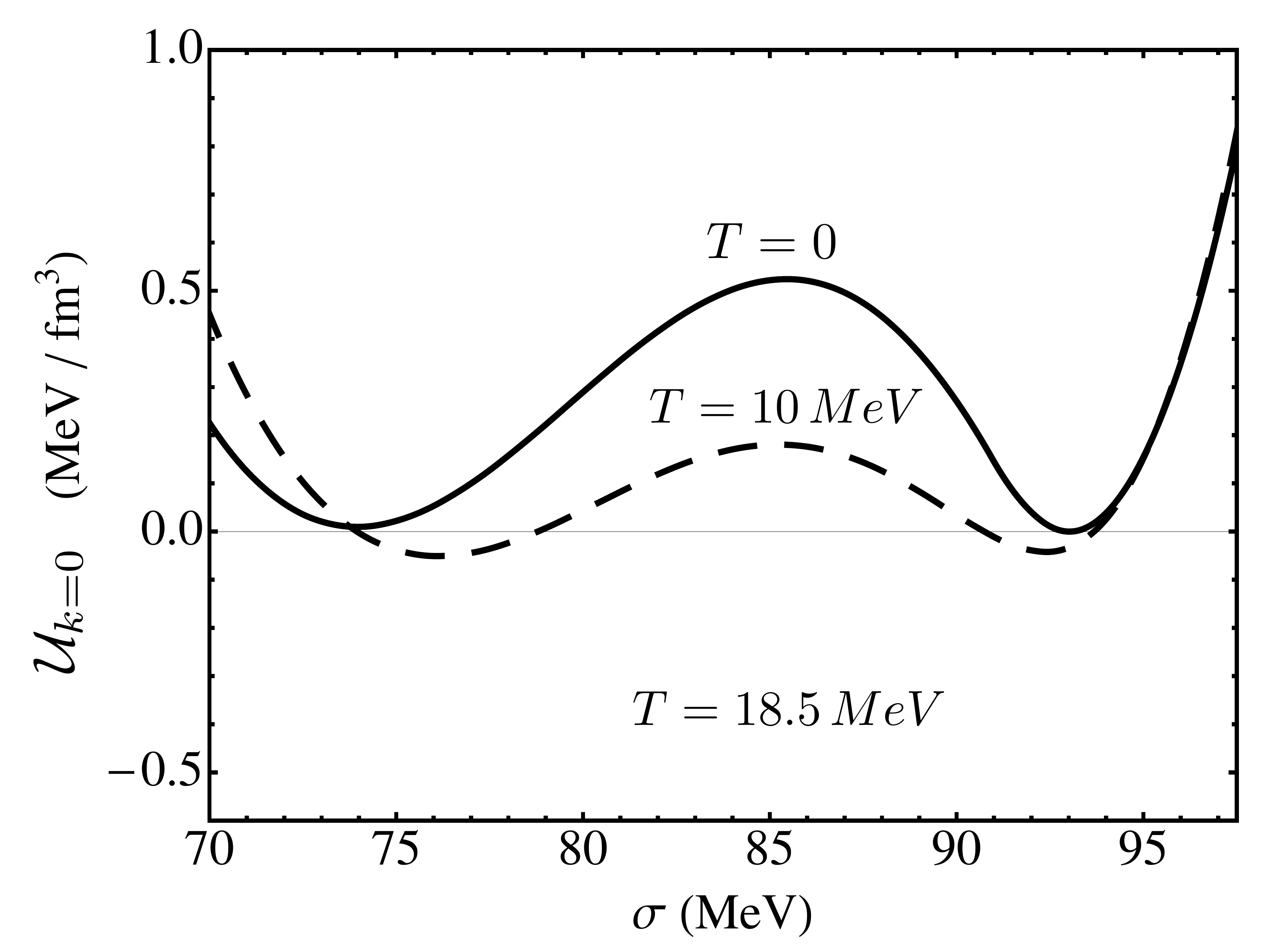}}
	\caption{The $\sigma$-dependent effective potential around the liquid-gas phase transition for three different temperatures.}
	\label{fig:U_T}
\end{figure}
Next, we study finite temperatures. Figure \,\ref{fig:U_T} shows the effective potential for three different temperatures in the range of the first-order liquid-gas transition. The chemical potential is adjusted such that the two minima are degenerate. One observes that the minima move closer towards each other as the temperature increases. At the {\it critical temperature},
\begin{align}\label{eq:Tcrit}
	T_c= 18.3\,{\rm MeV}~,
\end{align}
and for $\mu= 913$ MeV, the minima are no longer separated, and we have reached the second-order critical endpoint of the liquid-gas phase transition. For higher temperatures there is a unique minimum. 

The surface tension of nuclear droplets is related to the potential hill between the two minima. With increasing temperature, this surface tension decreases until it vanishes at the critical point so that liquid and gas phases are no longer separated.

\begin{figure}
	\centerline{\includegraphics[width=9cm] {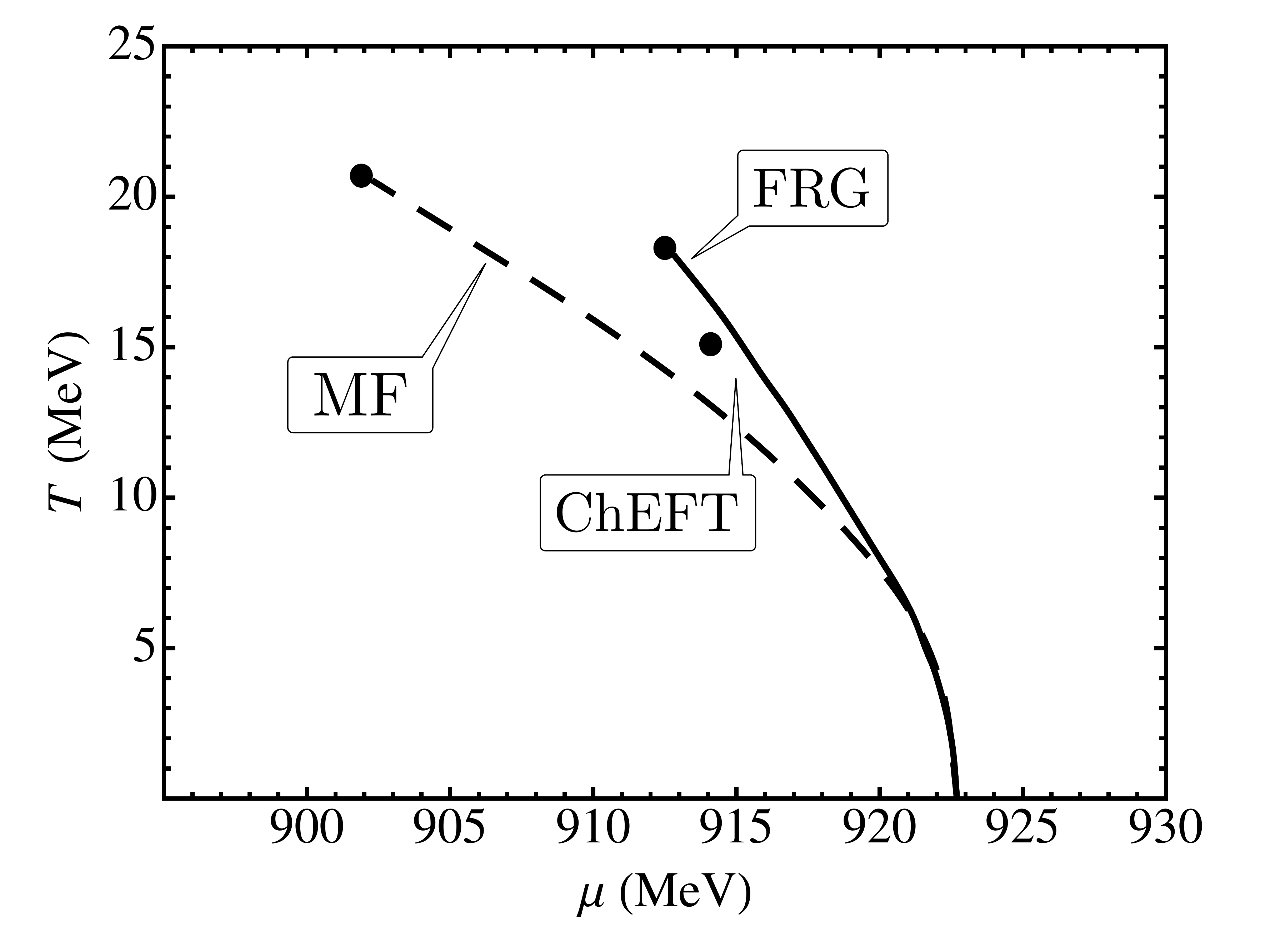}}
	\caption{Liquid-gas phase transition in a $T-\mu$ diagram. Dashed line: mean-field result of the ChNM model. Solid curve: FRG calculation including bosonic fluctuations and particle-hole excitations. Dotted curve: in-medium chiral effective field theory calculation of Refs.~\cite{Fiorilla2012}.}
	\label{fig:liquidgasphase}
\end{figure}

Fig.\,\ref{fig:liquidgasphase} shows the liquid-gas transition in a $T$--$\mu$ phase diagram. The left-bending of the transition curve can be understood from a Clausius--Clapeyron type relation. Along the first-order line, the minima (corresponding to the gas and the liquid phase) must be degenerate. Therefore, the total differentials of the effective potentials agree for both phases:
\begin{align}
	\frac{\partial {\cal U}_{\rm{liquid}}}{\partial\mu}d\mu+\frac{\partial {\cal U}_{\rm{liquid}}}{\partial T}dT=\frac{\partial {\cal U}_{\rm{gas}}}{\partial\mu}d\mu+\frac{\partial {\cal U}_{\rm{gas}}}{\partial T}dT\,.
\end{align}
The slope of the transition line can then be expressed as the ratio of differences between baryon number densitites, $n_{\rm{liquid}}-n_{\rm{gas}}$, and entropy densities, $s_{\rm{liquid}}-s_{\rm{gas}}$, i.e.,
\begin{align}
	\frac{dT}{d\mu}=-\frac{n_{\rm{liquid}}-n_{\rm{gas}}}{s_{\rm{liquid}}-s_{\rm{gas}}}\,.
\end{align}
In the limit of vanishing temperatures, the denominator vanishes as a consequence of Nernst's theorem. The slope diverges and the transition line hits the $\mu$-axis at a right angle. For non-zero temperatures, particle-hole excitations around the Fermi surface contribute to the entropy in the liquid phase. The entropy is therefore larger in the liquid phase than in the gas phase. Moreover, the liquid phase is denser than the gas phase and the slope of the transition line is negative, as observed.

For comparison, we show the result from the mean-field calculation of Ref. \cite{FW2012}. We will in the following distinguish between the MF-ChNM model (at the mean-field level) and the full FRG-ChNM model (with fluctuations included). The MF-ChNM critical endpoint is located at a higher temperature, $T_c\simeq 21$ MeV and a smaller chemical potential. Fluctuations beyond the mean-field approximation bend the phase-transition boundary towards higher chemical potentials. The curvature of the FRG-ChNM boundary line is in good agreement with a ChEFT result of Refs.\,\cite{Fiorilla2012}. This latter calculation was performed in a perturbative framework up to three-loop order in the free energy density, including all possible one- and two-pion exchange processes in the medium to that order. Moreover, three-body forces and $\Delta$-isobar excitations were included. In the FRG-ChNM framework pion and nucleon loops are resummed in a non-perturbative way. In addition, part of the effects that are treated explicitly in ChEFT are relegated to the parametrization of the effective potential. Nevertheless, the results are remarkably consistent with each other. In fact, an advanced ChEFT calculation \cite{WHKW2014} starting from low-momentum chiral $NN$ and $3N$ interactions, with a cutoff varied in the interval $400 - 500$ MeV, determines the critical temperature in the range $T_c = 17 - 19$ MeV which compares very well with the FRG-ChNM result, Eq.\,\eqref{eq:Tcrit}. We recall that the critical temperature deduced from various nuclear reactions and multifragmentation experiments is $T_c=17.9\pm0.4$ MeV \cite{Kar2008,ELMP2013}. 

It should be noted that Coulomb effects are absent in our infinite-matter model. Furthermore, effects from the formation of light clusters (d, t, $^3$He, $\alpha$) at low densities have to be considered. Using relativistic mean-field and microscopic quantum statistical models, studies of such effects \cite{Typel2010} showed that their influence becomes relevant at densities $n \lesssim 0.01$ fm$^{-3}$, but the resulting shift of the critical temperature is small, $\Delta T_c \simeq 1$ MeV. 

\begin{figure}[h!]
	\centerline{\includegraphics[width=9cm] {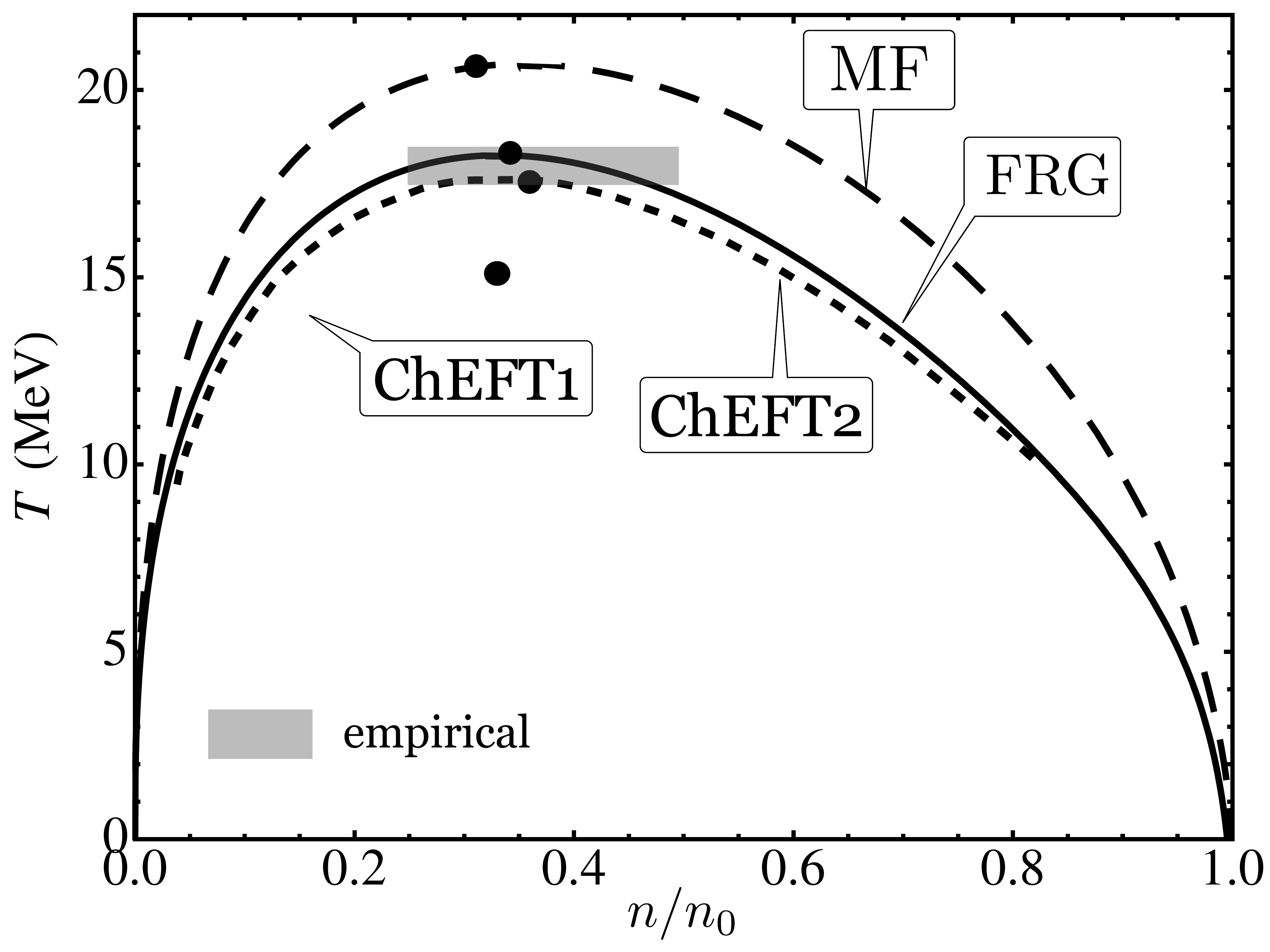}}
	\caption{Coexistence regions of the nuclear liquid-gas phase transition. Solid line: FRG-ChNM model. Dashed line: MF-ChNM model. Dotted line (ChEFT$_1$): in-medium ChPT at NLO with explicit $\Delta(1230)$ degrees of freedom~\cite{Fiorilla2012}. Short-dashed line (ChEFT$_2$): advanced ChEFT calculation using N$^3$LO chiral NN interaction together with N$^2$LO three-body forces in Kohn-Luttinger-Ward 2nd order many-body perturbation theory \cite{WHKW2014}. The dots indicate the respective critical endpoints. The empirical $T_c$ and $n_c$ range (shaded area) is taken from \cite{ELMP2013}.}
	\label{fig:Trho}
\end{figure}

Fig.\,\ref{fig:Trho} displays the coexistence region of the liquid-gas phase transition in the temperature-density plane. Mean-field and full FRG results of the ChNM model are compared with (perturbative) ChEFT computations: an earlier NLO computation including explicit $\Delta(1230)$ degrees of freedom \cite{Fiorilla2012}, and a more recent advanced calculation using N$^3$LO chiral NN interactions together with N$^2$LO three-body forces \cite{WHKW2014}. This latter ChEFT result agrees very well with that of the FRG-ChNM model once fluctuations are taken into account. The density at the critical point comes out consistently around $n_c =n_0/3$, again in good agreement with the empirical $n_c =0.06\pm0.01$ fm$^{-3}$ \cite{Kar2008,ELMP2013}. One may add at this point that the results presented here are also quite compatible with an early study using phenomenological Skyrme forces \cite{SCM1976}.

\begin{figure}[h!]
         \centerline{\includegraphics[width=9cm] {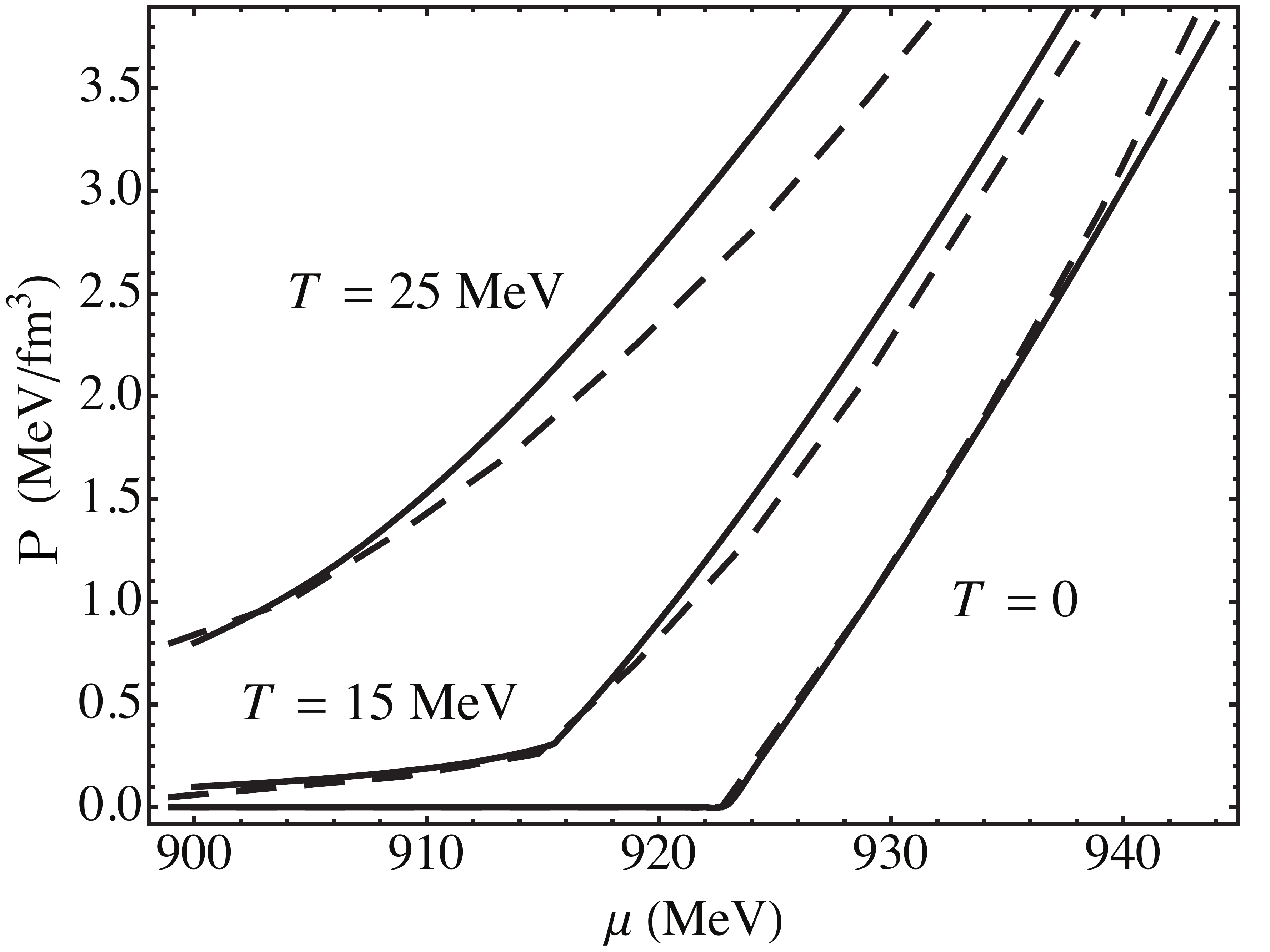}}
	\caption{Pressure as a function of baryon chemical potential in symmetric nuclear matter for three different temperatures. Solid line: FRG-ChNM model \cite{DHKW2013}. Dashed line: ChEFT result \cite{Fiorilla2012}.}
	\label{fig:p_mu}
\end{figure}
The grand-canonical potential is found by evaluating the effective potential at its minimum, and the pressure as a function of the chemical potential can then be computed. A comparison of the pressure $P(\mu)$ resulting from the FRG-ChNM model and the ChEFT calculations is shown in Fig.\,\ref{fig:p_mu}. Not surprisingly, since the effective potential is adjusted to reproduce the properties of nuclear matter at $\mu=\mu_c$ and $T = 0$, the equations of state in both FRG-ChNM and ChEFT calculations agree very well around this point. The equations of state match also for larger chemical potentials at $T=0$. As the temperature increases, some deviations appear between the two approaches, although they remain small for temperatures up to 15-20 MeV. These features reflect the similarity of the first-order transition lines in the phase diagram. Given the different treatments of the pionic physics (non-perturbative versus perturbative) in the FRG-ChNM and in-medium ChEFT approaches, the close similarity of these results is once again remarkable. 

\begin{figure}
	 \centerline{\includegraphics[width=9cm] {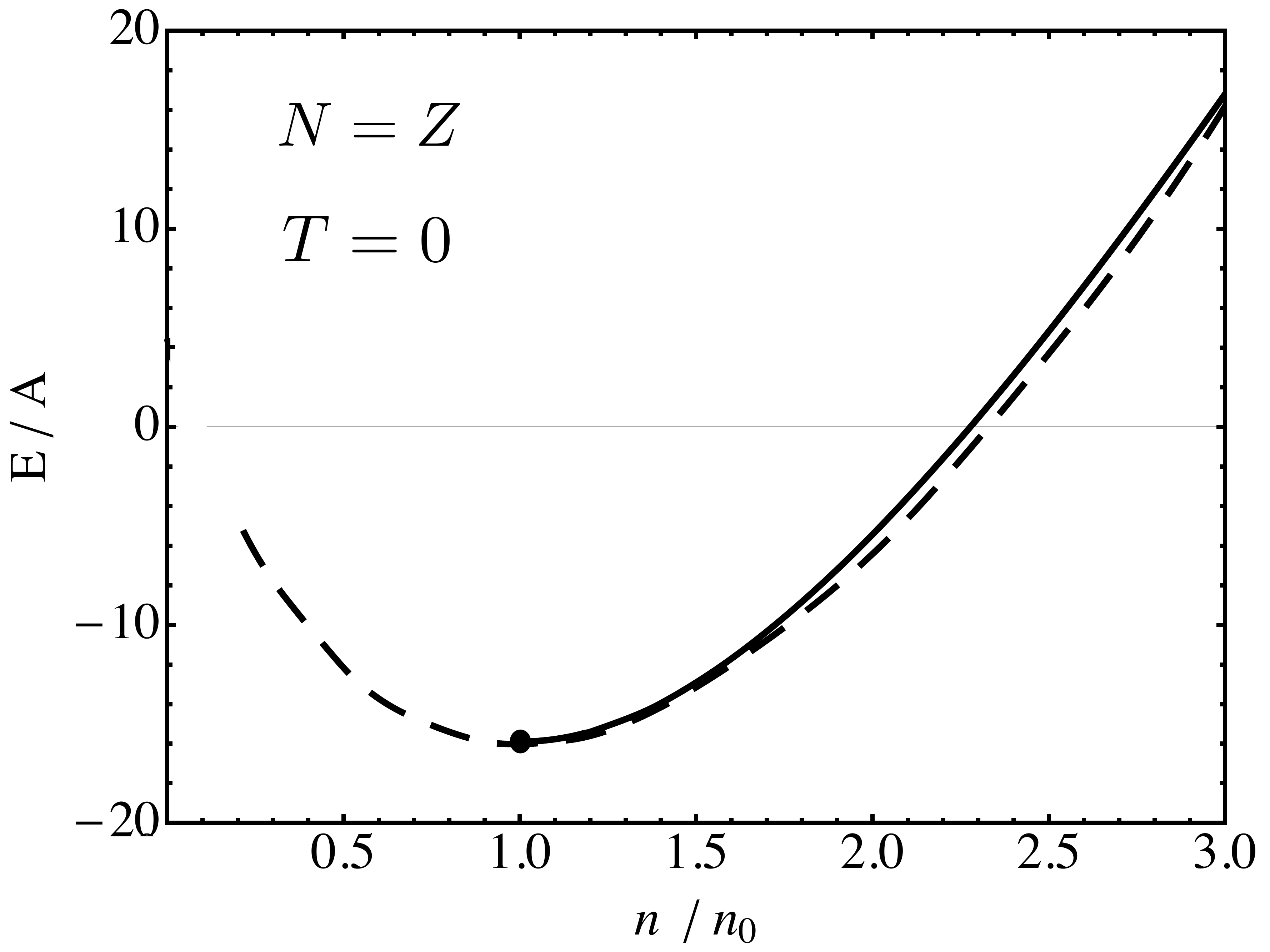}}
	\caption{The energy per particle of symmetric nuclear matter at zero temperature. Solid line: FRG-ChNM model \cite{DW2015}. Dotted line: Akmal-Pandharipande-Ravenhall EoS~\cite{Akmal1998}. Dashed line: AFDMC~\cite{Armani2011}.}
	\label{fig:e_a_sym}
\end{figure}

The energy per particle of symmetric nuclear matter is shown as a function of density in Fig.\,\ref{fig:e_a_sym}. It is of some interest to examine how the FRG calculations extrapolate to higher densities in comparison with two "realistic" equations of state (EoS) obtained in quite different ways. The first one is the Akmal-Pandharipande-Ravenhall EoS \cite{Akmal1998} which is based on the phenomenological Argonne~$v_{18}$ two-nucleon interaction together with the Urbana~IX three-nucleon force. Relativistic boost corrections were included in the calculation. The second EoS was computed using an auxiliary-field diffusion Monte-Carlo (AFDMC) framework \cite{Armani2011}. The FRG-ChNM model is in agreement with both EoS up to densities as large as three times nuclear saturation density, indicating among other features that the many-body forces generated non-trivially in the FRG method appear to be reliable.

\subsection{Asymmetric nuclear matter and neutron matter}

\begin{figure}
	 \centerline{\includegraphics[width=8cm] {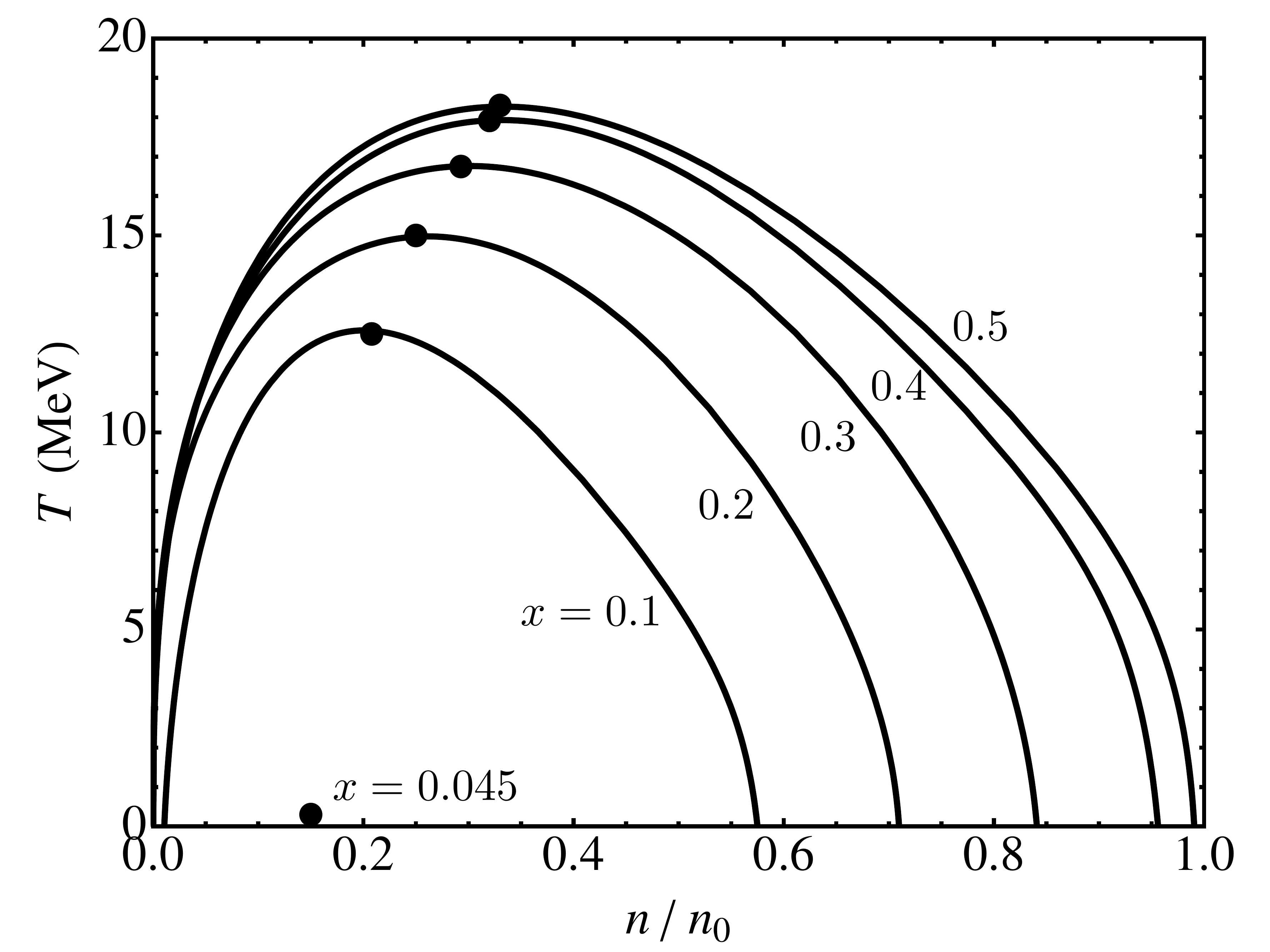}}
	\caption{The liquid-gas coexistence regions for different proton fractions $x = Z/A$. The trajectory of the critical point is shown as the dotted line.\label{fig:coexistence}}
\end{figure}

While symmetric nuclear matter is self-bound, pure neutron matter is unbound, and it is instructive to investigate the systematic change of the liquid-to-gas transition pattern as the fraction of protons, 
\begin{equation}
x =  {Z\over A} = {n_p\over n_p + n_n}~~,
\end{equation}
is reduced continuously. This is where the isospin dependence of the nuclear forces enters prominently, primarily from mechanisms involving the exchange of two (or more) pions, and from the short-range interaction $G_\tau(N^\dagger\boldsymbol\tau N)^2$.
\begin{figure}
	 \centerline{\includegraphics[width=7cm] {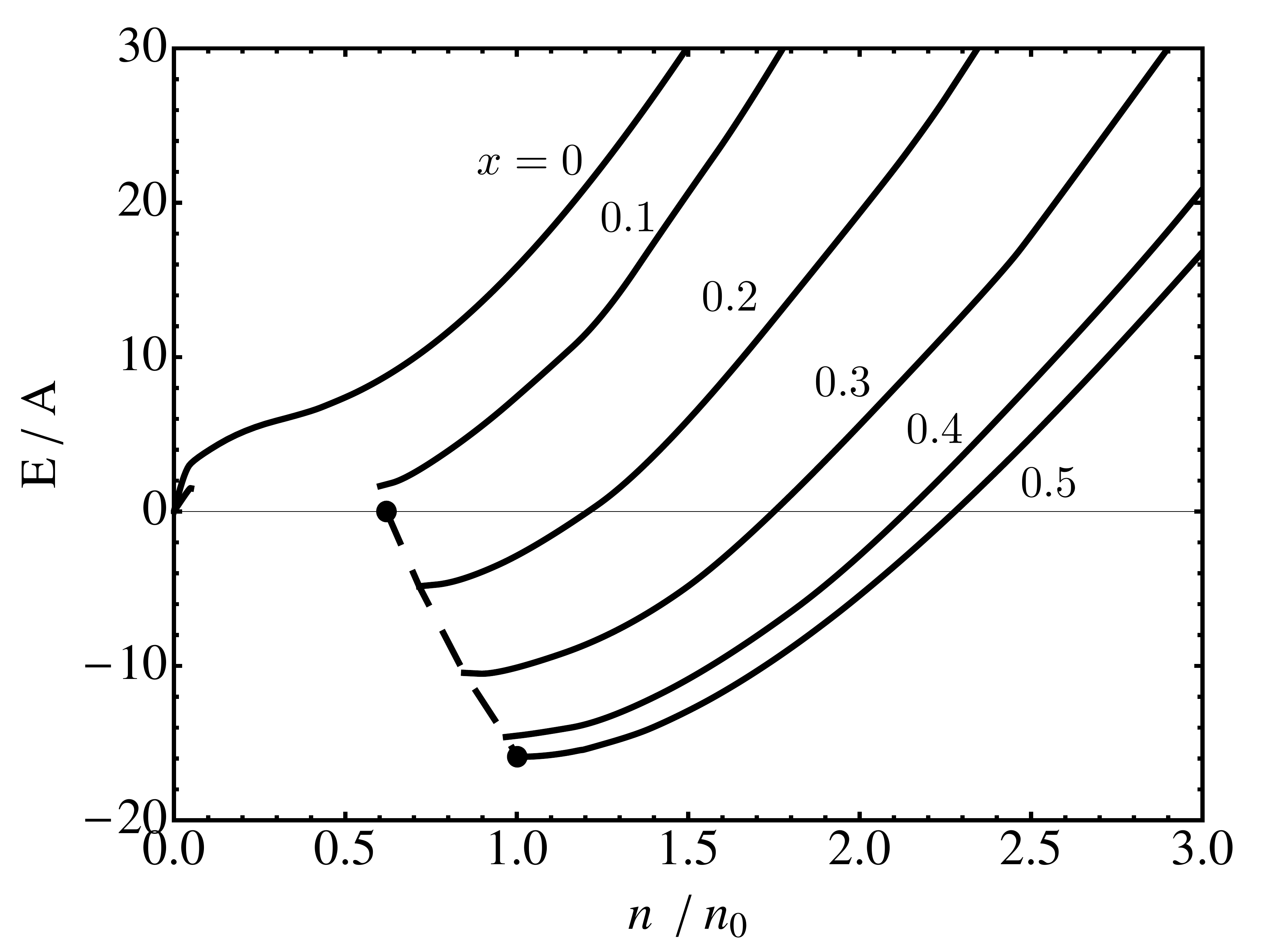}}
	\caption{The equation of state for different proton fractions $x= Z/A$ at vanishing temperature. The dashed curve denotes the absolute minimum of the energy per particle. The dotted line results from a Maxwell construction.}
	\label{fig:e_a}
\end{figure}
In Fig.\,\ref{fig:e_a} the energy per particle at $T = 0$ is shown as a function of baryon density for different proton fractions, $x = Z/A$, from symmetric nuclear matter ($x=0.5$) to pure neutron matter ($x=0$). As the proton fraction $x$ is lowered $E/A$ increases, until for $x\simeq0.11$ the energy per particle vanishes at the minimum (the upper endpoint of the dashed curve in Fig.~\ref{fig:e_a}). For even smaller values of $x$, the absolute minimum occurs at zero density: very neutron-rich matter is no longer self-bound. There is still a remnant of the first-order phase transition, and the density is still discontinuous as a function of the chemical potential. However, the coexistence region extends no longer down to vanishing density \mbox{$n=0$}. In Fig.\,\ref{fig:coexistence} the coexistence regions in a temperature/density-plot are shown for different proton fractions. For instance at $Z/A = 1/10$ the coexistence region starts at non-vanishing density determined by a Maxwell construction from the energy per particle, depicted as the dotted line in Fig.\,\ref{fig:e_a} for $x=0.1$. Finally, for $x$ smaller than a critical value of $x=0.045$ the energy per particle rises monotonously as a function of density. There is no longer a second minimum and the coexistence region vanishes altogether.
As the temperature increases, the phase coexistence region melts until it disappears at a certain $x$-dependent critical temperature characterized by a second-order critical endpoint. From the behavior of the coexistence regions one can again read off the critical endpoint for symmetric matter. The trajectory of the critical endpoint as it evolves with decreasing proton fraction $x$ is indicated by the dotted curve. 

Next we move to a more detailed discussion of pure neutron matter based on the FRG-ChNM approach. This is done in comparison with results from advanced many-body computations reported in the recent literature. Once again the emphasis is now on the isospin-dependence of the NN interaction. It arises prominently from pion exchange mechanisms (in particular, from isovector two-pion exchange), plus the short-distance term driven by the isovector-vector coupling strength $G_\tau$. We recall that $G_\tau$  is fixed to reproduce a symmetry energy  $E_{\rm{sym}}=32\,{\rm MeV}$. The $L$ parameter representing the slope of the symmetry energy (see Eq.~\eqref{eq:SL}) is $L=66.3\,{\rm MeV}$, close to the empirical range \mbox{$40\text{ MeV}\lesssim L \lesssim 62\text{ MeV}$} \cite{Lattimer2013}.
\begin{figure}
	 \centerline{\includegraphics[width=8cm] {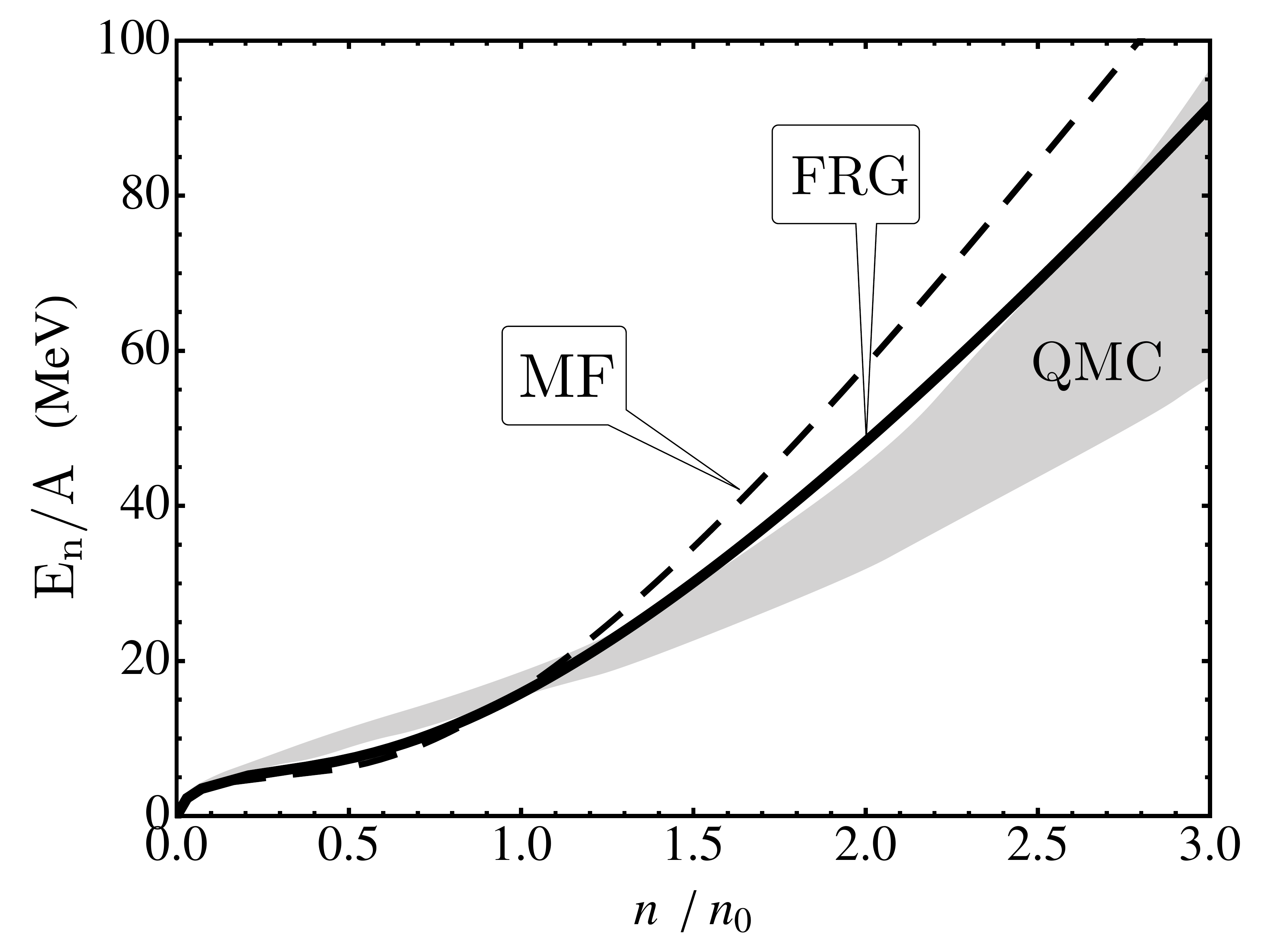}}
	\caption{The equation of state for pure neutron matter at $T=0$ with $E_{\text{sym}}=32\text{ MeV}$ in the mean-field approximation (MF) and with fluctuation (FRG) \cite{DW2015}. The grey band shows QMC results \cite{Gandolfi2012} with different spatial and spin structures of the three-neutron interaction, and with symmetry energies in the range $32.0\,{\rm MeV}\le E_{\rm sym} \le 33.7\,{\rm MeV}$.} 
\label{fig:n_e}
\end{figure}

Fig.\,\ref{fig:n_e} shows the energy per particle for neutron matter calculated in the FRG-ChNM model as a function of density (black line). In comparison, results obtained in a quantum Monte Carlo study with realistic two- and three-nucleon interactions \cite{Gandolfi2012} are also shown. Note that in contrast to the mean-field approximation which systematically deviates at higher densities, the FRG treatment of fluctuations improves significantly the comparison with ab-initio many-body calculations of $E_n/A$.

\begin{figure}
	 \centerline{\includegraphics[width=8cm] {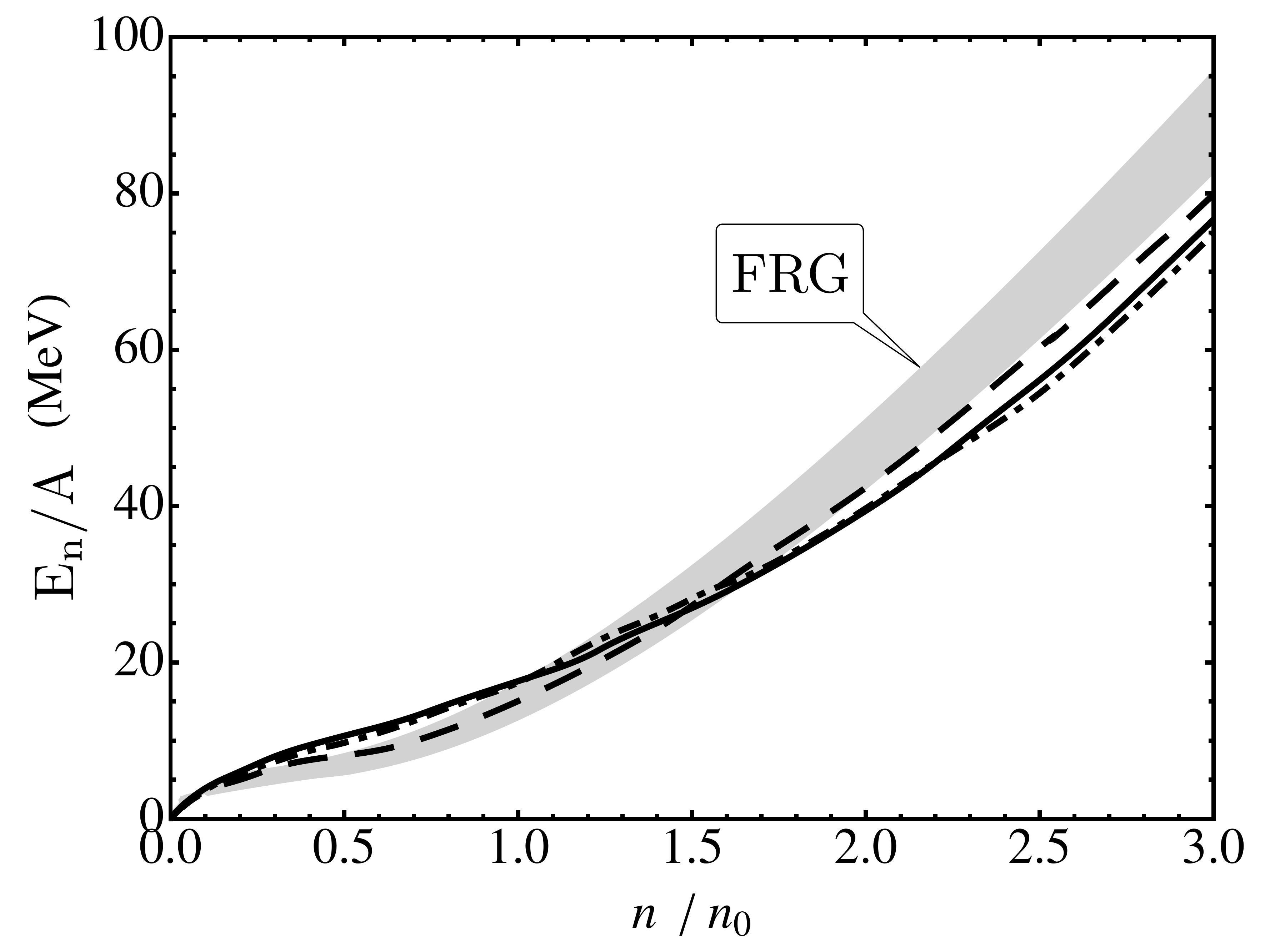}}
	\caption{The equation of state for pure neutron matter. The grey band encodes FRG results with symmetry energies in the range $29{\rm MeV}\le E_{\rm sym} \le 33\,{\rm MeV}$ \cite{DW2015}. Also shown for reference are predictions from ChEFT (full line, \cite{Hell2014}), QMC based on realistic potentials (dashed line \cite{Armani2011}), QMC based on chiral potentials (dotted line \cite{Roggero2014}), and the Akmal-Pandharipande-Ravenhall EoS (dashed-dotted line \cite{Akmal1998}).\label{fig:n_e_2}}
\end{figure}

Fig.\,\ref{fig:n_e_2} displays a band of calculated FRG-ChNM neutron matter equations of state covering a range of symmetry energies, $29\,{\rm MeV}\le E_{\rm sym} \le 33\,{\rm MeV}$, corresponding to an interval $0.9\,{\rm fm}^2\le G_\tau \le 1.2\,{\rm fm}^2$ of short-range isovector-vector couplings. Also shown for comparison are the APR equation of state based on phenomenological potentials \cite{Akmal1998}, results from chiral effective field theory \cite{Hell2014}, and different QMC computations using phenomenological \cite{Armani2011} and chiral potentials \cite{Roggero2014}. The equations of state obtained in the FRG-extended ChNM model agree quite well with all these results up to densities as high as $n=3\,n_0$. 

At this point it is of some interest to discuss remaining uncertainties in ChEFT calculations. Such uncertainties derive, for example, from limited quantitative knowledge of the three-body forces, from the range of momentum space cutoff scales used in defining the two-body NN input, and from missing higher-order multi-nucleon correlations. For symmetric and asymmetric nuclear matter (including thermodynamics), uncertainty estimates can be drawn from recent advanced perturbative ChEFT calculations such as those of Refs.\cite{KTHS2013,DHS2016,WHKW2014,WHK2015}.  For example, the energy per particle in neutron matter at $n \sim n_0$ is subject to an uncertainty of $\Delta E_n/A \simeq 4$ MeV, opening up to a band at higher densities that is comparable to the grey "QMC" area in Fig. \ref{fig:n_e}. For symmetric and asymmetric nuclear matter a detailed uncertainty analysis is reported in \cite{DHS2016} based on several N$^3$LO chiral nucleon-nucleon potentials combined with a variety of three-body interactions. It is demonstrated that the saturation energy and density for symmetric nuclear matter are correlated within a "Coester band", with a width of less than 3 MeV, that includes the empirically constrained $E/A(n_0) \simeq -(15.9\pm 0.4)$ MeV together with $n_0 = (0.16\pm 0.1)$ fm$^{-3}$. Calculated incompressibilties $K$ are in the range $K \sim 180 - 250$ MeV, and the typical uncertainty interval of the symmetry energy is $E_{\rm sym} \sim 30 - 35$ MeV.

\subsection{Constraints from neutron stars}

Matter in the interior of a neutron star is subject to charge neutrality and chemical beta equilibrium. Charge neutrality implies that electron and muon densities together with the proton density satisfy the condition 
\begin{align}
	n_p=n_e+n_\mu\,~.
\end{align}
Beta equilibrium implies that neutron, proton, and electron chemical potentials are related by
\begin{align}
	\mu_n = \mu_p+\mu_e~~.
\end{align}
Electrons and muons are thus added to the ChNM model and they are assumed to contribute as free Fermi gases to the energy density and pressure of the model. Charge neutrality and beta equilibrium leave only one single chemical potential as a remaining free parameter.  With these conditions, an equation of state applicable to the interior of a neutron star is readily calculated within the FRG-improved ChNM model. 
The resulting pressure versus energy density curve, $P(\varepsilon)$, falls well within the acceptable EoS band constrained by neutron star data as discussed in Refs. \cite{Ozel2016, Hell2014, Kurkela2014, Hebeler2013}.
Here we briefly examine how this figures in the mass-radius plot for neutron stars, especially in view of the existence of two heavy n-stars with accurately determined masses: J1614-2230 with $M=(1.97\pm0.04)\,M_\odot$ \cite{Demorest2010}, and J0348+0432 with $M=(2.01\pm0.04)\,M_\odot$ \cite{Antoniadis2013}.

Given the equation of state $P(\varepsilon)$, masses and radii of neutron stars\footnote{for a rcent review, see \cite{LP2016} and refs. therein} can be computed from the Tolman--Oppenheimer--Volkoff (TOV) equations,
\begin{align}
	\begin{aligned}
\frac{dP(r)}{dr}&=-\frac {\cal G}{r^2}\left[\varepsilon(r)+P(r)\right]{M(r)+4\pi r^3 P(r)\over 1-2{\cal G}M(r)/r}\,, \\
		&~~~~~~~\frac{dM(r)}{dr}=4\pi r^2\varepsilon(r)\,.
	\end{aligned}
\end{align}
where ${\cal G}$ is the gravitational constant and $r$ is the radial variable. With boundary conditions at $r = 0$, namely  $M(0)=0$ and $\varepsilon(0)=\varepsilon_c$ varying the central energy density $\varepsilon_c$, the mass-radius curve of the star, $M(R)$, is generated:
\begin{align}
	M=M(R)=4\pi\int_0^Rdr\;r^2\varepsilon(r)\,.
\end{align}
The outer crust of the neutron star consists of an iron lattice and hence the energy density in that region is $\varepsilon_{\rm Fe}=4.4\cdot 10^{-12}\,{\rm MeV/fm}^3$. The neutron star radius is then implicitly defined by the relation $\varepsilon(R)=\varepsilon_{\rm Fe}$. 

Moving inward from the crust, the nuclei become more neutron rich as the density increases and electrons are captured. The inner crust contains (possibly superfluid) neutrons. The crust is frequently parametrized by the Skyrme-Lyon (SLy) equation of state \cite{Baym1971,Douchin2001}. This SLy EoS is matched to the FRG-ChNM model at the point where the energy-density curves intersect, which happens at a density $n\simeq0.3\,n_0$. From there on to higher densities the FRG-ChNM equation of state is taken as a model for the neutron star core. 

We do not consider a possible crossover transition to quark matter (see, e.g., Refs.\,\cite{Hell2014,Masuda2016}), nor do we include other exotic types of matter such as kaon condensates. Hyperons are also not included as they would generally soften the equation of state unless strong additional repulsion is introduced for compensation \cite{Weissenborn2012, Lonardoni2015}. Recent developments in constructing hyperon-nucleon forces from chiral SU(3) effective field theory \cite{Haidenbauer2013, Petschauer2016a, Petschauer2016b} do indicate that such repulsive effects in $\Lambda$-nuclear interactions at high density can in fact emerge from the momentum dependence of the $\Lambda N$ interaction and possibly from $\Lambda NN$ three-body forces \cite{Petschauer2016c}.
\begin{figure}
	 \centerline{\includegraphics[width=8cm] {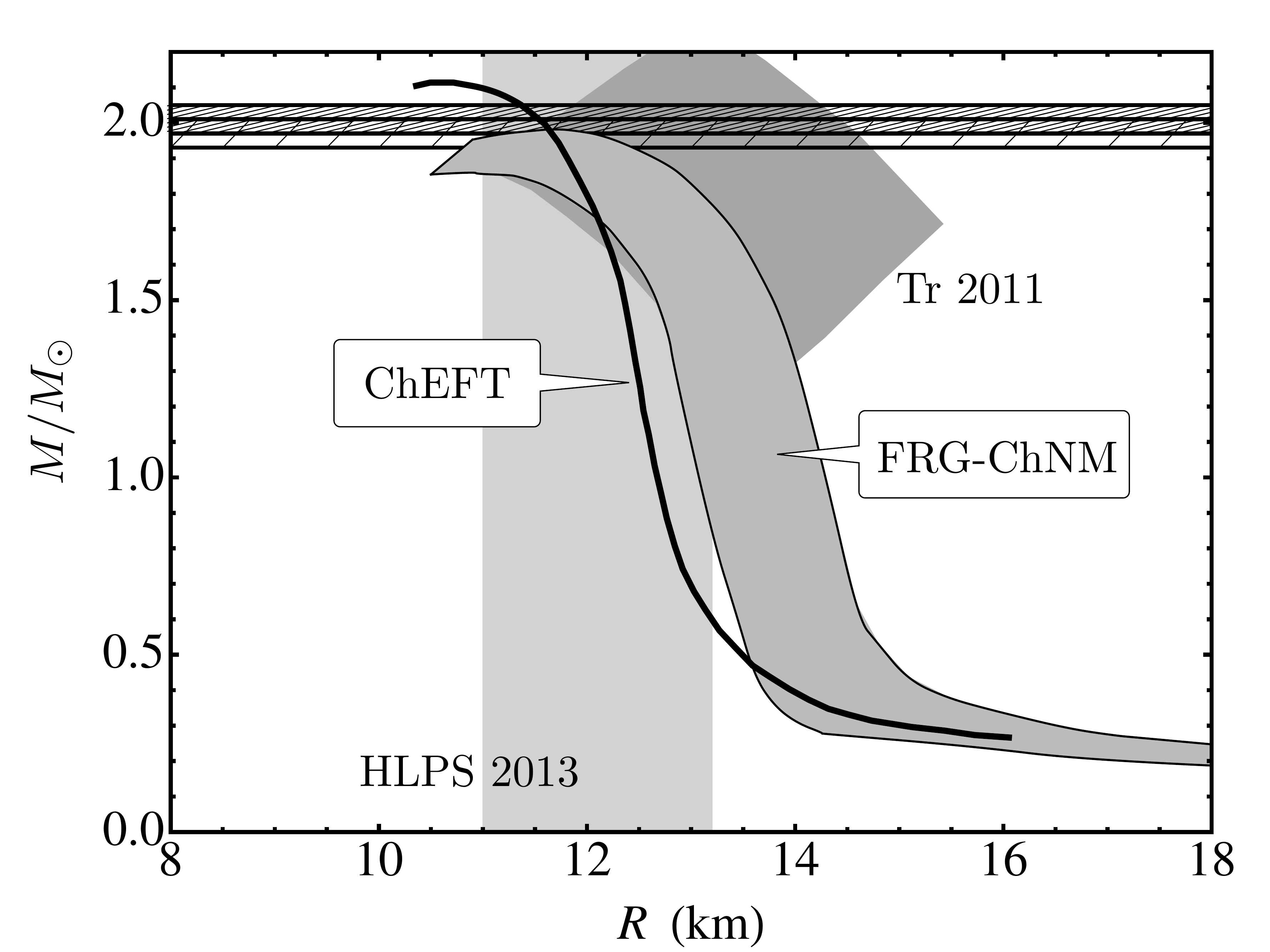}}
	\caption{Mass-radius relation of neutron stars. Grey band: FRG-ChNM calculation for a range of accptable symmetry energies \cite{DW2015}. Solid curve: ChEFT result \cite{Hell2014}. Constraints on radii, "Tr2011" from \cite{Truemper2011}, "HLPS2013" from \cite{Hebeler2013, Lattimer2014}, and from two-solar-mass neutron stars \cite{Demorest2010, Antoniadis2013} are shown for guidance.\label{fig:m_r}}
\end{figure}

Figure\,\ref{fig:m_r} shows the mass-radius relation solving the TOV equations with the FRG-ChNM equation of state. The grey band of mass-radius trajectories results from using symmetry energies in an empirically acceptable range $29\,{\rm MeV}\le E_{\rm sym}\le37$ MeV (or, correspondingly, a range of isovector-vector couplings $0.9\,\text{fm}^2\le G_\rho\le1.46\,{\rm fm}^2$). For not too small symmetry energies, the equation of state is found to be sufficiently stiff in order to support the observed two-solar-mass neutron stars. For comparison a ChEFT mass-radius trajectory is also shown in Fig.\,\ref{fig:m_r}. It is based on input yielding a symmetry energy of $E_{\rm sym} = 33.5$ MeV. Accepting a  wider range of symmetry energies, between 30 and 37 MeV, would introduce an uncertainty band similar to that of the FRG-ChNM case, with a band width of about $\Delta R \simeq 1$ km in the relevant mass range.

Unlike the precise $2\,M_{\odot}$ mass determinations, the empirical radius constraints for neutron stars are far less accurate. They are subject to model dependent assumptions. Nevertheless, limits on minimal and maximal radii can be inferred from different sources, such as X-ray burst oscillations, thermal emission, and stars with largest spin frequency. The result of such a detailed analysis \cite{Truemper2011} is a rhomboidal region (depicted in grey in Fig.~\ref{fig:m_r}), which should be intersected by a realistic equation of state. For comparison the acceptable radius interval according to Ref.~\cite{Lattimer2014} is also shown. The FRG-ChNM equation of state satisfies these constraints, as well as the EoS calculated using three-loop in-medium ChEFT \cite{Hell2014}. 

\begin{figure}
 \centerline{\includegraphics[width=7cm] {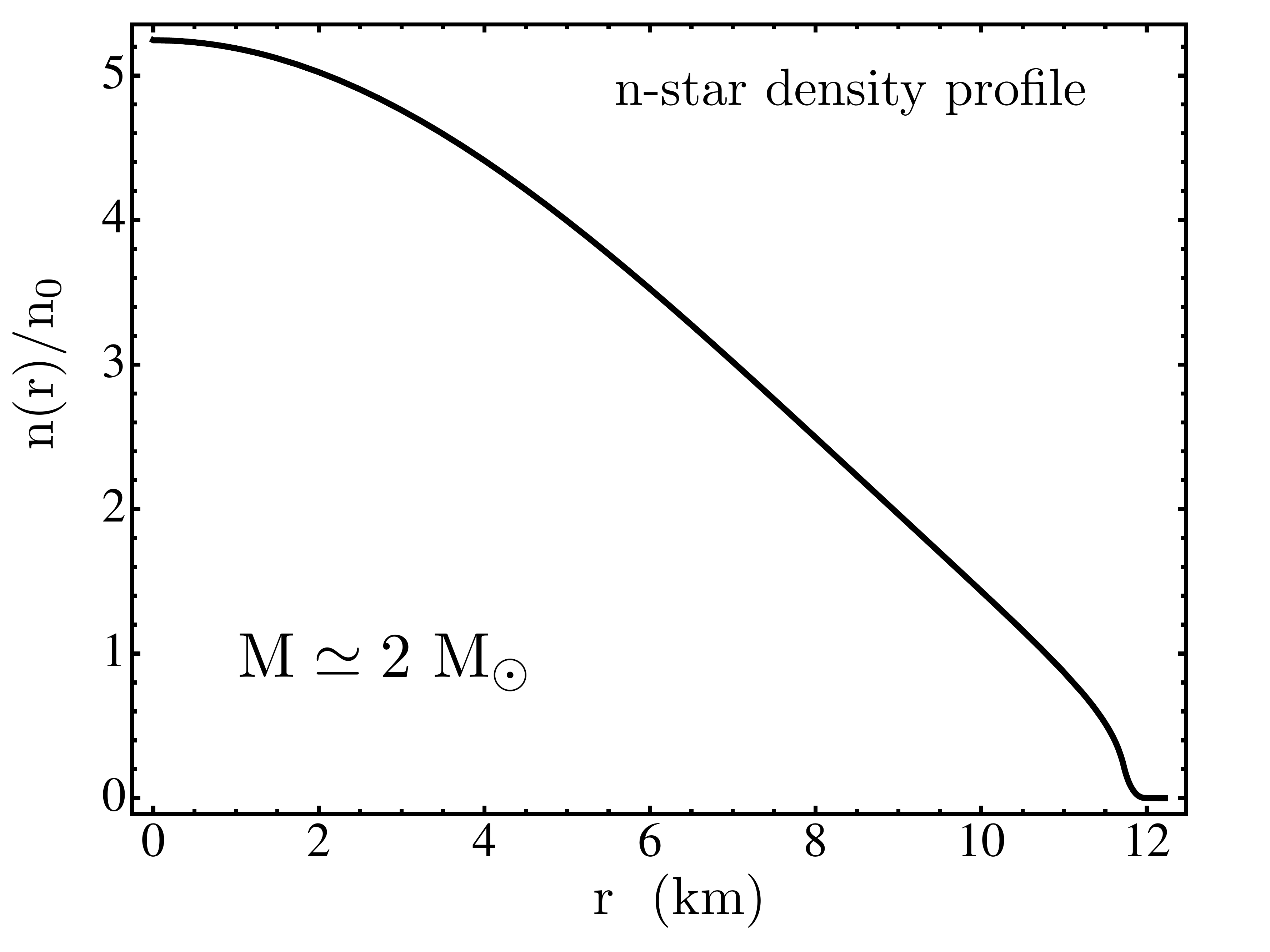}}	
	\caption{Density profile $n(r)$ (in units of nuclear matter equilibrium densiy $n_0$) for a two-solar-mass neutron star computed with the FRG-ChNM equation of state. \label{fig:n_r}}
\end{figure}

Figure\,\ref{fig:n_r} shows a typical calculated density profile of a neutron star with mass $M=1.97\,M_\odot$ using $G_\tau=1.46\,{\rm fm}^{-2}$ (i.e., a symmetry energy of 37 MeV). It is noteworthy that even in the center of the neutron star the density does not exceed about five times nuclear saturation density. The required stiffness of the EoS does not permit ultrahigh densities in the inner core of the star. These findings are in agreement with a corresponding ChEFT computation \cite{Hell2014}.

The question might nonetheless be raised whether approaches like the ChNM model, based on pion and nucleon degrees of freedom, are still applicable at densities as high as $5\,n_0$.
A necessary condition for this to work in such compressed matter is that the system remains in the hadronic phase of QCD with spontaneously broken chiral symmetry. This is the key question that will now be addressed in the following section.

\section{In-medium pion mass and chiral order parameter}

\subsection{Pion mass in nuclear matter}

\begin{figure}
 \centerline{\includegraphics[width=5cm] {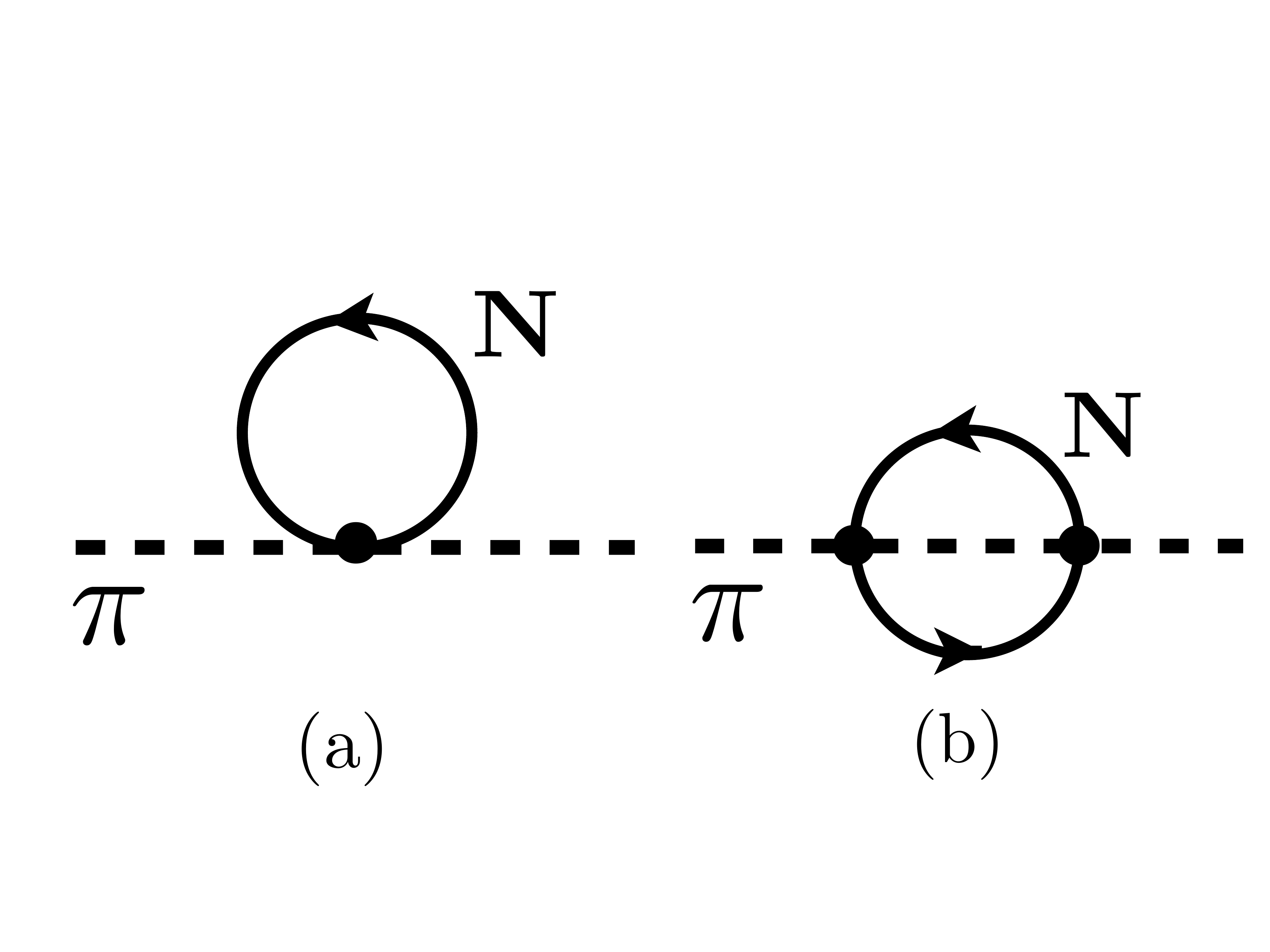}}	
	\caption{Processes contributing to s-wave pion-nucleon scattering in a nuclear medium. (a) Hartee (mean-field) term; (b) Next-to-leading order double-scattering term ("fluctuations"). \label{fig:pi-N}}
\end{figure}

\begin{figure}
	 \centerline{\includegraphics[width=8cm] {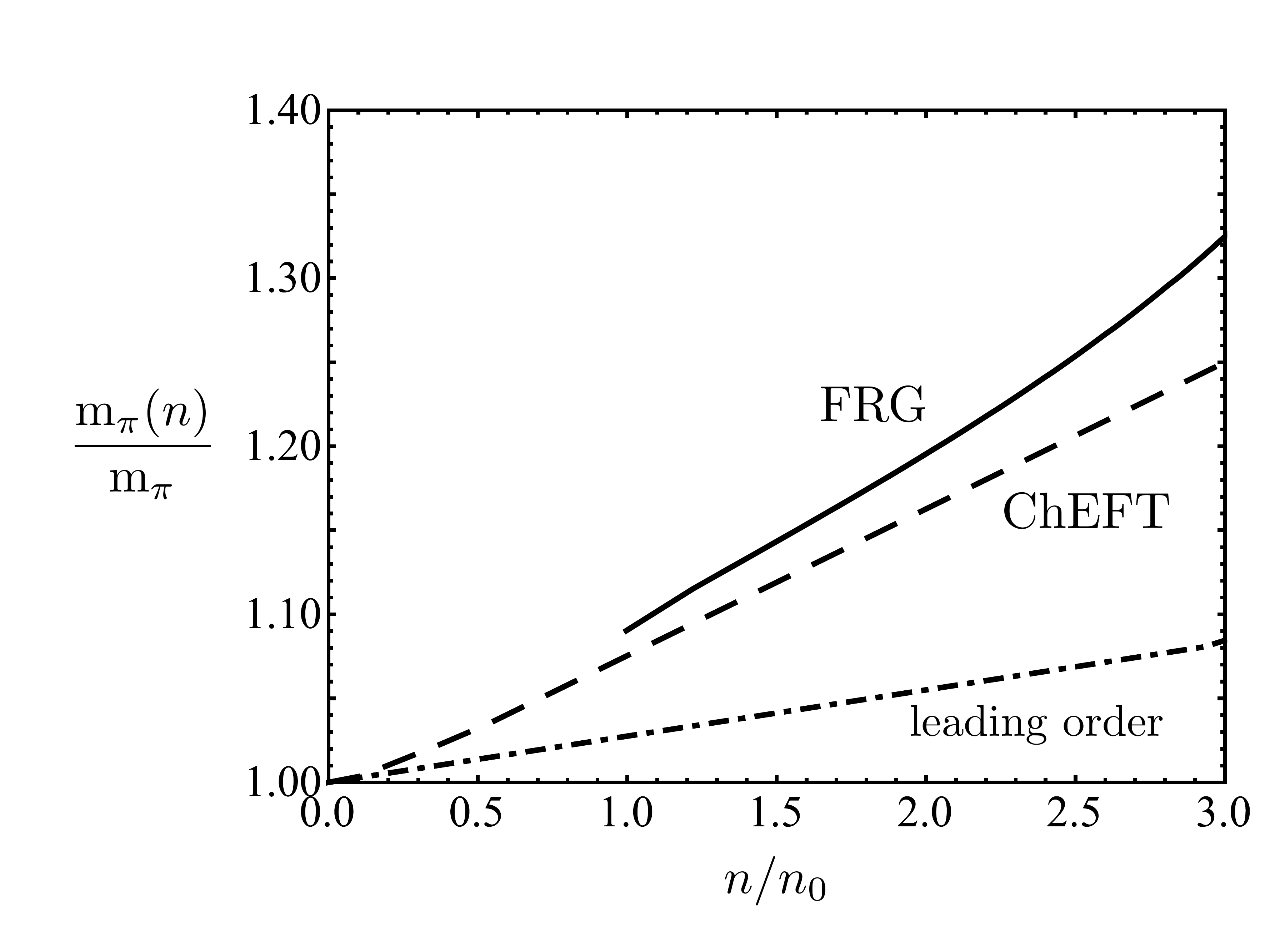}}
	\caption{In-medium pion mass (normalized to the vacuum mass) in symmetric nuclear matter at $T = 0$ as a function of baryon density. Solid line: FRG-ChNM calculation, dashed line: in-medium chiral perturbation theory (ChEFT) \cite{Goda2014}. Dash-dotted line: leading (linear) order in the density expansion.\label{fig:m_pi}}
\end{figure}

The pion mass can be extracted from Eq.~\eqref{eq:mpi} as 
\begin{align}
m_\pi^2=\frac{\partial {\cal U}_{k=0}}{\partial\chi}~,
\end{align}
where the right-hand side is evaluated at the minimum of the full effective potential at $k=0$. The density dependence of the pion mass plays an important role in low-energy pion-nuclear interactions \cite{EW1988}, e.g. in the analysis of deeply-bound pionic atoms based on the pion-nucleus optical potential \cite{Yamazaki1996,Kolomeitsev2003,KKW2005}. In the context of chiral effective theories and FRG this is an excellent test case for the role of pionic fluctuations. The threshold s-wave $\pi^-$-nuclear optical potential for isospin-symmetric nuclei is of the form
\begin{align}
V_{\rm opt} = -{2\pi\over m_\pi}\,b_0^{\rm eff}\,n~,
\end{align}
where the effective scattering length,
\begin{align}
b_0^{\rm eff}= b_0 - \left(b_0^2 + 2b_1^2\right)\langle 1/r\rangle~,
\end{align}
is dominated by the double scattering contribution involving the isospin-dependent s-wave parameter $b_1$ while the isospin-even parameter $b_0$ is small (in fact it vanishes in the chiral limit). The inverse correlation length associated with the propagating pion in the double scattering process is $\langle 1/r\rangle = 3p_F/2\pi$ for a gas of nucleons
with Fermi momentum $p_F$. Thus, the change of the pion mass in the medium, $\Delta m_\pi(n) \simeq V_{\rm opt}(n)$,
is governed by what the FRG scheme characterizes as pionic fluctuations (Fig.\,\ref{fig:pi-N}(b)), rather than being driven by the mean-field (Hartree) term (Fig.\,\ref{fig:pi-N}(a)) linear in the density $n$ and proportional to $b_0$. Empirically, $V_{\rm opt} \simeq 0.1\, m_\pi$ at $n \simeq n_0 = 0.17$ fm$^{-3}$ from the analysis of pionic atoms. 
The importance of the double-scattering contribution of order $n^{4/3}$ to the in-medium pion mass is, of course, realized also in the chiral effective field theory approach \cite{Ericson1966,Waas1997,Goda2014}. In Fig.\,\ref{fig:m_pi} we plot the FRG-ChNM model result for the pion mass as a function of density for symmetric nuclear matter at vanishing temperature. The non-trivial part of the corresponding curve starts at \mbox{$n = n_0$} because of the first-order liquid-gas transition. For comparison, the first-order (mean-field) approximation in the density expansion is shown, together with a recent in-medium chiral perturbation theory computation \cite{Goda2014}. In agreement with ChEFT and phenomenology, we find an enhancement of the pion mass by about ten percent at nuclear saturation density. In essence, this close correspondence between results of FRG-ChNM (a non-perturbative approach starting from a linear sigma model) and of ChEFT (a perturbative approach based on the non-linear sigma model) also demonstrates, a posteriori, that the FRG-ChNM produces the proper low-energy s-wave pion-nucleon scattering amplitude in vacuum.

Isospin breaking effects become significant when computing the in-medium pion mass for asymmetric nuclear matter. The masses of $\pi^+$, $\pi^-$ and $\pi^0$ split in such a medium \cite{Kaiser2001}. For example, the mass change for a $\pi^-$at leading order in the density (neglecting the small $b_0$ term) is now driven by the isospin-dependent parameter 
$b_1$: $\Delta m_\pi^{-}(n_n, n_p) \simeq -(2\pi/m_\pi)\,b_1\,(n_n-n_p)$, with $b_1 \simeq -0.1\, m_\pi^{-1}$. In neutron matter, the mass shift is repulsive for $\pi^-$ and attractive for $\pi^+$.

\subsection{Chiral symmetry restoration and order parameters}

At zero chemical potential chiral symmetry is restored in its Wigner-Weyl realisation at high temperatures $T > T_c \simeq 0.15$ GeV \cite{Borsanyi2010, Bazavov2012}. It is presumably also restored at large baryon chemical potential and low temperature, although the critical value of $\mu$ at which this transition takes place is unknown and may be extremely large. Calculations using Nambu--Jona-Lasinio or chiral quark-meson models commonly predict a first-order chiral phase transition  at vanishing temperature for quark chemical potentials, $\mu_q$, around 300\,MeV (see e.g., \cite{Fukushima2008,RRW2007,Schaefer2007,HRCW2009,Herbst2013}). Translated into baryonic chemical potentials, $\mu\simeq3\mu_q$, chiral symmetry would be restored close to the equilibrium point of normal nuclear matter, $\mu_c =923$ MeV. Nuclear physics with its well-established empirical phenomenology teaches us that this can obviously not be the case. However, these calculations -- apart from the fact that they operate with (quark) degrees of freedom that are not appropriate for dealing with the hadronic phase of QCD -- work mostly on the basis of the mean-field approximation. It is therefore of great importance to examine how fluctuations beyond mean-field can change this scenario. Using the FRG-ChNM approach, we shall indeed point out that the mean-field approximation cannot be trusted: it is likely that fluctuations shift the chiral phase transition to extremely high baryon densities.

Consider the chiral condensate, $\langle\bar{q}q\rangle$, as a function of temperature and baryon density. The Hellmann-Feynman theorem in combination with the Gell-Mann--Oakes--Renner relation gives the in-medium chiral condensate in the form
\begin{align}
	\frac{\langle\bar\psi\psi\rangle(n,T)}{\langle\bar\psi\psi\rangle_0} = 1-\frac{\partial\mathcal F(n,T)}{f_\pi^2\, \partial m_\pi^2}\,,
\end{align}
where ${\cal F}$ is the free-energy density. 
This condensate is proportional to the expectation value of the $\sigma$ field which takes over the role of a chiral order parameter in the ChNM model. Its vacuum value, $\sigma_0 = f_\pi$, turns into a function $f_\pi(n,T)$ of density and temperature\footnote{Here the in-medium pion decay constant refers to the one associated with the time component of the axial current}.

For isospin-symmetric matter, ChEFT calculations of the in-medium chiral condensate have been performed to three-loop order by directly computing the
dependence of ${\cal F}(n,T)$ on the pion mass \cite{Fiorilla2012,KHW2008}, while $\sigma/f_\pi$ as function of baryon density and temperature has been calculated in Refs.\,\cite{FW2012,DHKW2013} using the ChNM model. \begin{figure}
	\centering
	\begin{overpic}[width=0.48\textwidth]{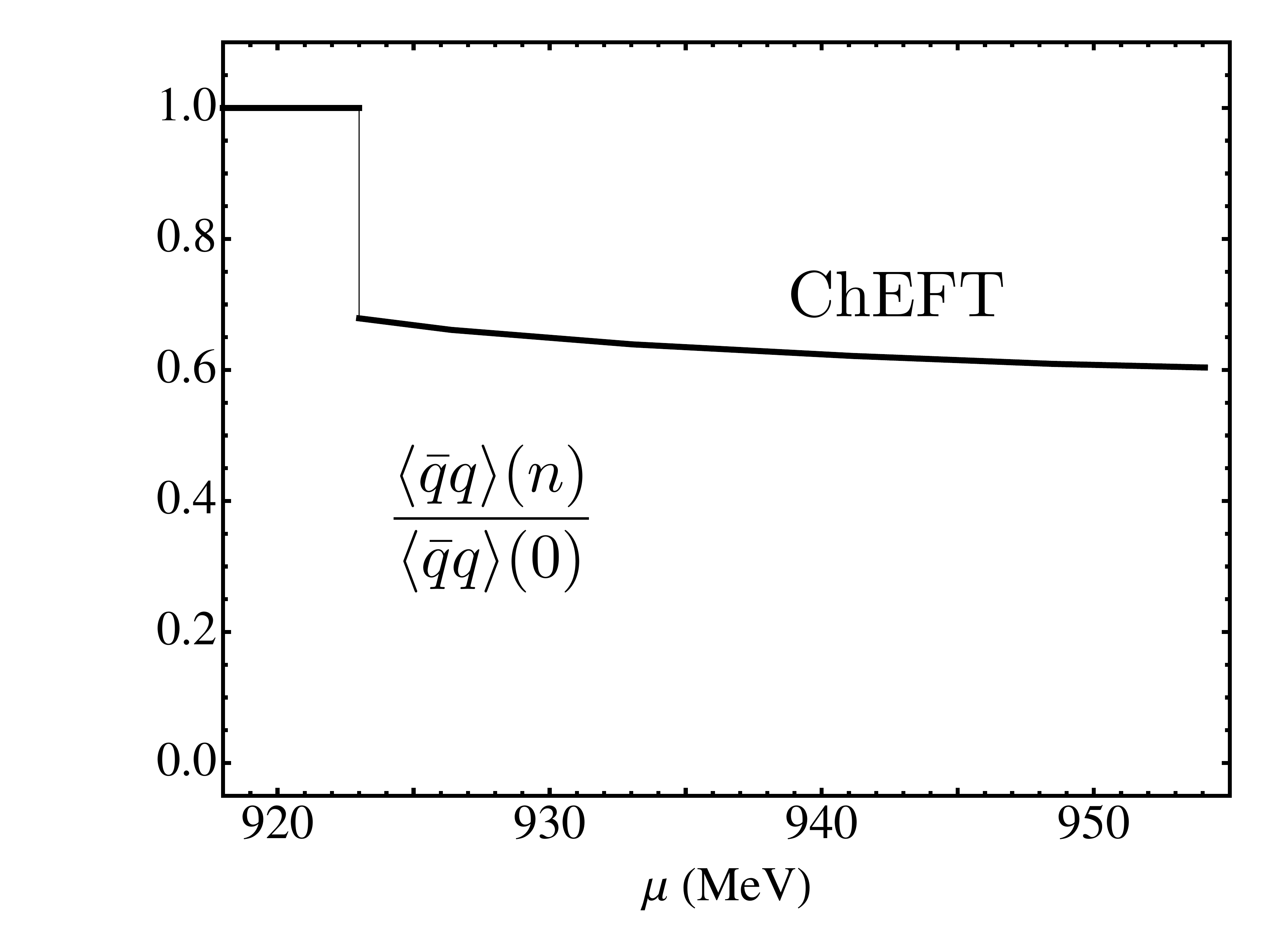}
	\end{overpic}
	\quad
	\begin{overpic}[width=0.48\textwidth]{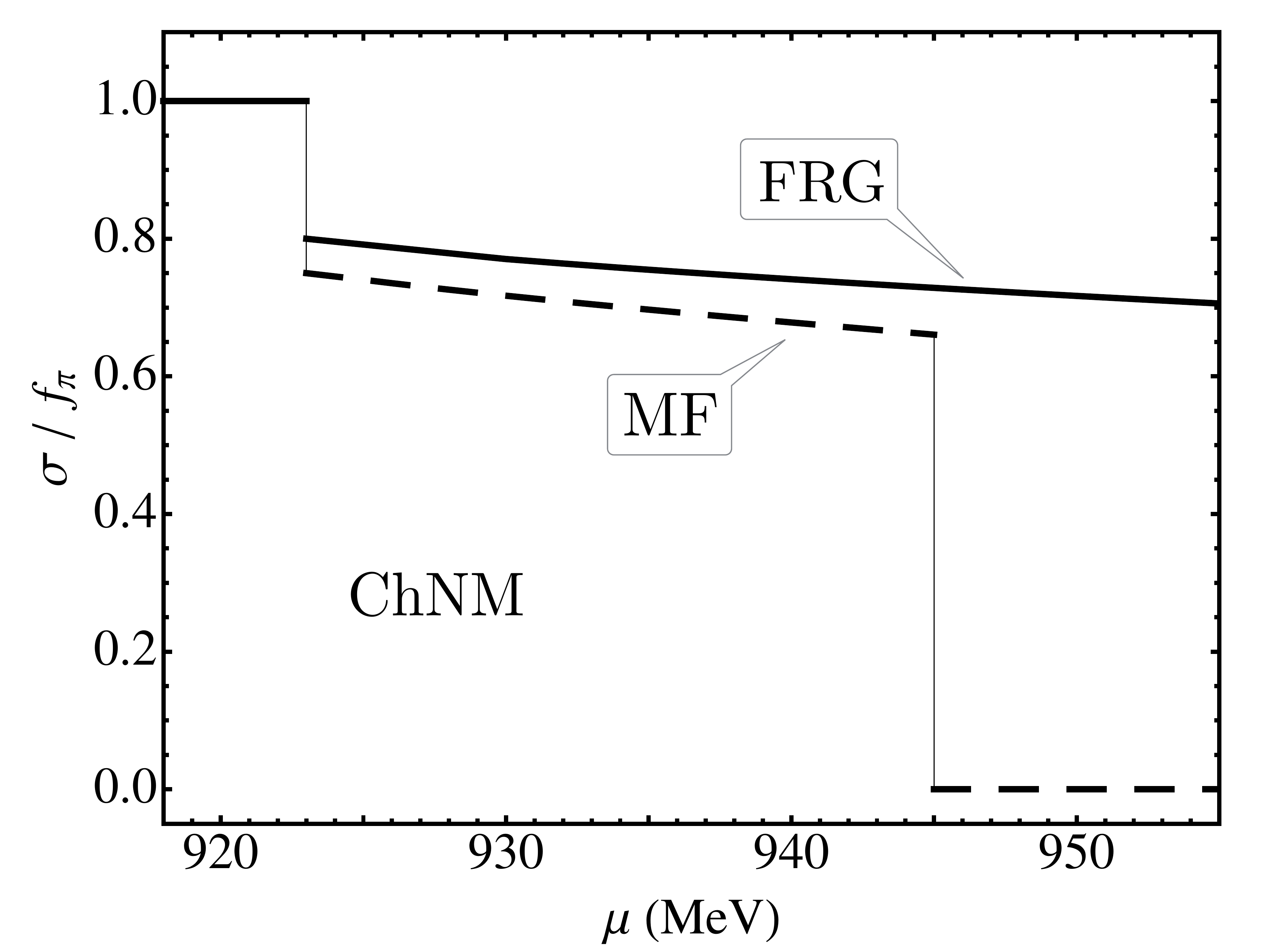}
	\end{overpic}
	\caption{Chiral condensate and sigma field in symmetric nuclear matter at $T=0$ as function of baryon chemical potential. Left: normalized chiral condensate computed in ChEFT\,\cite{Fiorilla2012}. Right: chiral order parameter of the ChNM model in mean-field approximation (MF, dashed line) and with fluctuations included (FRG, solid line) \cite{DHKW2013}.}
	\label{fig:SigmaMu}
\end{figure}
In mean-field approximation \cite{FW2012}, chiral symmetry gets restored in a first-order transition at a density as low as $n \simeq 1.6\,n_0$. Once fluctuations are taken into account \cite{DHKW2013}, the chiral restoration transition is shifted to chemical potentials well beyond $\mu \sim 1$ GeV or densities exceeding $3\,n_0$. Similar trends are found in ChEFT calculations of the in-medium chiral condensate. 

Results of both ChEFT and ChNM calculations are displayed in Fig.\,\ref{fig:SigmaMu}.
Note that the first-order liquid-gas transition leaves its trace also in the chiral order parameter, featuring a discontinuity at $\mu=\mu_c = 923$ MeV. The right-hand side of Fig.\,\ref{fig:SigmaMu} shows an instructive comparison between mean-field and full FRG results of the ChNM model. The first-order chiral transition seen in mean-field approximation at $\mu=945$ MeV (or $n=1.6\,n_0$) gets stabilized once fluctuations are included. On the left-hand side we show for comparison the chiral condensate as computed in ChEFT. One observes that both the chiral condensate in ChEFT and the chiral order parameter $\sigma$ in the FRG-ChNM model are in close correspondence and run almost in parallel.

\begin{figure}
	 \centerline{\includegraphics[width=7cm] {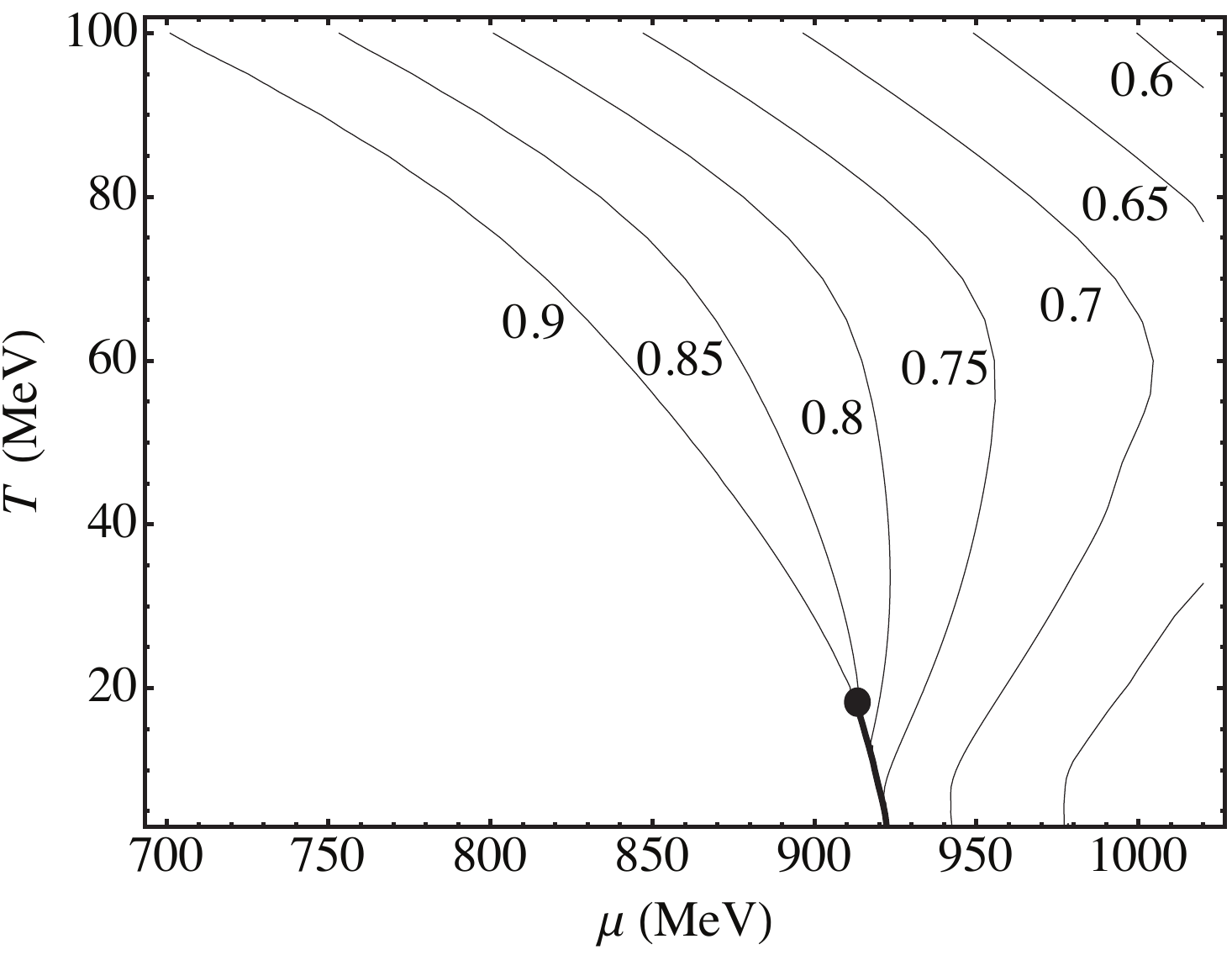}}
	\caption{Contour plots of the chiral order parameter $\sigma/f_\pi$ calculated in the FRG-ChNM model \cite{DHKW2013}.}
	\label{fig:sigmacontour}
\end{figure}

It is instructive to examine the order parameter $\sigma$ in the $T$--$\mu$ phase diagram in the neighbourhood of the liquid-gas transition. Figure\,\ref{fig:sigmacontour} shows contours for given values of $\sigma/f_\pi$ calculated in the FRG-ChNM model including fluctuations. It is evident that within the region of temperatures $T\lesssim 100$ MeV and baryon chemical potentials $\mu\lesssim 1$ GeV, the order parameter stays non-zero and there is no tendency towards a chiral phase transition.

\begin{figure}
	 \centerline{\includegraphics[width=7cm] {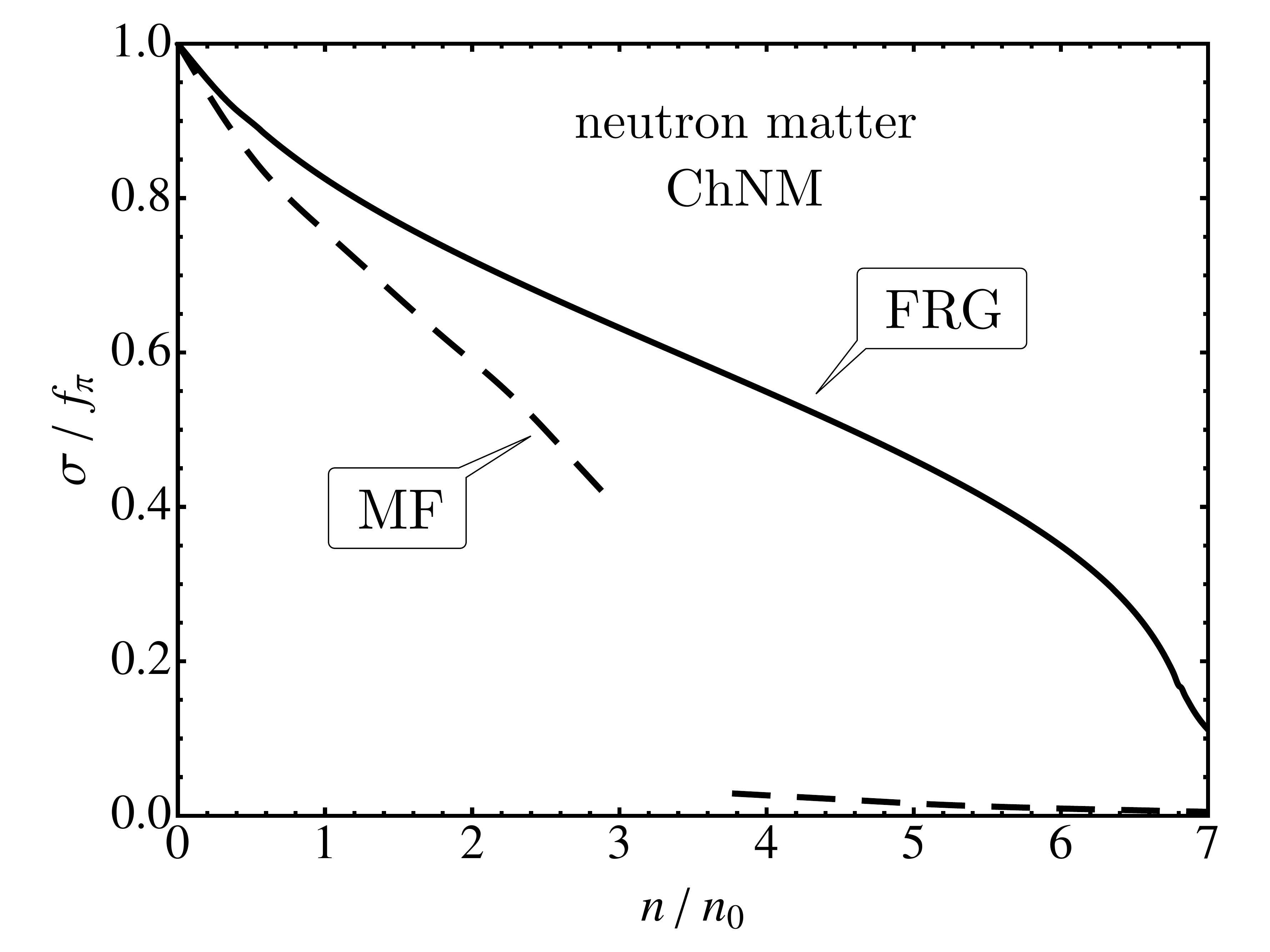}}
	\caption{Density dependence of the chiral order parameter for pure neutron matter at vanishing temperature \cite{DW2015}. Solid line (FRG): FRG-ChNM model calculation. Dashed line (MF): mean-field result.}
	\label{fig:chiral_restoration_NM}
\end{figure}

For pure neutron matter at $T=0$, the chiral condensate has been calculated previously within (perturbative) chiral effective field theory \cite{Kaiser2009,Krueger2013}. Chiral nuclear forces treated up to four-body interactions at N$^3$LO \cite{Krueger2013} were shown to work moderately against the leading linearly decreasing condensate with increasing density around $n \simeq n_0$, with a rate of stabilization less pronounced than in symmetric nuclear matter. It is instructive at this point to investigate the density dependence of the chiral order parameter $\sigma$ in neutron matter within the FRG-ChNM model. The non-perturbative FRG approach permits extrapolating to higher densities.  Results are presented in Fig.\,\ref{fig:chiral_restoration_NM}. In mean-field approximation, the order parameter $\sigma/f_\pi$ shows a first-order chiral phase transition at a density of about $3\,n_0$. However, the situation changes qualitatively when fluctuations are included using the FRG framework. The chiral order parameter now turns into a continuous function of density, with no indication of a phase transition. Even at five to six times nuclear saturation density the order parameter still remains at about forty percent of its vacuum value. Only at densities as large as $n \sim 7n_0$ does the expectation value of $\sigma$ show a more rapid tendency towards a crossover to restoration of chiral symmetry in its Wigner--Weyl mode. But this is even beyond the range of densities that may be reached in the cores of neutron stars. We thus observe a huge influence of higher order fluctuations involving Pauli blocking effects in multiple pion-exchange processes and multi-nucleon correlations at high densities. With neutron matter remaining in a phase of spontaneously broken chiral symmetry even up to very high densities, this encourages further-reaching applications and tests of the FRG-ChNM approach in constructing an equation of state for the interior of neutron stars. 

Both the present approach and the chiral effective field theory treatment rely on the assumption that, even in compressed baryonic matter, the proper baryonic degrees of freedom are nucleons rather than liberated quarks. This may in fact not be altogether unrealistic. Models based on chiral symmetry describe the nucleon as a compact valence quark core (typically with a radius of about 1/2 fm) surrounded by a mesonic cloud \cite{BRW1986}. The meson cloud determines most of the proton r.m.s. charge radius of 0.87 fm. For the neutron the description in terms of core and cloud is analogous except that the electric charges of quark core and meson cloud now add up to form an overall neutral object. A similar picture emerges in chiral soliton models of the nucleon \cite{KMW1987} which feature a compact baryonic core with a radius $\langle r^2\rangle_B^{1/2} \simeq 0.5$ fm. This half-fermi scale figures also in the repulsive core 
observed in Lattice QCD calculations of the nucleon-nucleon force \cite{HALQCD2012}. Now, even at densities of neutron matter as high as $n\sim 5\,n_0$, the average distance between two neutrons is still about 1 fm, hence the baryonic cores do not yet overlap appreciably at densities characteristic of neutron star interiors. Of course, the pionic fields surrounding the baryonic sources are highly inhomogeneous and polarized in compressed matter, but this effect is properly dealt with in the FRG treatment of pionic fluctuations. Important aspects of nucleon structure beyond the "pointlike" limit and respective changes in the strongly interacting medium enter in terms of nucleon polarizabilities involving low-lying $\pi N$ and $\pi\pi N$ excitations of the nucleon. It is therefore perhaps not so surprising that an equation of state based entirely on nucleons and pionic degrees of freedom works well for neutron stars, once repulsive correlations and many-body forces jointly contribute to producing a sufficiently stiff EoS .

\section{Summary and conclusions}  

In this report, a chiral model of pions interacting with nucleons (ChNM) has been combined with the functional renormalization group (FRG) as an efficient approach to the thermodynamics of nuclear and neutron matter. The effective potential of this model, non-linear in the chiral fields, has altogether five parameters that are constrained by known ground state properties of nuclear matter. Solving the FRG flow equations, a non-perturbative treatment of fluctuations beyond mean mean-field approximation is possible. Fluctuations of primary importance involve pionic degrees of freedom and nucleonic particle-hole excitations around the filled Fermi sea. These fluctuations incorporate one- and two-pion exchange mechanisms plus emergent three-body forces as they also appear in perturbative treatments using chiral effective field theory. The non-linear structure of the effective potential is even richer in content as it generates iterations to all order of such mechanisms, together with multi-nucleon interactions, when solving the FRG equations. The following is a list of key results: 
\begin{itemize}
\item{The behaviour of the nuclear liquid-gas phase transition has been studied in detail, including the systematic evolution of the phase diagram from symmetric to asymmetric nuclear matter and to neutron matter. The ChNM model produces a nuclear phase diagram that is remarkably close to the one found in perturbative calculations of the free-energy density using in-medium chiral effective field theory (ChEFT). } 
\item{A test case for the role of pionic and nuclear particle-hole fluctuations is the density dependence of the pion mass. In symmetric nuclear matter, the shift of the pion mass through s-wave pion-nuclear interactions is dominated by a charge-exchange double-scattering process whereas the leading (mean-field) term is suppressed. The FRG-driven ChNM calculation shows a close correspondence with an advanced calculation based on chiral perturbation theory (ChPT) in reproducing the emprical mass shift deduced from pionic atom data.}
\item{A further point of interest is the behavior of the chiral condensate, or equivalently, the expectation value of the $\sigma$ field (the in-medium pion decay constant) as a function of baryon density or chemical potential. Chiral models treated in mean-field approximation often display a first-order chiral phase transition not too far from the density of equilibrium nuclear matter. Fluctuations shift this transition upward to very high densities. For symmetric nuclear matter up to about three times the equilibrium density, the results of the non-perturbative FRG approach are  to in-medium ChPT: fluctuations involving Pauli principle effects and many-body forces  stabilize the chiral order parameter against restoration of chiral symmetry at too low densities. This effect is even more pronounced in compressed neutron matter. Whereas a mean-field calculation would suggest the appearance of a first-order chiral phase transition at about $n\sim 3\,n_0$, the full FRG treatment, with inclusion of fluctuations, stabilizes the Nambu-Goldstone phase of spontaneously broken chiral symmetry up to densities exceeding $5\,n_0$, indicating that hadronic degrees of freedom may persist even in the deep interior of neutron stars.}
\end{itemize}
Altogether, the present study underlines the crucial importance of respecting well-established nuclear physics constraints in the investigation of the QCD phase diagram in regions of compressed baryonic matter.
 
\vspace{1cm}
\noindent
{\bf Appendix: Computational details}
\vspace{0.2cm}

Here we briefly examine different methods  for solving Wetterich's equation numerically. We focus on the leading order in the derivative expansion in the local potential approximation. Even in this case, infinitely many operators are allowed in principle and therefore an infinite system of coupled differential equations would have to be solved. A truncation is necessary. One possible truncation is to perform a Taylor expansion of the potential up to a certain order around the renormalization-scale dependent minimum. Consider as a generic example the $O(N)$ model from Eq.~\eqref{eq:gamma_k_LPA} (with $Z_k\equiv1$ and $Y_k\equiv0$). The ansatz for the effective potential is
\begin{align}
	U_k(\chi)=\sum_{n=1}^{N_{\rm max}}\frac{a_{n,k}}{n!}(\chi-\chi_{0,k})^n\,.
\end{align}
with $\chi=\frac 12\phi_a\phi_a$ $(a = 1,\dots,N)$ the $O(N)$-invariant square of the fields $\phi_i$ and $\chi_{0,k}$ its value at the $k$-dependent minimum of the potential. The right-hand side of the flow equation~\eqref{eq:Wetterich} can also be expanded in a power series around $\chi_{0,k}$. Comparing the coefficients of both sides leads to flow equations for the couplings $a_{n,k}$ and for $\chi_{0,k}$. The convergence of this expansion can be checked by going to higher orders, i.e., by increasing $N_{\rm max}$.

The Taylor-expansion method is numerically fast. However, this method relies on the existence of a unique minimum. It breaks down at a first-order phase transition, such as the liquid-gas transition in nuclear matter.

An alternative approach that also works for first-order transitions is the {\it grid method} \cite{Adams1995}. This method has been used for producing the results reported in section \ref{sec:nuclFRG}. On the $\chi$-axis, a set of $N_{\#}$ grid points, $\chi_i$, is chosen. The grid points are distributed with constant spacing, $d=\chi_{i+1}-\chi_i$. Around each of these grid points, a local effective potential $U_{k,i}$ is defined on the interval $\chi_i-\frac d2\le\chi\le\chi_i+\frac d2$. Each potential is expanded up to the third power in $\chi$ (i.e., sixth powers in the fields) around its respective grid point, 
\begin{align}
	U_{k,i}(\chi) = \sum_{n=0}^3\frac{a_i^{(n)}(k)}{n!}(\chi-\chi_i)^n\,.
\end{align}
The full effective potential $U_k(\chi)$ is then obtained by gluing together these piecewise defined potentials. The flow equations for $a_i^{(0)}(k)$ are given by Wetterich's equation evaluated at $\chi_i$. Wetterich's equation involves the second derivative of the potential with respect to the fields, so the flow equation will be a function $f$ that involves first and second $\chi$-derivatives of the potential. The flow equation for $a_i^{(1)}(k)$ is obtained by differentiating the flow equation for $a_i^{(0)}(k)$ with respect to $\chi$. The emerging system of differential equations is
\begin{align}\label{eq:flow_grid}
		&\partial_ka_i^{(0)}(k)=f\big(k,\chi_i,a_i^{(1)}(k),a_i^{(2)}(k)\big)~~, \\
		&\partial_ka_i^{(1)}(k)=(\partial_\chi f)\big(k,\chi_i,a_i^{(1)}(k),a_i^{(2)}(k),a_i^{(3)}(k)\big)~~,
\end{align}
and so forth. This is so far a system of $2N_\#$ uncoupled differential equations, because the equation at grid point $i$ contains no information about the neighboring grid points. However, there are $4N_\#$ unknowns and the system is underdetermined. Additional constraints are introduced by matching the potentials at different grid points. The continuity condition for the grand canonical potential and its derivative imply that the potential $U_k(\chi)$ and its derivative $U_k^{(1)}(\chi)$ must be continuous as well, following the Gibbs-Duhem relation. Therefore the functions $U_{k,i}(\chi)$ and $U_{k,i}^{(1)}(\chi)$ must be matched between any two grid points. In total, there are $2(N_\#-1)$ matching conditions. In order to get a closed system of equations, two more equations are needed. These additional constraints are provided by matching the second derivative $U_{k,i}^{(2)}(\chi)$ for the outmost grid points, between $i=1$ and $i=2$ and between $i=N_\#-1$ and $i=N_\#$. In total, the complete matching conditions are:
\begin{align}
&\sum_{n=0}^3\frac 1{n!}\left(\frac d2\right)^n\Big(a_i^{(n)}(k)-(-1)^na_{i+1}^{(n)}(k)\Big)=0\,, ~{\rm for~ }i=1,\ldots,N_{\#}-1\,,\\
		&\sum_{n=0}^2\frac 1{n!}\left(\frac d2\right)^n\Big(a_i^{(n+1)}(k)-(-1)^na_{i+1}^{(n+1)}(k)\Big)=0\,, ~{\rm for~ }i=1,\ldots,N_{\#}-1\,, \\
		&\sum_{n=0}^1\frac 1{n!}\left(\frac d2\right)^n\Big(a_i^{(n+2)}(k)-(-1)^na_{i+1}^{(n+2)}(k)\Big)=0\,,~{ \rm for~ }i=1 ~{\rm and~ } i=N_{\#}-1\,.
\end{align}
The set of constraint equations can be written as a system of linear equations,
\begin{align}\label{eq:omegamatrix}
	A\begin{pmatrix} a_1^{(0)}(k) \\ \vdots \\ a^{(0)}_{N_{\#}}(k) \\ a_1^{(1)}(k) \\ \vdots \\ a_{N_{\#}}^{(1)}(k) \end{pmatrix}=B\begin{pmatrix} a_1^{(2)} (k)\\ \vdots \\ a^{(2)}_{N_{\#}}(k) \\ a_1^{(3)}(k) \\ \vdots \\ a_{N_{\#}}^{(3)} \end{pmatrix},
\end{align}
where $A$ and $B$ are $(2N_\#\times2N_\#)$-matrices. The matrices $A$ and $B$ are complicated for larger values of $N_\#$ and we do not write them down explicitly. However, the matrix $B$ can easily be inverted using computer algebra. The second and third derivatives of the potential, $a^{(2)}_i(k)$ and $a^{(3)}_i(k)$, can then be expressed as functions of zeroth and first derivatives, $a^{(0)}_i(k)$ and $a^{(1)}_i(k)$, respectively. We call these functions $g_i$ and $h_i$ for each $i=1,\ldots,N_\#$:
\begin{align}
		a_i^{(2)}(k)&=g_i\big(\big\{a_j^{(0)}(k),\,a_j^{(1)}(k)\big\}_{j=1,\ldots,N_\#}\big), \\
		a_i^{(3)}(k)&=h_i\big(\big\{a_j^{(0)}(k),\,a_j^{(1)}(k)\big\}_{j=1,\ldots,N_\#}\big).
\end{align}
The functions $g_i$ and $h_i$ depend only linearly on their arguments. The coefficients take their maximal values at grid points $j=i$ and adjacent points, whereas the influence of more distant grid points decreases drastically. The $2N_\#$ functions $g_i$ and $h_i$ allow us to eliminate $a_i^{(2)}(k)$ and $a_i^{(3)}(k)$ from the flow equations~\eqref{eq:flow_grid}. The resulting set of coupled differential equations for $a_i^{(0)}(k)$ and $a_i^{(1)}(k)$ can be solved using efficient numerical routines such as Runge-Kutta algorithms. As an initial condition one must provide the potential and its first derivative at the UV scale, i.e., $a_i^{(n)}(k = \Lambda)$, for $n=1,2$ and $i=1,\ldots,N_{\#}$.

The equations are to be integrated from $k=\Lambda$ down to $k=0$. In practice it is not possible to integrate exactly down to $k=0$ because of numerical instabilities. An infrared cutoff $k_{\rm IR}$ needs to be introduced. If $k_{\rm IR}$ is chosen sufficiently small, the location of the minimum does not change any more and the physical predictions are left unaltered.

The grid method has the advantage that it determines the potential not only around its minimum as in the Taylor-expansion method, but globally as a function of $\chi$.  As a downside the numerical evaluation is considerably more costly	 compared to the Taylor-expansion approach. Nonetheless, the grid method is mandatory for studying situations in which a first-order phase transition can occur, and we have used it throughout for the explicit calculations presented in this rewiew.

\vspace{1cm}
\noindent
{\bf Acknowledgements}
\vspace{0.1cm}

This work was supported in part by the DFG Cluster of Excellence Origin and Structure of the Universe. One of the authors (W. W.) gratefully acknowledges the hospitality of the Kavli Institute for Theoretical Physics where this manuscript was finalized, with partial support by the National Science Foundation under grant No. PHY11-25915. He thanks Evgeny Epelbaum, Barry Holstein and Andreas Wirzba for stimulating discussions.





\end{document}